\documentclass[iop]{emulateapj}

\newcommand{\Mm}{{\mathrm{\, Mm}}}
\newcommand{\kms}{{\mathrm{\, km \; s^{-1}}}}

\newcommand{\totalevents}{{196}}
\newcommand{\nflare}{{72}}
\newcommand{\npe}{{11}}
\newcommand{\njet}{{1}}
\newcommand{\nmoreton}{{1}}
\newcommand{\nunk}{{111}}
\newcommand{\neruption}{{9}}
\newcommand{\nlao}{{90}}
\newcommand{\nsao}{{106}}

\newcommand{\nsaosexcited}{{32}}
\newcommand{\nlong}{{163}}
\newcommand{\ntrans}{{33}}

\newcommand{\listampli}{{6, 7, 38, 65, 108, 134, 152, 156,  \text{and} \,171}}

\newcommand{\nar}{{45}}
\newcommand{\nin}{{99}}
\newcommand{\nhe}{{52}}

\usepackage{amsmath}
\usepackage{float}
\usepackage[colorlinks = true,
            urlcolor  = blue,
            citecolor = blue,
            anchorcolor = blue]{hyperref}

\usepackage{xcolor}
\definecolor{olive}{rgb}{0,0.6,0}

\shorttitle{GONG Catalog of Solar Filament Oscillations}
\shortauthors{Luna et al.}

\begin{document}

\title{GONG Catalog of Solar Filament Oscillations Near Solar Maximum}

\author{M. Luna\altaffilmark{1,2}, J. Karpen\altaffilmark{3}, J. L. Ballester\altaffilmark{4,5}, K. Muglach\altaffilmark{3,6}, J. Terradas\altaffilmark{4,5}, T. Kucera\altaffilmark{3} and H. Gilbert\altaffilmark{3}}

\altaffiltext{1}{Instituto de Astrof\'{\i}sica de Canarias, E-38200 La Laguna, Tenerife, Spain}
\altaffiltext{2}{Departamento de Astrof\'{\i}sica, Universidad de La Laguna, E-38206 La Laguna, Tenerife, Spain}
\altaffiltext{3}{NASA Goddard Space Flight Center, Greenbelt, MD 20771, USA}
\altaffiltext{4}{Departament de F{\'i}sica, Universitat de les Illes Balears (UIB), E-07122 Palma de Mallorca, Spain}
\altaffiltext{5}{Institute of Applied Computing \& Community Code (IAC$^3$), UIB, Spain}
\altaffiltext{6}{Catholic University of America, Washington, DC 20064, US}

\begin{abstract}
We have catalogued $\totalevents$ filament oscillations from the GONG $H{\alpha}$ network data during several months near the maximum of solar cycle 24 (January - June 2014). Selected examples from the catalog are described in detail, along with our statistical analyses of all events. Oscillations were classified according to their velocity amplitude:  $\nsao$ small-amplitude oscillations (SAOs), with velocities  $<10\kms$, and $\nlao$ large-amplitude oscillations (LAOs), with velocities $>10\kms$. Both SAOs and LAOs are common, with one event of each class every two days on the visible side of the Sun. For nearly half of the events we identified their apparent trigger. The period distribution has a mean value of 58$\pm$15 min for both types of oscillations. The distribution of the damping time per period peaks at $\tau/P=1.75$ and $1.25$ for SAOs and LAOs respectively. We confirmed that LAO damping rates depend nonlinearly on the oscillation velocity. The angle between the direction of motion and the filament spine has a distribution centered at $27^\circ$ for all filament types. This angle agrees with the observed direction of filament-channel magnetic fields, indicating that most of the catalogued events are longitudinal (i.e., undergo field-aligned motions). We applied seismology to determine the average radius of curvature in the magnetic dips, $R\approx89$ Mm, and the average minimum magnetic-field strength, $B\approx16$ G. The catalog is available to the community online, and is intended to be expanded to cover at least 1 solar cycle.

\end{abstract}

\section{Introduction}\label{sec:intro}

Filament oscillations were first observed visually \citep[Greaves, Newton, \& Jackson, reported by][]{dyson1930, newton1935, bruzek1951}, followed by  photographic observations that revealed a significant relationship with flares  (\citet{dodson1949,bruzek1957} and \citet{becker1958}). \citet{moreton1960} confirmed that wave disturbances initiated during the impulsive phase of flares were responsible for triggering prominence oscillations both near and far from the flare. \citet{ramsey1966} determined  periods and damping times for several oscillating filaments, but did not find any correlation between the period or damping time and  the dimensions of the filament, the distance to the associated flare, or its size. In these early observations, some events were called ``winking filaments" because these filaments were visible in $H{\alpha}$ when they were at rest, but disappeared while oscillating. Because these observations were made with  narrow-band $H{\alpha}$ filters, Doppler-shifted absorption from prominence material traveling at sufficiently large line-of-sight (LOS) velocities ($>23\kms$) fell outside the 0.5\AA\ bandpass of the filter and thus became invisible in $H{\alpha}$. 

Nowadays, thanks to both space- and ground-based instruments, observations of large-amplitude filament oscillations (LAOs: $v > 10\kms$) have become common. (The terms ``large" and ``small" amplitude are defined later in this Section.) The exciters identified thus far include Moreton or EIT waves \citep{eto2002,okamoto2004,gilbert2008,asai2012}, EUV waves \citep{liu2012, shen2014a,xue2014,takahashi2015}, shock waves \citep{shen2014b}, nearby jets, subflares and flares \citep{jing2003, jing2006,vrsnak2007,Li2012}, and the eruption of the filament \citep{isobe2006,isobe2007, pouget07,chen2008,foullon2009,bocchialini2011}. 

Many of the observed flare-induced LAOs in filaments exhibit motions in different directions relative to the axial magnetic field (polarization). For instance, the material can undergo vertical \citep{eto2002,okamoto2004, shen2014a}, horizontal  \citep{kleczek1969,hershaw2011,gosain2012,liu2012,shen2014b}, or longitudinal (field-aligned)  \citep{jing2003,jing2006,vrsnak2007,Li2012,zhang2012,luna2014,shen2014b} motions. Oscillations with a mixed character \citep{gilbert2008} have also been observed.

The first theoretical models proposed to explain the excitation, restoring forces, and damping mechanisms of large-amplitude longitudinal oscillations were purely analytical \citep{hyder1966,kleczek1969}. One-dimensional, hydrodynamic, numerical models have been employed successfully to describe longitudinal oscillations \citep{vrsnak2007,luna2012b,luna2012c,zhang2012,zhang2013,ruderman2016,zhou2017}, while 2D and 3D MHD models have described more completely the features of observed longitudinal and transverse oscillations \citep{terradas2013,terradas2015,terradas2016,luna2016}.

Spectroscopic techniques have revealed oscillations with much smaller peak velocities than those of LAOs, with amplitudes from the noise level of $0.1\kms$ to $10\kms$. \citet{harvey1969} first measured oscillatory periods between 1 and 17 min, while later observations yielded characteristic periods ranging from a few to 90 min. Although the triggering mechanisms of these small-amplitude oscillations (SAOs) have not been clearly identified, they are generally believed to be excited by the periodic motions of filament magnetic fields driven by photospheric or chromospheric oscillations \citep[see review by][]{arregui2012}.

A variety of approaches has been used to categorize and understand filament oscillations. The simplest are based on a single property, such as the peak velocity \citep[e.g.,][]{arregui2012}, the nature of the trigger \citep[e.g.,][]{oliver1999,oliver2002}, or the period \citep[e.g.,][]{arregui2012}. The apparent tendency of periods to group below 10 min, in the range 10 - 40 min, or 40 - 90 min \citep{arregui2012} led to classifications denoted as short-, intermediate- and long-period oscillations, respectively. Very short-periods below 1 min \citep{balthasar1993}, very long-periods above 5 hours \citep{foullon2004,pouget2006}, and even periods longer than 20 hours \citep{efremov2016} have been reported. Classification based only on the period does not reflect the nature, origin, or exciter of the oscillations, however.  More complex schemes have proven to be difficult to employ consistently \citep[e.g.,][]{vrsnak1993}.

Because oscillation velocities have been measured from the observable threshold to 100 $\kms$, the velocity amplitude alone is not the most definitive criterion by which oscillation events can be categorized. In spite of these limitations, a widely accepted, velocity-based division between small-amplitude and large-amplitude oscillations has proven to be both convenient and physically justifiable. We can relate the observed oscillation amplitudes to their linear or nonlinear character by considering the characteristic Alfv\'en and sound speeds in prominences, which are of the order of 100$\kms$ and 10$\kms$, respectively.  Therefore, oscillations with velocity amplitudes above 10$\kms$ exceed the local sound speed, and hence can be considered nonlinear oscillations, while smaller velocity amplitudes would be linear. In general, small-amplitude oscillations (SAOs) exhibit amplitudes below 10$\kms$, are not related to flare activity, are local, and can be appropriately analyzed or modeled using methods of linear perturbations. LAOs are usually associated with energetic events, are of global character, and as the velocity amplitude is $\ge$10-20$\kms$, require a nonlinear approach. As we demonstrate in the present work, however, exceptions to these ``rules'' exist. 

Prominence seismology aims to determine physical parameters that are difficult to measure by direct means in these magnetized plasma structures. This remote diagnostics method combines observations of oscillations and waves in these structures with theoretical results from the analysis of oscillatory properties of given prominence models, as first suggested by \citet{tandberg-hanssen95}. The first seismological determinations of magnetic field strength in winking filaments used a simple model of longitudinal motions based on a harmonic oscillator \citep{hyder1966,kleczek1969}.
\citet{vrsnak2007} analyzed large-amplitude longitudinal oscillations (LALOs) in a prominence to infer the Alfv\'en speed; assuming the mass density of the prominence plasma, they also determined the azimuthal and axial magnetic field strengths. Our theoretical investigation of oscillations in simulated prominence threads strengthened the foundations of the damped harmonic oscillator model for LALOs, providing a basis for applications to observations \citep{luna2012b,luna2012c}. Subsequent seismological analyses of LALOs in prominences have derived the radius of curvature of dipped field lines supporting prominence threads, the minimum magnetic field strength, the energy injected by the triggering jet, and the mass accretion rate according to the thermal nonequilibrium model \citep{Li2012,bi2014,luna2014,luna2016,zhang2017}. Using the same seismological techniques, we determined the curvature radius of the magnetic field dips and the minimum field strength from the largest prominence oscillation ever reported in the literature; these results were validated by reconstructing the filament magnetic field from the photospheric field in combination with the flux-rope insertion method \citep{luna2017}. 

To interpret observed prominence LAOs directed transverse to the magnetic field, an MHD approach is required. Some observations of oscillatory behavior have been interpreted and analyzed as global or standing kink modes \citep[e.g.,][]{hershaw2011,liu2012,xue2014}. A theoretical analysis predicted a linear relationship between the damping time ($\tau$) and the period ($P$) that could be compatible with resonant absorption as the damping mechanism \citep{ruderman2002,ofman2002,arregui2008}. However, this interpretation must be considered with care because the use of scaling laws to discriminate between damping mechanisms is questionable, at least for resonant absorption \citep{arregui2008a}. Much work remains before the physical models of both longitudinal and transverse LAOs are sufficiently detailed and comprehensive to adequately link theory and simulations with observed prominence motions. 

To date, all studies of oscillating prominences have been focused on one or, at most, a few episodes. In order to understand this phenomenon thoroughly and derive key physical characteristics via seismology of all types of prominences over the solar cycle, we have begun to compile a systematic, large data set of oscillation events. Thus far we have identified and analyzed $\totalevents$ events during several months close to the maximum of solar cycle 24, using GONG $H{\alpha}$ data. We found that LAOs are very common on the Sun (one event every two days on the visible hemisphere), and that the frequency of SAOs is similar to that of LAOs, yielding one SAO or LAO per day. Our large sample of prominence oscillations has enabled the first statistically significant study of filament oscillations and their pertinent properties, including their apparent triggers, damping times, periods, filament type, filament dimensions, peak velocities, directionality with respect to the filament spine, and maximum displacements. With the information in this catalog, one can derive minimum field strength and other unobservable characteristics through seismology, and begin to explore the implications of longitudinal and transverse oscillations for prominence stability, evolution, and eruption. We have made the catalog available to the community at the following URL: \url{http://www.iac.es/galeria/mluna/pages/gong-catalogue-of-laos.php}   

This paper presents both individual examples of interest and statistical analyses that explore potential relationships among the derived parameters. In \S \ref{sec:gong-description} the GONG data used in the catalog are described, while in \S \ref{sec:catalog} the GONG catalog of prominence oscillations is introduced. \S \ref{sec:selection-events} presents the method used to detect oscillations and select events for the catalog. The criteria used to classify prominence types are introduced in \S \ref{sec:filament-classification}. \S \ref{sec:filament-parameters} explains how we identified the triggering mechanism and derived the filament parameters.  \S \ref{sec:slit-direction-motion} and \S \ref{sec:analysis-oscillations} discuss the time-distance approach and analysis methods used to characterize the oscillations, respectively. \S \ref{sec:description-examples} describes selected events in detail, while in \S \ref{sec:statistics} we present the results of our statistical study of filament oscillations.  A seismological analysis of  selected events comprises \S \ref{sec:seismology}, and the results are summarized in \S \ref{sec:conclusions}. A full list of events and their oscillation parameters is in Appendix \ref{annex:list-events}. We describe our new method for constructing time-distance diagrams with data from curved slits in Appendix \ref{annex:curved-slits}.

\section{Description of the NSO GONG network data}\label{sec:gong-description}

Nowadays, it is possible to monitor the full Sun nearly continuously with the space-based Solar Dynamics Observatory \citep[SDO;][]{Lemen2012a} or the ground-based network of telescopes of the Global Oscillation Network Group (GONG) (\url{http://gong2.nso.edu}). Continuous coverage of the full Sun is needed for a complete study of filament oscillation events. SDO offers the best spatial resolution and temporal cadence, and the observations are independent of the local conditions of the Earth's atmosphere, in contrast to the GONG telescopes. However, the filaments and their periodic movements are not easy to detect in SDO data. In some situations the oscillation is clear in the GONG H$\alpha$ data, but it is not possible to see the filament in absorption in the SDO EUV images because of foreground emission. In addition, the structures seen by SDO are complex and very dynamic, making the detection of periodic movements very difficult. Therefore we use GONG data to perform our survey of filament oscillations. The GONG network telescopes offer sufficiently good spatial resolution and temporal cadence to detect prominence oscillations with periods of a few tens of minutes.  

The GONG H$\alpha$ images allow us to identify filaments easily and to follow their motions. We interpreted the filament motions as displacements of the prominence plasma in the plane of the sky. However, $H\alpha$ intensity depends on LOS velocities. It is worth to mention that exists the possibility that this effect may produce a disappearance of parts of the filament giving the impression that the remaining visible filament is moving. With the H$\alpha$ GONG data we can study the massive set of oscillations observed since August 2010, the date when the network started to operate. Here we focus on an analysis of GONG data from several months close to the maximum of solar cycle 24, from 1 January 2014 to 30 June 2014. Cycle 24 started in 2008 and reached minimum in early 2010, with a double-peaked maximum in 2013 and 2014.

The GONG network telescopes are of identical design and construction and are placed around the world at the following locations: Learmonth (L), Udaipur (U), El Teide (T), Cerro Tololo (C), Big Bear (B) and Mauna Loa (M). The telescope locations were selected to follow the diurnal motion of the Sun in the sky, in order to collectively ensure full-day coverage \citep{harvey1996}. Each telescope takes data daily, weather permitting, with some temporal overlap of coverage between telescopes. The temporal cadence of the GONG data is 1 min with a pixel size of $\sim$1 arcsec. For each data sequence of each telescope we compensate the solar differential rotation using the \emph{drot\_map.pro solarsoft} routine. The reference time to de-rotate the images is the central time of each temporal data sequence for each telescope and day.

\section{GONG catalog of prominence oscillations}\label{sec:catalog}
The objective of our GONG catalog is to completely describe the oscillations detected in solar filaments between 1 January 2014 and 30 June 2014. The catalog contains information about the properties of the oscillating filaments, the apparent triggers of the oscillation, and the oscillation parameters. With this information we construct a comprehensive global picture of the filament oscillations close to the maximum of solar cycle 24.

In the following sections we describe the methods we used to construct the catalog (\S\ref{sec:selection-events} to \S\ref{sec:analysis-oscillations}). 
The full results of the survey are shown in Tables \ref{table:table_filaments0} to \ref{table:table3} in Appendix \ref{annex:list-events}. The first group (Tables \ref{table:table_filaments0} to  \ref{table:table_filaments3}) displays data  describing the observations and the filaments. The first column corresponds to the number of the oscillation event, ordered in time starting 1 January 2014. 
The second column lists the telescope where the event is detected (L, U, T, C, B, M). The third column lists the central time of the temporal sequence associated with each telescope used to analyze the event (see \S \ref{sec:gong-description}). The fourth column shows the averaged position of the filaments at the reference time (see detailed description in \S \ref{sec:description-examples}).
The fifth column indicates the filament type (AR, IT, QS) described in \S \ref{sec:filament-classification}. The sixth and seventh columns contain the length, $L$, and width, $W$ of the filament measured as described in \S \ref{sec:filament-parameters}. In the eighth column we indicate the possible triggering agent described in \S \ref{sec:filament-parameters}. The last column shows whether the filament erupted in the temporal sequence analyzed, indicated by a Y (Yes).

The second group, Tables \ref{table:table0} to \ref{table:table3}, shows the oscillation parameters resulting from the fitting method described in detail in \S \ref{sec:analysis-oscillations} with Equation (\ref{eq:oscillation-fit}). The columns indicate (1) the event number; (2) the initial time of the sequence used for the fit; (3) the angle $\alpha$ between the oscillation direction and the filament spine; (4) the period ($P$); (5) damping time ($\tau$); (6) damping time per period ($\tau/P$); (7) maximum displacement  ($A$); and (8) velocity amplitude ($V$).

In \S \ref{sec:filament-classification} to \S \ref{sec:analysis-oscillations} we will use Event 1 from Table \ref{table:table0} as our reference event to describe the methodology.  Although the figures are specific to this event,  the results and explanations are valid for all events listed in the Tables.

\section{Event Selection}\label{sec:selection-events}
Our first action was to detect the filaments that may oscillate by visual inspection of the GONG H$\alpha$ data (\url{http://gong2.nso.edu}), in which the filaments are seen as dark absorption structures (see Figure \ref{fig:gong_ha_example}). The oscillations were identified as periodic displacements of a part of the filament.  The GONG webpage shows very good quality movies with full cadence for all six network telescopes. We analyzed daily observations from each telescope, and selected data that showed a clear or suspected oscillatory event for in-depth analysis. In this initial inspection we identified 408 potential cases. We initially identified each event to be associated with one day and one telescope. For cases where the oscillation continued at the end of the observing period of the selected telescope, we did not utilize the subsequent telescope observations in order to extend the oscillation data. In addition, we checked the data carefully to avoid double counting the same oscillation observed by two telescopes with overlapping data. For cases in which a second oscillation appeared during a given observing period, we defined a new event with the same location and telescope as the preceding event and we marked it with an asterisk next to the event number.

Once we identified the filaments that might oscillate, we downloaded the reduced H$\alpha$ data in the form of FITS files from the GONG server. We de-rotated the images in order to compensate for solar rotation and to study the proper motion of the filaments over the solar surface. All images were de-rotated using a reference time that corresponds to the central time of the temporal sequence as described in \S \ref{sec:gong-description}. This de-rotation algorithm only works on the solar disk, so we discarded events in prominences seen at the limb and focused exclusively on filaments seen in absorption on the disk. The coordinates are given in the usual Heliocentric-Cartesian coordinates \citep{thompson2006}.

\section{Filament classification}\label{sec:filament-classification}
\begin{figure}[!ht]
\centering\includegraphics[width=0.5\textwidth]{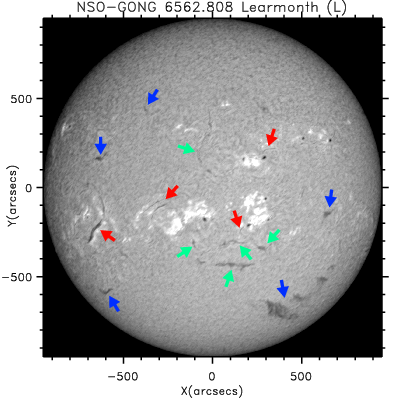}
\caption{GONG H$\alpha$ image from the Learmonth telescope illustrating the 3 types of filaments, which are seen as dark structures against the bright chromosphere. The red arrows point to  active region (AR) filaments; green arrows point to intermediate filaments (IT) between ARs; blue arrows point to quiescent (QS) filaments.
\label{fig:gong_ha_example}
}
\end{figure}

In the catalog we assigned filament types exclusively based on GONG H$\alpha$ data, according to the position scheme of \citet{engvold2014}, as active region (AR), intermediate (IT), or quiescent (QS). In Figure \ref{fig:gong_ha_example} we have marked the three types by colored arrows (AR - red, IT - green, QS - blue). AR filaments are located close to sunspots and plages with a prominent spine and few or no barbs. ITs have one end close to an active region (AR) and the other end far from an AR; they exhibit both a spine and barbs. The QSes are far from any AR or plage region, with no clear spine. For filaments whose type was difficult to determine, we used SDO HMI magnetograms to distinguish whether the filament is close to a strong magnetic field or a quiet region. The catalog includes $\nar$ AR, $\nin$ IT and $\nhe$ QS filaments.

Following this classification we identify our reference case 1 as \emph{intermediate} or IT (see Table \ref{table:table_filaments0}), because the filament has one end located in plage  associated with the active region NOAA 11938 and the other end  in a quiet region (see Fig. \ref{fig:case1_temporal_evolution}).

\section{Triggering and Filament Parameters}\label{sec:filament-parameters}

We constructed a movie with the FITS data showing the region of interest surrounding each filament, enabling us to identify the most likely triggering agent and to study the filament motion. For more than half of the events we could not identify what triggered the oscillations, so we left column 4 empty in Tables \ref{table:table_filaments0} to \ref{table:table_filaments3}. Those cases for which we found a trigger were marked FLARE when a sudden H$\alpha$ brightening was detected nearby just before the oscillation onset; prominence eruption (PE) when a nearby filament erupted before oscillation onset; and JET when the trigger was a jet of plasma that hit the filament. In one case, 91, we  clearly observed a Moreton Wave (MW) emanating from a flare and hitting the filament, triggering its oscillation as described in \S \ref{subsec:case82}. 

\begin{figure*}[!ht]
\includegraphics[width=1\textwidth]{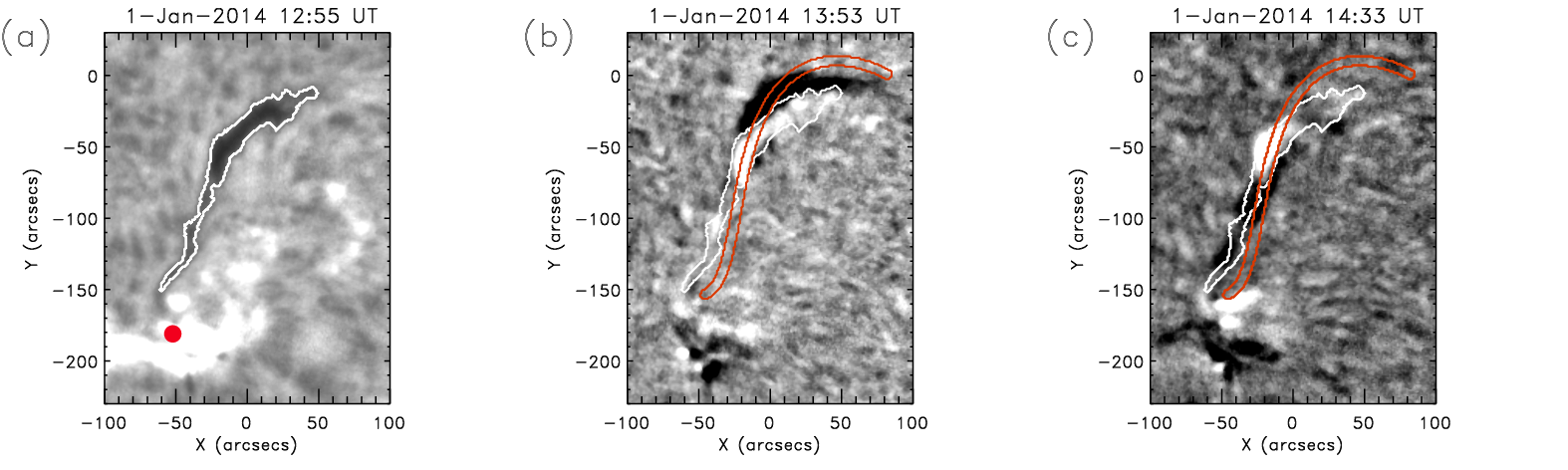}
\caption{Temporal sequence of the triggering and oscillations in event 1. (a) H$\alpha$ image showing the dark filament in its equilibrium position (12:55 UT, before oscillation onset), outlined by a white contour. The brightening associated with the triggering flare and the averaged flare position (red dot) are also visible. (b) Base difference H$\alpha$ image (13:53 UT - 12:55 UT) showing the northward displacement of the filament. The dark and white regions correspond to negative and positive differences, respectively. The contour of the equilibrium filament is overplotted in white. (c) Base difference H$\alpha$ image (14:33 UT - 12:55 UT) showing the southern displacement of the central part of the filament. In (b) and (c) the orange contour marks the slit used to track the filament motion.  \label{fig:case1_temporal_evolution}}
\end{figure*}

In Figure \ref{fig:case1_temporal_evolution} the sequence of events is shown for our reference case 1. The trigger was identified as a flaring region located south of the filament. In order to parametrize the flare position, we averaged the measured positions of several bright regions in the flare; this averaged position is marked by a red dot in Figure \ref{fig:case1_temporal_evolution}(a).  
This panel also shows the equilibrium position of the filament before the trigger perturbed the filament. Panels (b) and (c) show the difference images at the given times with the image shown in (a) subtracted, thus visualizing the displacements with respect to the equilibrium configuration. The initial motion was in the northwest direction (Fig. \ref{fig:case1_temporal_evolution}(b)), then the motion was reversed to travel toward the southeast (Fig. \ref{fig:case1_temporal_evolution}(c)).

\begin{figure}[!ht]
\centering\includegraphics[width=0.3\textwidth]{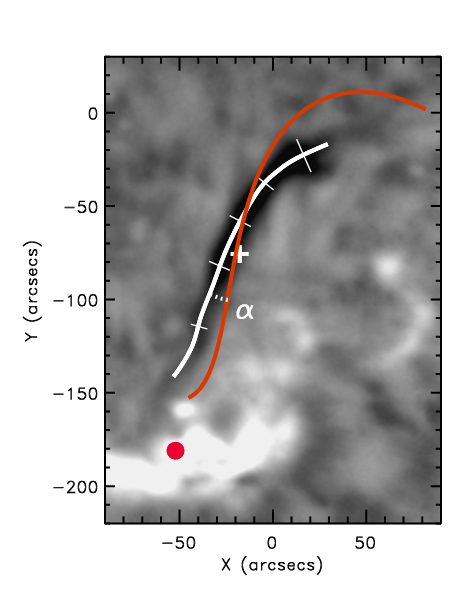}
\caption{H$\alpha$ image of event 1 averaged over 10 equally spaced times from the full observing period, showing the spine position (white line), the spine length ($L$), and the average filament width ($W$). The 5 thin line segments were used to calculate the average width of the dark band. The orange contour corresponds to the slit used to follow the motion and to construct the time-distance diagrams. $\alpha$ measures the angle between the direction of motion, i.e. the slit, and the filament spine.  In this example  $\alpha= 16^\circ$, $L=260$ Mm, and $W=15$ Mm. The red dot marks the averaged flare position; the white cross is the averaged filament position on the solar disk. \label{fig:paramfil}}
\end{figure}

Because the filaments were very dynamic and their shapes changed considerably during the observation intervals, we first generated an average image of the region of interest, as shown in Figure \ref{fig:paramfil} for case 1. This image was constructed by averaging 10 equally spaced images from the data sequence of the day and telescope selected (i.e., Cerro Tololo or C in this case). From this averaged image we determined the position of the spine following the dark filament (thick white line in Fig. \ref{fig:paramfil}), the length of the spine, $L$, and the average width of the filament, $W$. We defined the width of the filament at 5 equidistant positions along the spine as the length of the 5 segments plotted in the figure as thin lines. The average width and length characterize the size of the filament, to be used later in our statistical study. For this example the length is $L=269$ Mm and width $W=15$ Mm (see  event 1 in Table \ref{table:table_filaments0}). Using the positions of the filament spine we also obtained the averaged position of the filament on the solar disk, marked with a cross in the figure. 

\section{Time-distance diagrams and direction of the motion}\label{sec:slit-direction-motion}
We used the time-distance approach to analyze the filament oscillations. Because many oscillations reported in this work did not follow straight trajectories, but rather moved along curved paths, we could not apply the technique described by \citet{luna2014} based on straight slits. In addition, due to the relatively low resolution of the GONG data, we needed to generate time-distance diagrams with minimal reduction of the effective resolution in curved slits. To define the curved slit in the H$\alpha$ images for each event, we tracked the path of the oscillations by visual inspection. In order to generate the slit, we first traced the motion of the filament segment with the clearest and largest displacement. The slit, of length $l$ and width $w$ pixels, was placed lengthwise along the curved path of the motion as described in Appendix \ref{annex:curved-slits}. With the technique shown in the Appendix, we averaged the intensity over the transverse pixels, $w$, resulting in an intensity distribution along $l$. The time-distance diagrams display this intensity along the slit as a function of time. Figure \ref{fig:case1_temporal_evolution} shows that the slit matches the trajectory of the cool plasma for case 1. The vertical coordinate in the time-distance diagrams (e.g. Fig. \ref{fig:case1_temporal_evolution}(a)) corresponds to the distance along the slit in Mm, with the origin set to coincide approximately with the equilibrium position of the filament. This distance is measured in the plane of the sky in $x-y$ coordinates (i.e., Heliocentric-Cartesian coordinates); thus the displacements are projections of the actual motions onto the plane of the sky. Similarly, the velocities measured in the time-distance diagrams are also projections, so the actual values are probably larger.

The angle $\alpha$ between the direction of the oscillatory motion and the filament spine was measured at the intersection of the spine curve and the slit (orange curve in Fig. \ref{fig:paramfil}). This angle (dotted arc) characterizes the polarization of the oscillations in terms of longitudinal or transverse movements. In this example  $\alpha=16^\circ$ (see Table \ref{table:table0}), so the oscillation is longitudinal. In case 1, we found that the triggering location was aligned with the slit used to track the motion (Fig. \ref{fig:paramfil}), suggesting that the perturbation from the flare followed the same direction as the slit to reach the filament. The angle $\alpha$ is a projection onto the sky plane of the actual angle. The difference between these angles depends on the filament's position and orientation on the solar disk.

\begin{figure}
\centering\includegraphics[width=0.47\textwidth]{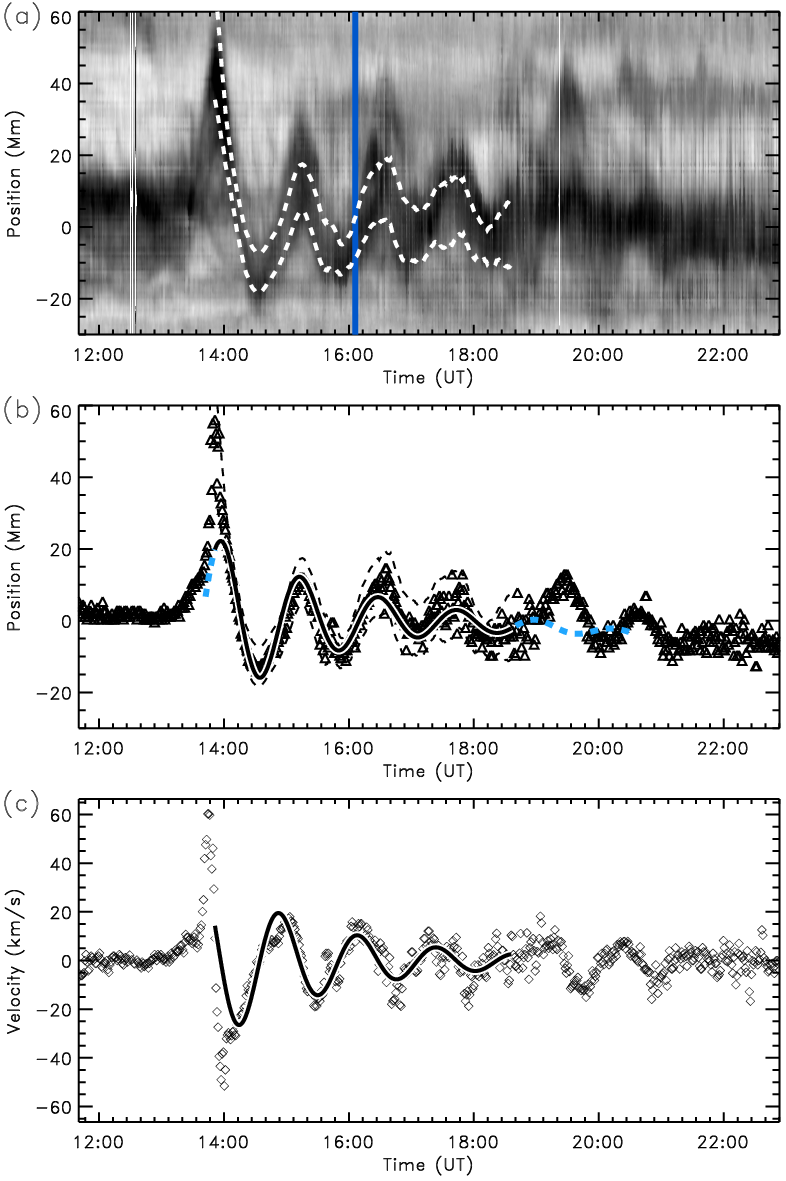}
\caption{Oscillation diagnostics of event 1. (a) Time-distance H$\alpha$ diagram. The dark band is the filament seen in absorption, surrounded by bright emission from the adjacent chromosphere. The two dashed lines mark the 1$\sigma$ level as discussed in \S \ref{subsec:errors}. (b) Triangles show the central position of the filament, $s_0(t)$, as a function of time $t$ as determined from the Gaussian fit along the slit using Eq.(\ref{eq:gaussfit}). The 1$\sigma$ region is delimited by two thick dashed lines. The thick solid line is the best fit to the triangles using Eq. (\ref{eq:oscillation-fit}). The blue dashed line is the same fitted function extrapolated to times outside the temporal range used to construct the fit. (c) The velocity as a function of time, computed as the time derivative of $s_0(t)$.  The velocity obtained with the fitted function (\ref{eq:oscillation-fit}) is overplotted as a solid curve.} \label{fig:timedistance-event1}
\end{figure}

Using the technique explained in Appendix \ref{annex:curved-slits}, we constructed time-distance diagrams for each event. In the resulting time-distance diagram for our reference case 1 (Fig. \ref{fig:timedistance-event1}(a)), the filament appears as a dark band surrounded by bright emission from the adjacent chromospheric plasma, and the filament oscillations are clearly visible. 

The slit was traced visually following the motion of the cool plasma in the H$\alpha$ images, introducing a subjective factor in the determination of the slit path. In addition, the relatively low spatial resolution of GONG data could produce a misalignment of the slit with the actual trajectory of the cool plasma, reducing the measured displacements over the slit and yielding another source of error in the measurements of the displacements. We have not quantified explicitly the error introduced by the misalignment, however, because we have overestimated the errors in the displacements as discussed in \S \ref{subsec:errors}. Additionally, the misalignment introduces an error in $\alpha$, but it is unclear how to assess the uncertainties in this parameter. We are currently developing automated techniques to track the motion of the filament, which will enable us to quantify and reduce the errors in the displacement and $\alpha$.

\section{Oscillation analysis}\label{sec:analysis-oscillations}
To determine the central position of each filament as a function of time, we plotted the intensity along the slit and fitted a Gaussian function to the intensity for each image of the observing sequence. Because the absorption of the filament depends on the column depth of cool plasma along the line of sight (LOS), we assumed that the intensity minimum corresponds to the central position in the direction along the slit. This Gaussian function also enabled us to determine the uncertainties of the oscillatory parameters (\S\ref{subsec:errors}). We used the {\em gaussfit.pro} IDL routine with a functional form
\begin{equation}\label{eq:gaussfit}
I(s)=g_0 \, e^{-\frac{1}{2}\left(\frac{s-s_0}{g_1} \right)^2} + g_2 +g_3\,s   + g_4\, s^2~ ,
\end{equation}
where $s$ is the coordinate along the slit, $g_0 < 0$ is the intensity amplitude, $s_0$ is the central position of the Gaussian, $g_1=\sigma_\mathrm{G}$ is the standard deviation, and the remaining terms are the background chromospheric emission.

\begin{figure}
\centering\includegraphics[width=0.5\textwidth]{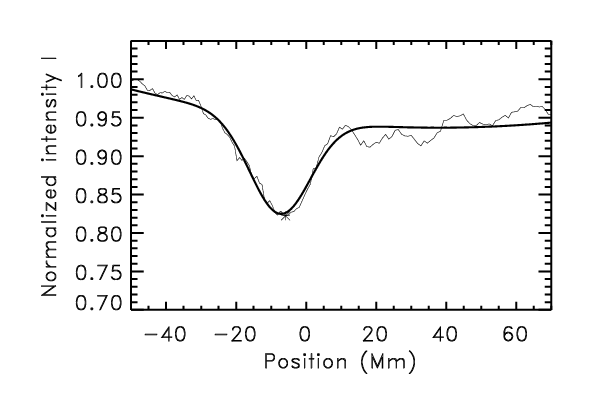}
\caption{H$\alpha$ intensity along the slit for case 1 at 16:06 UT (thin line), corresponding to the blue vertical line in Fig. \ref{fig:timedistance-event1}(a). An asterisk marks the intensity minimum. The thick line is the Gaussian fit to the observed intensity profile using Eq. (\ref{eq:gaussfit}).} \label{fig:gaussprofile}
\end{figure}

Figure \ref{fig:gaussprofile} shows the intensity along the slit at $\sim$16:06 UT (blue vertical line in Figure \ref{fig:timedistance-event1}(a)) in the reference case, with the position of the local minimum indicated by an asterisk. To avoid errors due to noisy data, we consistently used the central position of the Gaussian function, $s_0(t)$, to track the filament motion (see Figure \ref{fig:timedistance-event1}(b) for case 1). The measured velocity for all events was derived from the observations by computing the numerical derivative of $s_0(t)$. 
The function used to fit the oscillation is an exponentially decaying sinusoid, plus a third-order polynomial function to de-trend the proper motions of the filament:
\begin{eqnarray}\nonumber
y(t)&=&A_0 e^{-A_1 (t -t_0)} \cos \left[ A_2 (t - t_0) +A_3 \right] + \\ \label{eq:oscillation-fit}
&& A_4 + A_5 \, (t-t_0) +A_6 \, (t-t_0)^2+A_7 \, (t-t_0)^3\, , ~~
\end{eqnarray}
where $A_i$ are the coefficients of the fit. Sometimes the beginning of the oscillation is not well described by Equation (\ref{eq:oscillation-fit}), so we performed the fit in a selected time interval when the oscillation is clear in the time-distance diagram for each event.  In Equation (\ref{eq:oscillation-fit}), $t_0$ is the initial time of the fitted function (column 2, Tables \ref{table:table0} to \ref{table:table3}). The first few coefficients of the fit are associated with the oscillation in the following way: $A_0$ is the fitted displacement amplitude; $A_1=1/\tau$, where $\tau$ is the damping time; $A_2= 2 \pi /P$, where $P$ is the period; and $A_3$ is the initial phase. The remaining terms are the coefficients of the polynomial function that fits the background motion of the filament. This trend function filters out motions associated with long-period oscillations. Very long periods have been observed in a few oscillating filaments \citep{foullon2004,foullon2009,efremov2016}, but these motions are not clear in our data. Hence we focus our attention on more rapid oscillations. 

Figure \ref{fig:timedistance-event1}(c) shows the measured velocity (diamonds) and the best fit obtained from Equation \ref{eq:oscillation-fit}(solid curve) for case 1. The filament remained essentially stationary before 13:00 UT, then the velocity increased slowly between 13:00 and 13:30 UT  In the following $\approx$30 minutes the velocity suddenly jumped up to $60 \kms$, returned to zero at the time of maximum displacement (see Fig. \ref{fig:timedistance-event1}(b)), and increased again in the opposite direction to $-55 \kms$. In this phase the acceleration reached $140 \, \mathrm{m \, s^{-2}}$. The velocity does not resemble a sinusoidal oscillation until after $\sim$14:00 UT. The measured velocity and the fitted function agree very well.

\subsection{Errors}\label{subsec:errors}
\begin{figure}[!ht]
\centering\includegraphics[width=0.5\textwidth]{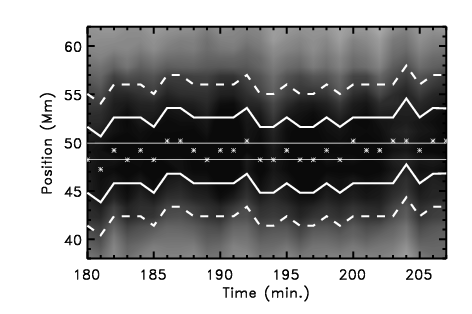}
\caption{Time-distance diagram of an interval with no clear oscillations. The asterisks mark the central positions of the Gaussian fit of Eq. (\ref{eq:gaussfit}), $g_1(t)$. The standard deviation of these points is $\sigma_\mathrm{noise}$; the two thin horizontal lines correspond to $\pm \, \sigma_\mathrm{noise}$ with respect to the mean value of these positions. The two dashed curves are at $g_1 \pm \sigma_\mathrm{G}$,  containing the dark region in the diagram. We use $\sigma = 0.5 \sigma_\mathrm{G}$ as the uncertainty in the filament position. Here the uncertainty region $g_1 \pm \sigma$ is delimited by two solid lines.} \label{fig:noise}
\end{figure}

Several sources of uncertainty in the measured oscillation parameters are attributable to the relatively low spatial resolution of the GONG images, including jitter, the uncertainties in the exposure time, and the abovementioned misalignment of the slit with the cool plasma trajectories. We assumed that the uncertainty in the oscillation parameters comes mainly from the errors in determining the position of the filament along the slit. 
In Figure \ref{fig:noise} we have plotted a time-distance diagram in an interval where no oscillation was evident. The asterisks mark the central positions of the Gaussian fit, $s(t)$. We clearly see a noisy signal in the figure, where the $1 \sigma$ standard deviation (thin straight lines) around the mean value is $\sigma_\mathrm{noise}=1.4 \Mm$. The figure also shows the width of the Gaussian fit, $\sigma_\mathrm{G}$, to the intensity (Eq. \ref{eq:gaussfit}) as dashed lines, which coincide with the borders of the dark band. We consider $\sigma_\mathrm{G}$ to overestimate the uncertainty of the central position because this uncertainty is comparable to the filament width. In fact, $\sigma_\mathrm{G} \gtrsim 3 \, \sigma_\mathrm{noise}$ in all analyzed events. Thus, we decided somewhat arbitrarily to set the uncertainty of the filament position as $\sigma=0.5 \, \sigma_\mathrm{G} \gtrsim \sigma_\mathrm{noise}$ (indicated by two solid curves in Figure \ref{fig:noise}). This overestimates the error because $\sigma > \sigma_\mathrm{noise}$ and the uncertainty in the position is larger than the variations seen in the figure. This error $\sigma$ will be considered the only source of uncertainty in the estimated oscillation parameters.

In Table \ref{table:table0} we have tabulated the best-fit parameters from Equation \ref{eq:oscillation-fit} for all oscillation events. Event 1 has a period of $P=76 \pm 1$ min, damping time $\tau=121\pm15$ min, maximum displacement  $A=23\pm2$ Mm, and peak velocity amplitude $V = 26\pm4~\kms$. The ratio $\tau/P=1.6\pm0.2$, indicating that the damping was strong and very efficient. Because the velocity amplitude was larger than $10\kms$, event 1 is a LAO.  Large velocities exceeding $40\kms$ occurred early in this event, but this phase did not produce a fit compatible with subsequent motions (Fig. \ref{fig:timedistance-event1}(c)). The best fit derived from Equation (\ref{eq:oscillation-fit}) has a much lower peak value of $26\kms$ and agrees well with most of the subsequent velocity oscillations. At the end of the fitted range, around 18:30 UT, the first oscillation ceased and a new one appeared. This new oscillation, case 2*, had a different phase and amplitude than case 1, which can be seen by comparing the observed triangles with the blue dashed line from the fit of case 1 in Figure \ref{fig:timedistance-event1}(b). Cases 1 and 2* have comparable periods, indicating that they are characteristic oscillations of the filament.
The direction of the motion is offset by $16^\circ$ from the filament spine (see Fig. \ref{fig:paramfil}(b) and Table \ref{table:table0}), suggesting  that the oscillation is longitudinal as well as large-amplitude (i.e., a LALO). This angle is comparable to the typical angle between the magnetic field and the filament spine, according to direct measurements \citep{Leroy1983a,Leroy1984a,tandberg1995,trujillo2002,casini2003,Lopez-Ariste2006a}, suggesting that the oscillation is aligned with the filament magnetic field.

\section{Selected Case Studies}\label{sec:description-examples}

In the previous Sections we used case 1 as a representative oscillation example from our event catalog. Here we describe selected additional events from the catalog to illustrate the intriguing variety of behaviors and oscillation characteristics encountered in our survey.  
\subsection{Event 58: LALO triggered by a two-ribbon flare}\label{subsec:case51}
\begin{figure*}[!ht]
\includegraphics[width=1\textwidth]{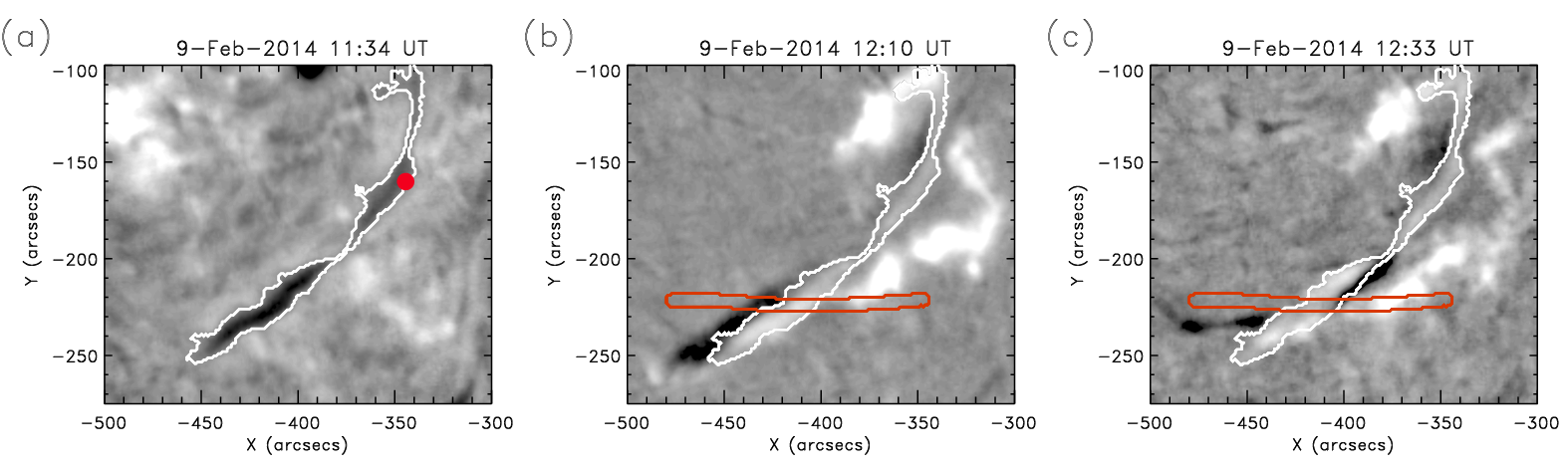}
\caption{Temporal sequence of the triggering and oscillations in event 58. Panels and annotations are as in Fig. \ref{fig:case1_temporal_evolution}. (a) Pre-oscillation H$\alpha$ image at 11:34 UT. The white contour outlines the equilibrium filament position at flare onset (11:24 UT). The triggering two-ribbon flare is not visible in this initial frame. The red dot marks the average position of the flaring region (evident in (b)). (b) Base difference H$\alpha$ image (12:10 UT - 11:34 UT). The two flare ribbons appear as bright patches north and south of the filament. (c) Base difference H$\alpha$ image (12:33 UT - 11:34 UT).\label{fig:case3_temporal_evolution}}
\end{figure*}

\begin{figure}[!ht]
\centering\includegraphics[width=0.48\textwidth]{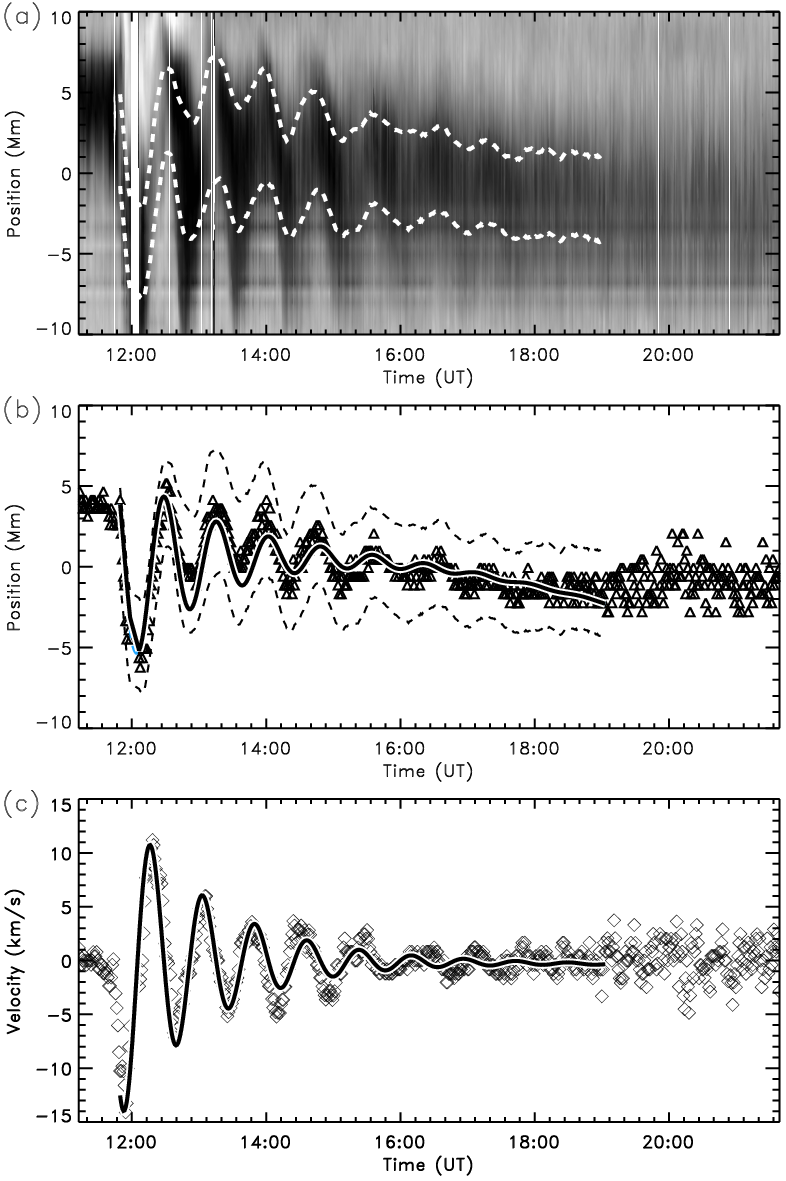}
\caption{Oscillation diagnostics of event 58. Panels and annotations are as in Fig. \ref{fig:timedistance-event1}} \label{fig:timedistance-event51}
\end{figure}

Event 58 of the survey (Table \ref{table:table0}) is a LAO with a peak velocity of $14\kms$, triggered by a two-ribbon flare that straddled the AR filament (see Figure \ref{fig:case3_temporal_evolution}(a)). This flare produced an oscillation in which the cool plasma was displaced initially in the west-east direction from its pre-flare equilibrium position (white contour in Figure \ref{fig:case3_temporal_evolution}(a)). Then the motion was reversed, so the filament reached maximum western elongation (Fig. \ref{fig:case3_temporal_evolution}(b)), followed by another reversal that produced a peak eastward displacement of smaller amplitude (Fig. \ref{fig:case3_temporal_evolution}(c)). The Figure also demonstrates that the slit used to track the motion and construct the time-distance diagram closely follows the trajectory of the cool filament plasma. Very clear oscillations in the H$\alpha$ intensity are evident in Figure \ref{fig:timedistance-event51}(a), triggered around 11:50 UT and lasting for more than 7 periods from $\sim$12:00 to 18:00 UT. Before flare onset the dark filament band is almost horizontal, so the filament was almost at rest. In Figure \ref{fig:timedistance-event51}(b), the central position $s(t)$ and its best fit from Eq.~\ref{eq:oscillation-fit} agree well until $\sim$16:00 UT. Thereafter the fit is more damped than $s(t)$ and a slight phase difference is also evident. After 19:00 UT the measured $s(t)$ is very noisy because the quality of the H$\alpha$ images was reduced. 

The fitted velocity matches the measured velocity well (Fig. \ref{fig:timedistance-event51}(c)). There were no large velocities during the initial phase associated with the triggering, in contrast with case 1  (Fig. \ref{fig:timedistance-event1}(c)). Although the initial velocity exceeded $10\kms$, indicating that this event is a LAO, in less than a period the velocity fell below this threshold. The direction of motion was offset by $32^\circ$ from the filament spine, identifying this event as a possible LALO.

\subsection{Event 63: LALO in a large quiescent filament}\label{subsec:case56}
\begin{figure*}[!ht]
\centering\includegraphics[width=1\textwidth]{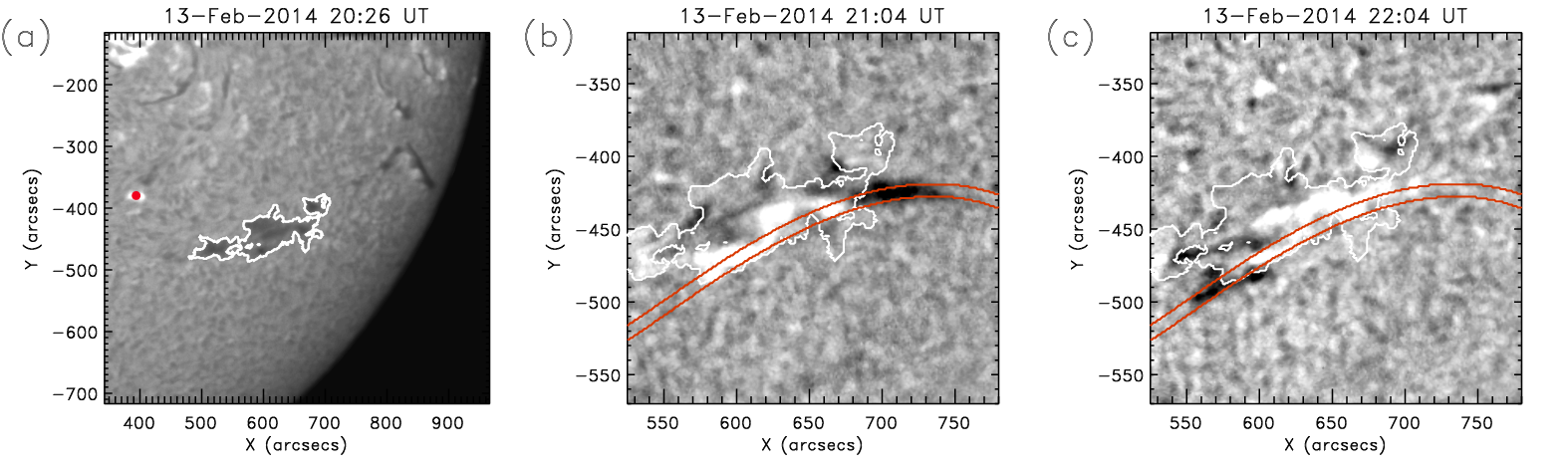}
\caption{Temporal sequence of the triggering and oscillations in event 63. Panels and annotations are as in Figure \ref{fig:case1_temporal_evolution}. Here panels (b) and (c) show a smaller region centered at the filament. (a) Pre-oscillation H$\alpha$ image at 20:26 UT. (b) Base difference H$\alpha$ image (21:04 UT - 20:26 UT), showing the initial northwestward displacement of the filament along the slit. (c) Base difference H$\alpha$ image (22:04 UT - 20:26 UT) showing the subsequent southeastward displacement. \label{fig:case56_temporal_evolution}}
\end{figure*}

In case 63 the oscillation occurred in a very large, fragmented \emph{quiescent} filament (QS) in the southern hemisphere. The oscillation appeared in the southern segment, possibly triggered by the flare centered at the red dot in Figure \ref{fig:case56_temporal_evolution}(a). The H$\alpha$ flare brightened slightly before the oscillation began at $\sim$18:26 UT. Part of the prominence plasma moved toward the northwest or west along a curved trajectory. Difference images in Figures \ref{fig:case56_temporal_evolution}(b) and (c) reveal the maximum elongation at two times after the initiation: a dark region along the slits appeared, first on one side and then on the other side of the pre-oscillation position of the filament (Fig. \ref{fig:case56_temporal_evolution}(c)).

\begin{figure}[!ht]
\centering\includegraphics[width=0.5\textwidth]{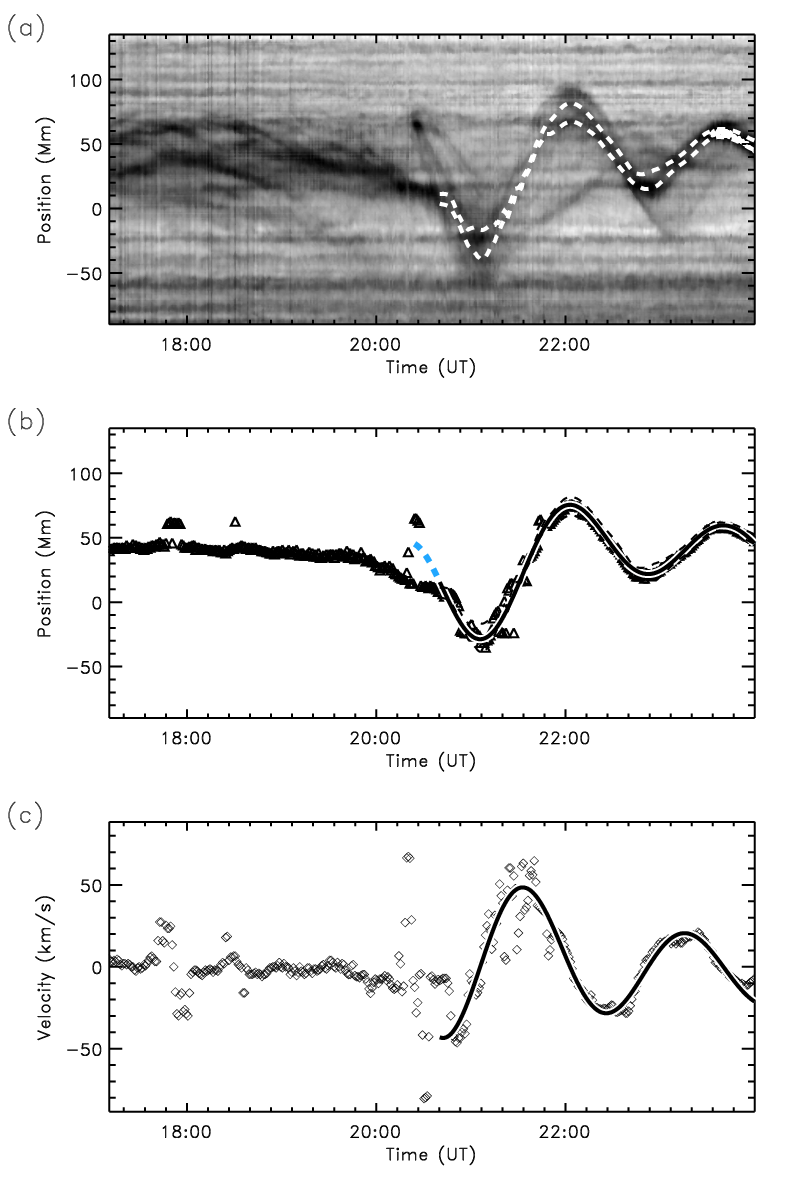}
\caption{Oscillation diagnostics of event 63. Panels and annotations are as in Fig. \ref{fig:timedistance-event1} } \label{fig:timedistance-event56}
\end{figure}

Several threads of cool plasma moved in a complex way before oscillation onset at 20:26 UT (Fig. \ref{fig:timedistance-event56}(a)). Thereafter the oscillation was very clear but only was observed for 2 cycles. Apparently the whole filament was displaced to the southeast before the oscillation, possibly indicating that the structure of the filament changed during the pre-flare phase. In Figure \ref{fig:timedistance-event56}(a) the positional uncertainty, $\sigma$, is indicated by two dashed lines. This uncertainty is very small where the dark band is very narrow. Initially some threads moved towards the centroid of the oscillation, but our tracking code ignored these outliers and only followed the motion of the main body, the darkest band at $s\sim$10 Mm. After 20:39 UT the entire filament oscillated, and Equation (\ref{eq:oscillation-fit}) provided a good fit to $s(t)$ (Fig. \ref{fig:timedistance-event56}(b)). The measured velocity amplitude, $V$, reached very large values $>50\kms$ (Fig. \ref{fig:timedistance-event56}(c)) .

The results of the fit are shown in Table \ref{table:table0}. The oscillation had a velocity amplitude of $V=48.5 \pm 2.4\kms$, $P= 103 \pm 1$ min, a damping time $\tau=175 \pm 12$ min, and $\tau/P=1.7\pm 0.1$ indicating very strong damping. The motion was only $\alpha=2^\circ$ misaligned with the filament spine, indicating that this event was probably a LALO. 

\subsection{Event 91: LALO triggered by a Moreton wave}\label{subsec:case82}
\begin{figure}[!hb]
\centering\includegraphics[width=0.5\textwidth]{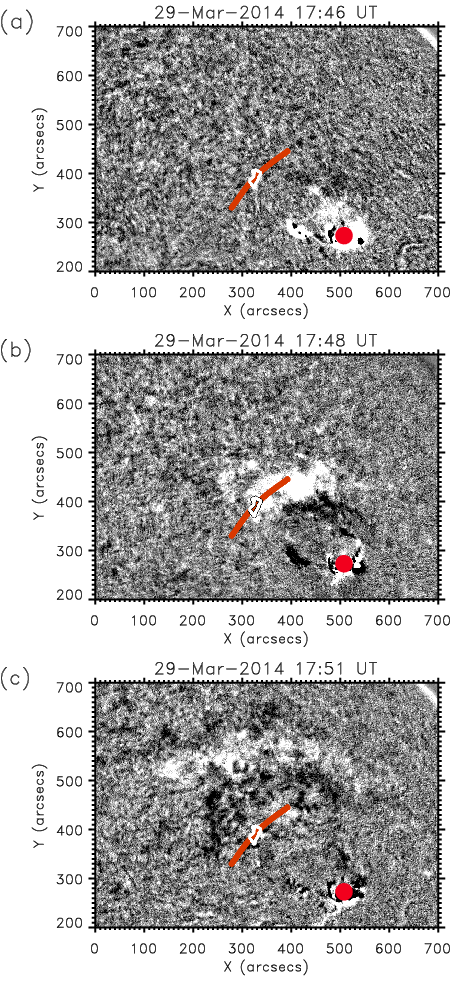}
\caption{Running difference H$\alpha$ images showing the propagating Moreton wave at selected times. In all panels the filament equilibrium position is outlined by a white contour, the slit is marked by a red arc, and the averaged flare position is marked by a red dot. (a) 17:46 UT, shortly after the wave was generated at the flaring region. (b) 17:48 UT, when the wave (white patch) reached the filament.  (c) 17:51 UT, as the wave (white arc) continued to expand and travel northward. } \label{fig:rd82-moretonwave}
\end{figure}

\begin{figure*}[!ht]
\includegraphics[width=1\textwidth]{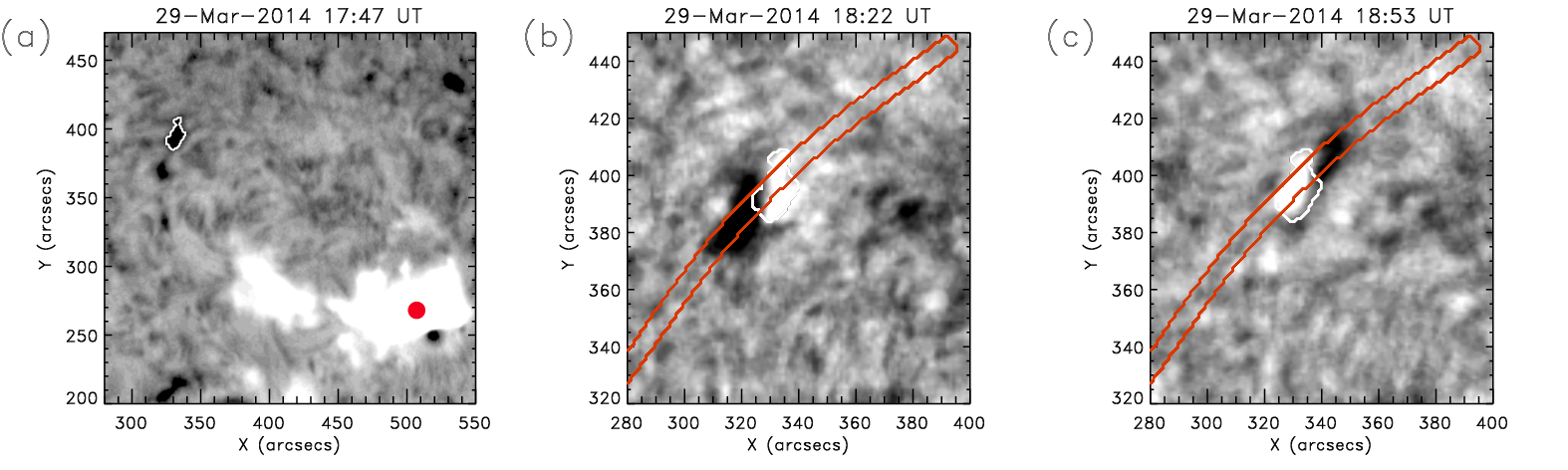}
\caption{Temporal sequence of the triggering and oscillations in event 91. Panels and annotations are as in Figure \ref{fig:case1_temporal_evolution}. (a) H$\alpha$ image of the flaring region and the filament at oscillation onset (17:47 UT). The red dot indicates the approximate position of the flare that produced the Moreton wave that triggered the oscillation. (b) and (c) Running difference H$\alpha$ images of a smaller region centered on the filament. The filament was displaced initially to the southeast (b), then toward the northwest (c). \label{fig:case82_temporal_evolution}}
\end{figure*}

\begin{figure}[!ht]
\centering\includegraphics[width=0.5\textwidth]{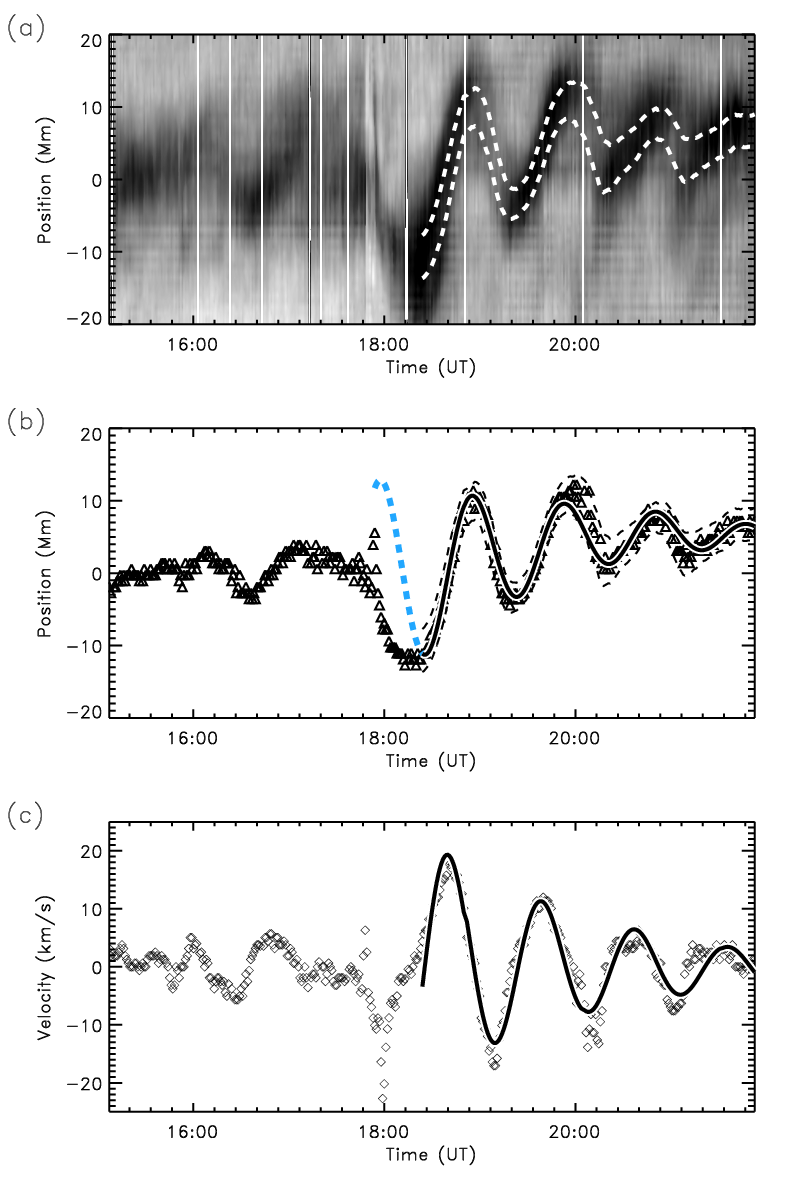}
\caption{Oscillation diagnostics of event 91.  Panels and annotations are as in Fig. \ref{fig:timedistance-event1} }\label{fig:timedistance-event82}
\end{figure}

Event 91 occurred in an intermediate filament (IT) located close to AR NOAA 12017. Around 17:41 UT a major flare occurred in the AR followed by a Moreton wave, which is visible in Figure \ref{fig:rd82-moretonwave}. This is the only event in the survey for which we could identify a Moreton wave connected to the filament oscillation. The wave emanated from the flare region (Fig. \ref{fig:rd82-moretonwave}(a)), hit the filament (Fig. \ref{fig:rd82-moretonwave}(b)) then continued to propagate northward (Fig. \ref{fig:rd82-moretonwave}(c)). Once the wave encountered the filament (Fig. \ref{fig:case82_temporal_evolution}(b)), the filament started to oscillate perpendicular to the propagation direction of the wave front.  At 18:53 UT (Fig. \ref{fig:case82_temporal_evolution}(c)), the motion was reversed and the oscillatory motion was fully established. The time-distance diagram in Figure \ref{fig:timedistance-event82}(a) reveals significant motions of the filament before and after flare onset. Around oscillation onset, at 17:47 UT, a white vertical region appears in Figure \ref{fig:timedistance-event82}(a), signalling the arrival of the Moreton wave at the filament throughout the slit. The Moreton wave initially produced complex filament dynamics, best seen in a movie of the event (not shown here). The oscillation became distinct around 18:00 UT, with a maximum southeastward displacement of 12 Mm from the equilibrium position around 18:22 UT (Fig \ref{fig:timedistance-event82}(b)) and a smaller opposing displacement at 18:53 UT (Fig \ref{fig:timedistance-event82}(b)). Interestingly, when the filament moved in the northwest direction, it was darker than when the motion was in the opposite direction (Fig. \ref{fig:timedistance-event82}(a)). The central position of the dark band oscillated clearly (Fig. \ref{fig:timedistance-event82}(b)), and was fitted very well with Equation (\ref{eq:oscillation-fit}) after 18:23 UT. However, between the triggering at 17:47 UT and 18:23 UT the motion does not fit the sinusoidal function (blue dashed line), probably due to the very complex motions prior to and during the passage of the wave through the filament.

Before the oscillations were triggered at 17:47 UT the filament reached velocities above 5$\kms$, but this motion was not periodic (Fig. \ref{fig:timedistance-event82}(c)). The oscillation velocity peaked at 18:00 UT; after $\sim$18:23 UT the measured velocity was very well fitted by the derivative of Equation (\ref{eq:oscillation-fit}), yielding a peak amplitude $V=19.3 \pm 2.3 \kms$, a peak displacement  $A=11 \pm 1$ Mm, period $P=58 \pm 1$ min, damping time $\tau=108 \pm 12$ min, and $\tau/P=1.9 \pm 0.2$. The angle between the motion and the filament spine was $\alpha=26^\circ$, again consistent with the typical direct measurements of the orientation of the prominence magnetic field relative to the spine \citep{Leroy1983a,Leroy1984a,tandberg1995,trujillo2002,casini2003,Lopez-Ariste2006a}. Therefore the motion may be aligned with the local magnetic field and the event is probably a LALO.

\subsection{Event 31: SALO with a very small velocity amplitude}\label{subsec:case26}
In the survey we also detected oscillations with very small amplitudes of only a few $\kms$, below the LAO lower limit of $10\kms$. Event 31 has the smallest velocity amplitude of the survey: $V=1.6 \pm 9 \kms$. The filament is an IT and the oscillation was triggered by a nearby flare. In this example the displacements were too small to be visible in a figure similar to Figure \ref{fig:case1_temporal_evolution}. Therefore we refer the reader instead to the movie of this event at the URL: \url{http://www.iac.es/galeria/mluna/pages/gong-catalogue-of-laos.php}.

\begin{figure}[!ht]
\centering\includegraphics[width=0.5\textwidth]{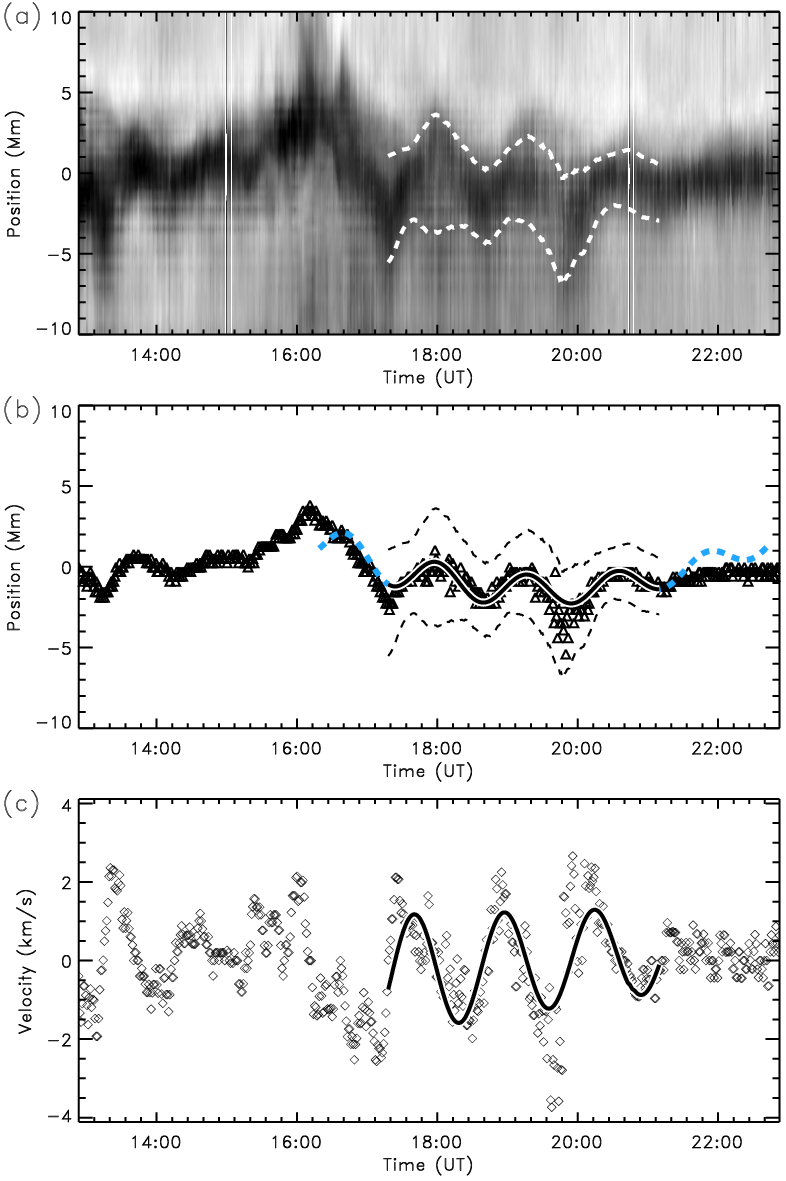}
\caption{Oscillation diagnostics of event 31. Panels and annotations are as in Fig. \ref{fig:timedistance-event1}.} \label{fig:timedistance-event26}
\end{figure}

Although the displacements are very small in this event, the oscillatory pattern is very clear (Fig. \ref{fig:timedistance-event26}(a)). Before 17:00 UT the filament appears to be moving, but clear oscillations started at 17:17 UT and ended at 21:06 UT. The displacement appears more or less constant throughout the event (Fig. \ref{fig:timedistance-event26}(b)).  However, the extrapolated oscillation some time before and after the fitted oscillation (blue dashed line in Figure \ref{fig:timedistance-event26}(b)) does not follow the filament motion, and the oscillation apparently ended without strong damping.  In general, the velocity has a large uncertainty for small-amplitude events. In this case the velocity error is $9 \kms$, much greater than the velocity $V=1.6 \kms$. Figure \ref{fig:timedistance-event26} exhibits distinct oscillatory motions, so we are probably overestimating the positional errors (see \S\ref{subsec:errors}). The displacement was very small ($A=1 \pm 1$ Mm), the oscillation period $P=77 \pm 4$ min, and the damping time $\tau=662 \pm 1546$ min, so the damping was very weak with a large uncertainty (see Table \ref{table:table0}). The angle between the slit and the spine $\alpha=26^\circ$, suggesting that the oscillation was longitudinal.

\subsection{Events 145 and 146*: double event}\label{subsec:case135}
\begin{figure*}[!ht]
\includegraphics[width=1\textwidth]{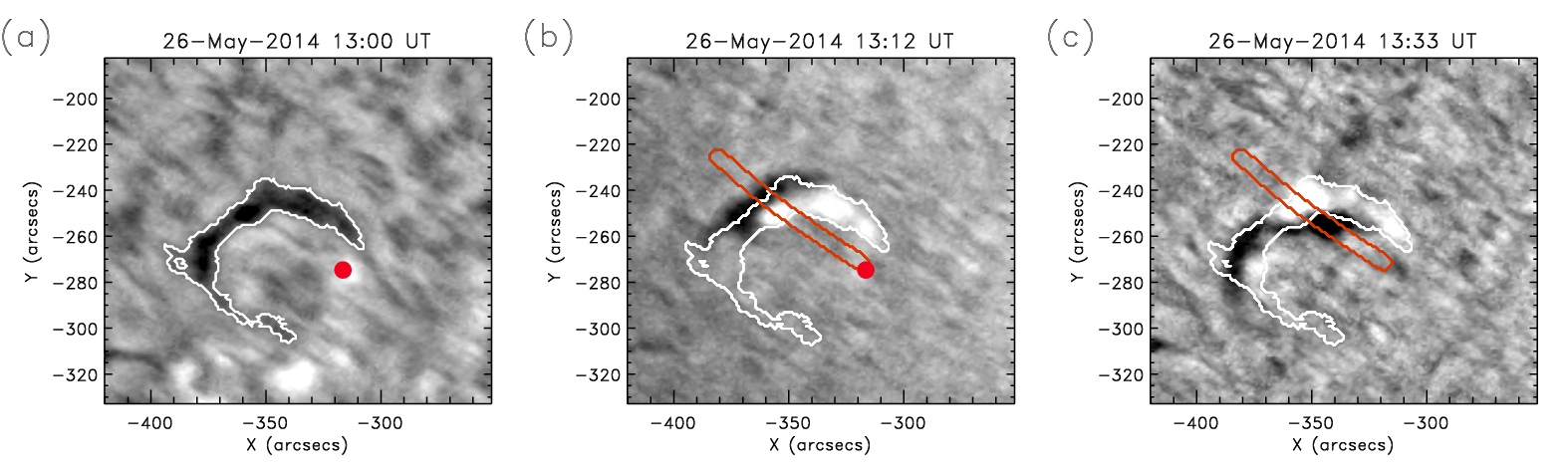}
\caption{Temporal sequence of the triggering and oscillations in event 145. Panels and annotations are as in Fig. \ref{fig:case1_temporal_evolution}. In (a) and (b) the red dot indicates the approximate position of the flare that probably triggered the oscillation. \label{fig:case135_temporal_evolution}}
\end{figure*}

In these events, two clear oscillations occurred in the same filament within one observing interval; therefore we labeled them as separate events. In our example the filament is of the IT type with a curved initial structure (Fig. \ref{fig:case135_temporal_evolution}(a)). Both oscillations involve motions mainly transverse to the filament spine, with $\alpha=50^\circ$.  A nearby flare, marked by the red dot in the figure, is the most likely trigger. At onset the entire filament was displaced laterally toward the northeast (Fig. \ref{fig:case135_temporal_evolution}(b)), then after a half period the filament moved to the other side of its equilibrium position (Fig. \ref{fig:case135_temporal_evolution}(c)). The slit (red outline) is over the region of the filament that oscillated with the largest displacement, which protruded from the rest of the structure. 

\begin{figure}[!ht]
\centering\includegraphics[width=0.45\textwidth]{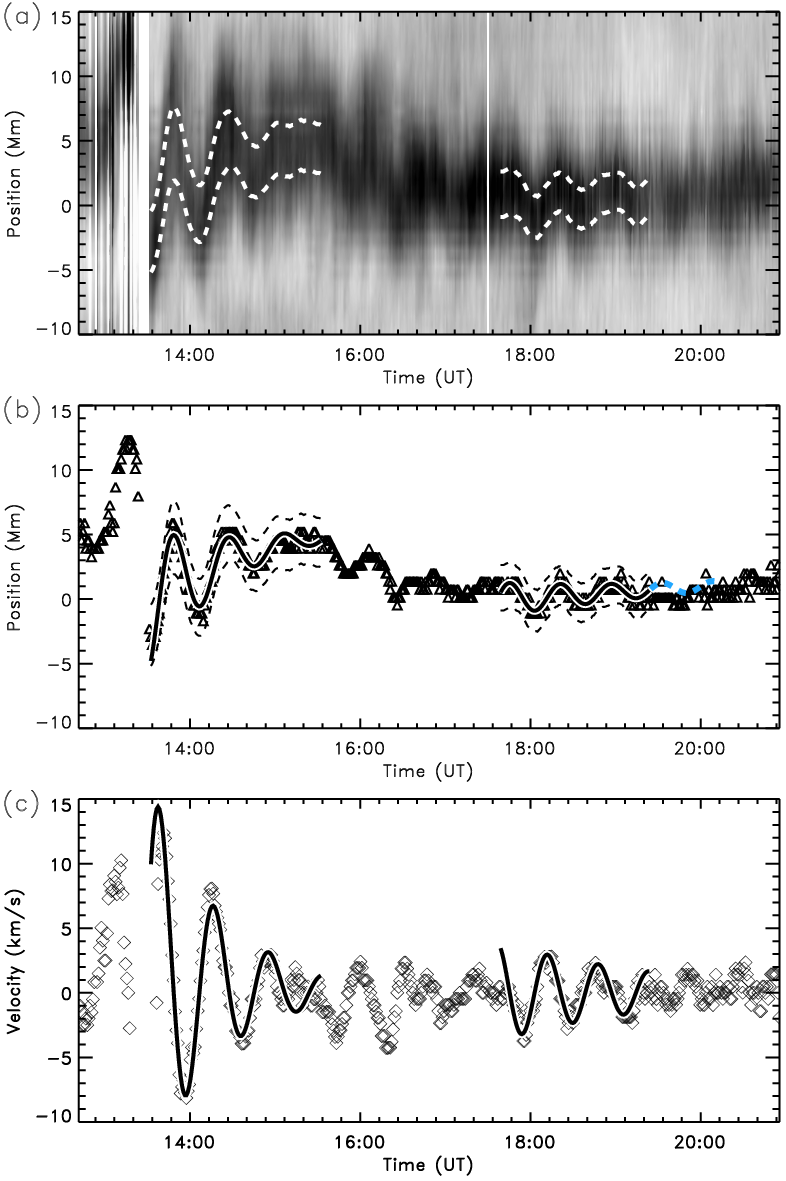}
\caption{Oscillation diagnostics of events 145 and 146*. Panels are as in Fig. \ref{fig:timedistance-event1} .} \label{fig:timedistance-event135}
\end{figure}

There were many data gaps, which appear as white vertical bands at the beginning of the time-distance diagram (Fig. \ref{fig:timedistance-event135}(a)). 
The first event, 145, oscillated most visibly between 13:31 and 15:33 UT. The oscillation period was $P=39 \pm 2$ min, $V=14.4 \pm 6.1 \kms$ and $\tau=48 \pm 13$ min (see Table \ref{table:table2}), indicating very strong damping with $\tau/P=1.2 \pm 0.4$.  Event 146*, which started at 17:38 UT and ended at 19:20 UT, had $P=36\pm 2$ min, $V=3.4 \pm 9.4$ and $\tau=119 \pm 146$ min with $\tau/P=3.3 \pm 4$. Figure \ref{fig:timedistance-event135}(a) leaves the impression that both events were part of a long oscillation starting at 13:31 UT and continuing throughout the temporal sequence. However, the velocity evolution demonstrates that case 145 ended at 15:33 UT (Fig. \ref{fig:timedistance-event135}(c)), followed by an interval of complex oscillation. Event 146* started at 17:38 UT with a very clear oscillation that was out of phase with previous motions. We conclude that repetitive triggering occurred in this filament.

It is interesting that both events had similar periods, suggesting that the filament oscillated with a characteristic frequency of the system (e.g., \citet{hyder1966}). However, the damping times were very different, indicating that either the damping mechanisms were different or the damping efficiency changed between events. The first case had a peak velocity $\sim 14 \kms$, much larger than the second event with $V \sim 3 \kms$. In \S \ref{subsec:dampingtime-per-period} we show that the damping time decreases with $V$ for the entire set of events. This nonlinear effect might explain the different damping times for cases 145 and 146*. Both oscillations are very clear in Figure \ref{fig:timedistance-event135}, and both fits according to Equation (\ref{eq:oscillation-fit}) are good. Although the second oscillation continued after 19:20 UT, the extrapolated fitted function (blue dashed line in Fig. \ref{fig:timedistance-event135}(b)) does not follow the center of the dark band (triangles) well during this interval, probably because the dark band is very light and the signal-to-noise ratio is very low in the time-distance diagram. Event 145 is a transverse LAO with $V=14.4 \kms$, and 146* is a transverse SAO with $V=3.4 \kms$ (see Figure \ref{fig:timedistance-event135}(c)).

\subsection{Event 151 and 152*: double event and amplified oscillation} \label{subsec:case141}
\begin{figure*}[!ht]
\includegraphics[width=1\textwidth]{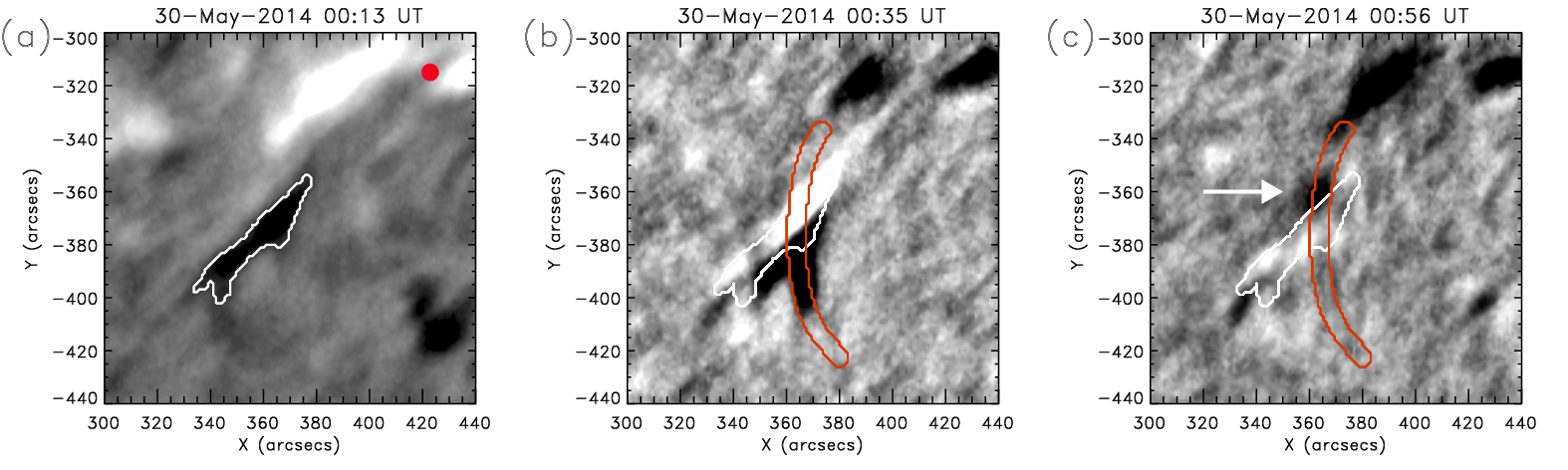}
\caption{Temporal sequence of the triggering and oscillations in events 151 and 152*. Panels and annotations are as in Fig. \ref{fig:case1_temporal_evolution}.  In (c) the white arrow points to the part of the prominence that oscillates.  \label{fig:case141_temporal_evolution}}
\end{figure*}

Unlike the double event described in the previous section, the cases presented here exhibit very different oscillation periods: $P=52\pm 2$ min for case 151 and $P=66 \pm 3$ min for case 152*. 
Figure \ref{fig:case141_temporal_evolution} shows the first stages of event 151; the initial motion is southward, followed by a reversal toward the north. The direction of motion for event 152* is similar to 151. Both events were apparently triggered by nearby flaring (red dot in Figure \ref{fig:case141_temporal_evolution}(a)).

\begin{figure}[!ht]
\centering\includegraphics[width=0.45\textwidth]{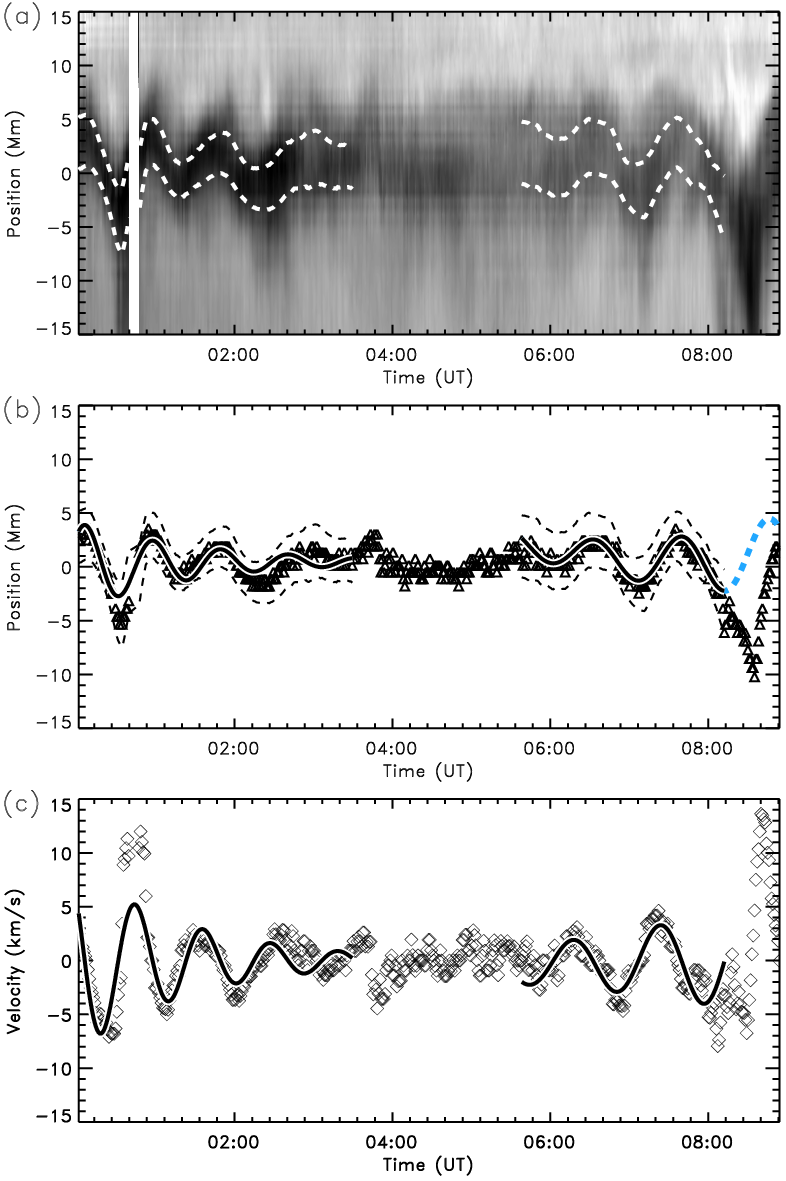}
\caption{Oscillation diagnostics of events 151 and 152*. Panels are as in Fig. \ref{fig:timedistance-event1} .}  \label{fig:timedistance-event141}
\end{figure}

Event 151 occurred between 00:00 UT and 3:33 UT (Fig. \ref{fig:timedistance-event141}(a)). The best-fit solution to the central position tracks the oscillation well, except for a small discrepancy in the last period (Fig. \ref{fig:timedistance-event141}(b)). The measured and fitted velocities plotted in panel (c) also agree well. The oscillation parameters are $P=52 \pm 2$ min, $\tau=89 \pm 23$ min, and $\tau/P=1.7 \pm 0.5$ (strong damping). This event is a SAO with $V=6.8  \pm 5.1 \kms$. In both events $\alpha=36^\circ$, suggesting longitudinal oscillations.

Figure \ref{fig:timedistance-event141} also shows the unusual oscillation of event 152*, which started at 5:37 UT, increased in amplitude, and ended at 8:53 UT. The best fit agrees well with $s(t)$ between 5:37 and 8:15 UT but not after this interval, suggesting that the plasma motion of this amplified oscillation is more complex than Equation (\ref{eq:oscillation-fit}) (Fig. \ref{fig:timedistance-event141}(b)). The oscillation parameters are $P=66 \pm 3$ min, $\tau=-163 \pm 105$ min, and $\tau/P=-2.5 \pm 2$, indicating very strong amplification. The large error in the damping time is probably overestimated, because the fitted function agrees very well with the oscillation. The fitted maximum velocity amplitude is $V=4.0 \pm 7.8 \kms$, but the measured velocity reached a maximum of $14 \kms$ at 08:40 UT. Later H$\alpha$ data reveals that the oscillation ceased and the filament became stationary again. Similar behavior was found by \citet{molowny-horas1999} for an amplified filament oscillation.

In order to amplify the oscillation the cool plasma must gain energy. Recently \citet{zhou2017} and \citet{zhang2017} found LALOs with an amplified oscillation followed by a damped phase, which they explain as a beating phenomenon between two interacting oscillators \citep[see, e.g,][]{luna2006,luna2008}. In this scenario the oscillations of two regions of the filament are coupled: an active oscillator that transfers energy to the other part of the filament (the passive oscillator). The passive oscillator gains energy with time so its oscillation is amplified, while the active portion loses energy, reducing its amplitude. The section of the filament oscillating in event 152* should be the passive oscillator because it is gaining energy. However, the active oscillator should be the other region of the filament, but it does not oscillate with a larger amplitude. Therefore this hypothesis does not explain events 151 and 152*.
\cite{ballester2016} found that cooling the prominence plasma could amplify its oscillations, but we can't test this hypothesis for lack of relevant temperature diagnostics. Alternatively, repetitive nearby flares could produce both oscillation events and possibly amplify the second. The H$\alpha$ data shows flaring activity close to the filament (Fig. \ref{fig:case141_temporal_evolution}(a)). In this situation the amplified oscillation would be driven by external forcing, which could explain why the period is different from that of the non-amplified previous event.

\subsection{Events 107 and 108*: double event with amplified oscillation and eruption}\label{subsec:case99}

\begin{figure*}[!ht]
\includegraphics[width=1\textwidth]{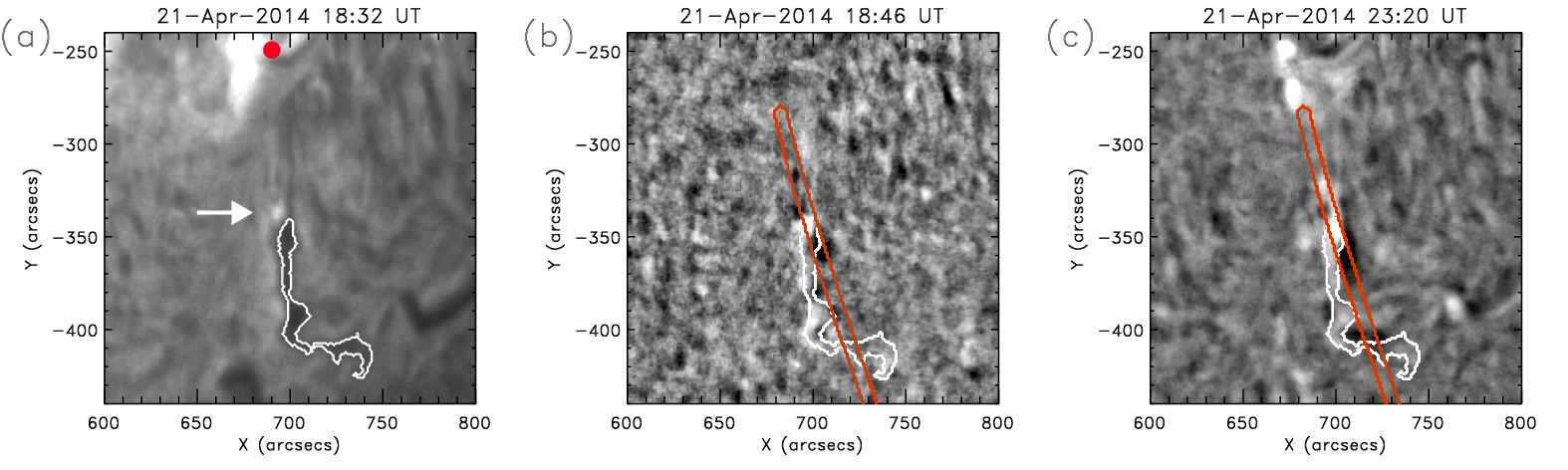}
\caption{Temporal sequence of the triggering and oscillations in events 107 and 108*. Panels and annotations are as in Fig. \ref{fig:case1_temporal_evolution}. \label{fig:case99_temporal_evolution}}
\end{figure*}

This double event is similar to that discussed in \S \ref{subsec:case135} --- a damped oscillation (case 107) immediately followed by an amplified oscillation (case 108*) --- but with a final eruption. Both events occurred in the same IT filament, and both events were triggered by flaring in an active region north of the filament (Fig.~\ref{fig:case99_temporal_evolution}(a)). For the first event $P=50\pm1$ min, $V=6.6 \pm 2.2 \kms$, and $\tau/P=3.1\pm 0.7$; for the second event $P=40\pm3$ min, $V=5.6 \pm 9.3 \kms$, and $\tau/P=-2.4\pm 2.0$. Both events are SAOs with $\alpha=20^\circ$, denoting longitudinal polarization. As in events 151 and 152*, the periods are different.

\begin{figure}[!ht]
\centering\includegraphics[width=0.45\textwidth]{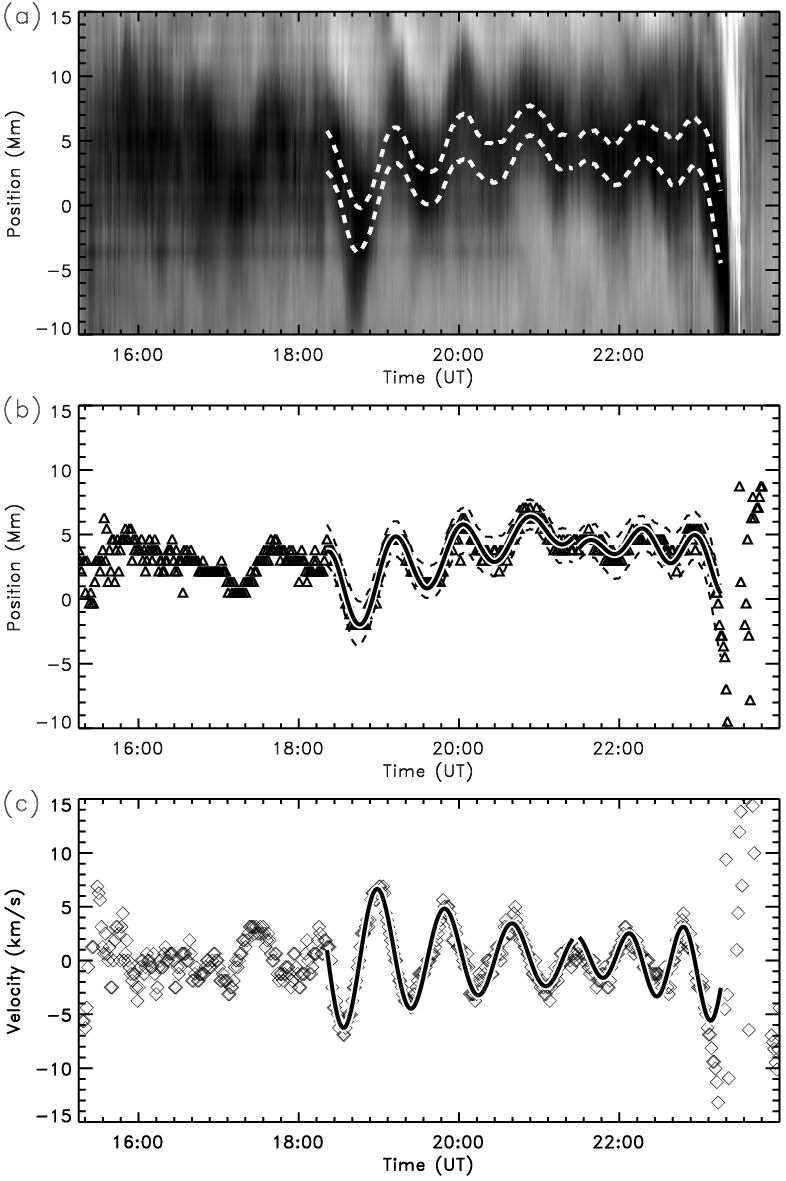}
\caption{Oscillation diagnostics of events 107 and 108*. Panels and annotations are as in Fig. \ref{fig:timedistance-event1} .}  \label{fig:timedistance-event99}
\end{figure}

Simultaneous with the oscillation onset for event 107, a white spot appeared north of the filament (marked with a white arrow in Figure \ref{fig:case99_temporal_evolution}(a)) and continued almost to the end of event 107.  Because the slit in Figure \ref{fig:case99_temporal_evolution}(b,c) passes over the white spot, we can also see this brightening in the resulting time-distance diagram (Fig. \ref{fig:timedistance-event99}(a)). Base difference images show the maximum elongation of the cool plasma in event 107 (Fig.~\ref{fig:case99_temporal_evolution}(b)) and for event 108* (Fig. \ref{fig:case99_temporal_evolution}(c)). At the end of event 108* the prominence erupts ($\sim$ 23:20 UT, not shown). The white spot that appeared north of the filament appears as a bright region at the top of the dark band in Figure \ref{fig:timedistance-event99}(a). This bright emission apparently followed the motion of the threads from 18:20 to 21:20 UT (end of event 107).

As with events 145 and 146*, the oscillation seems to be continuous between 18:20 UT and 23:26 UT. However, the event 107 and 108* oscillations differ significantly in phase, period, and damping time. Figure \ref{fig:timedistance-event99}(b) shows this discrepancy clearly: one fit (Eq.~(\ref{eq:oscillation-fit})) is very good in the first event and another is good in the second, except at the end of the event when the motion was obviously affected by the eruption. The differences between two events are equally evident in the measured velocities (Fig. \ref{fig:timedistance-event99}(c)). Before 18:20 UT the motions were small and disorganized, while after this time the damped oscillation is very clear. At 23:26 UT the period changed and the oscillation started to grow, ending in an eruption.

The main difference between events 107-108* and the amplified oscillation of \S \ref{subsec:case141} is that the filament erupts. However, similar explanations might apply for the amplification. The potential relationship between the amplified oscillation and the eruption is an intriguing topic for further study.

\section{Statistics}\label{sec:statistics}

The catalog consists of $\totalevents$ oscillation events which we found by analyzing six months of GONG data in cycle 24 (see Tables \ref{table:table_filaments0} to \ref{table:table3}). In about 43\% of the cases we identified the apparent trigger of the oscillation: $\nflare$ events were triggered by flares, $\npe$ by prominence eruptions, $\njet$ by a jet, and $\nmoreton$ by a Moreton wave. However, in $\nunk$ cases the triggering agent was not identified. In $\neruption$ cases the filament erupted during the temporal range analyzed.

As discussed in \S\ref{sec:intro} we classified the oscillations according to their maximum velocity amplitude as SAOs ($V < 10\kms$) and LAOs ($V > 10\kms$). Of the $\totalevents$ oscillation events there are $\nsao$ SAOs and $\nlao$ LAOs. Over the six months of the survey this averages to one oscillation event per day on the visible solar disk. The occurrence rate of one LAO event every two days implies that LAOs are a common phenomena on the Sun, in contrast to previous statements that LAOs are scarce \citep[e.g.,][]{tripathi2009}. We also found a similar rate for SAOs.

\begin{figure*}[!ht]
\centering\includegraphics[width=\textwidth]{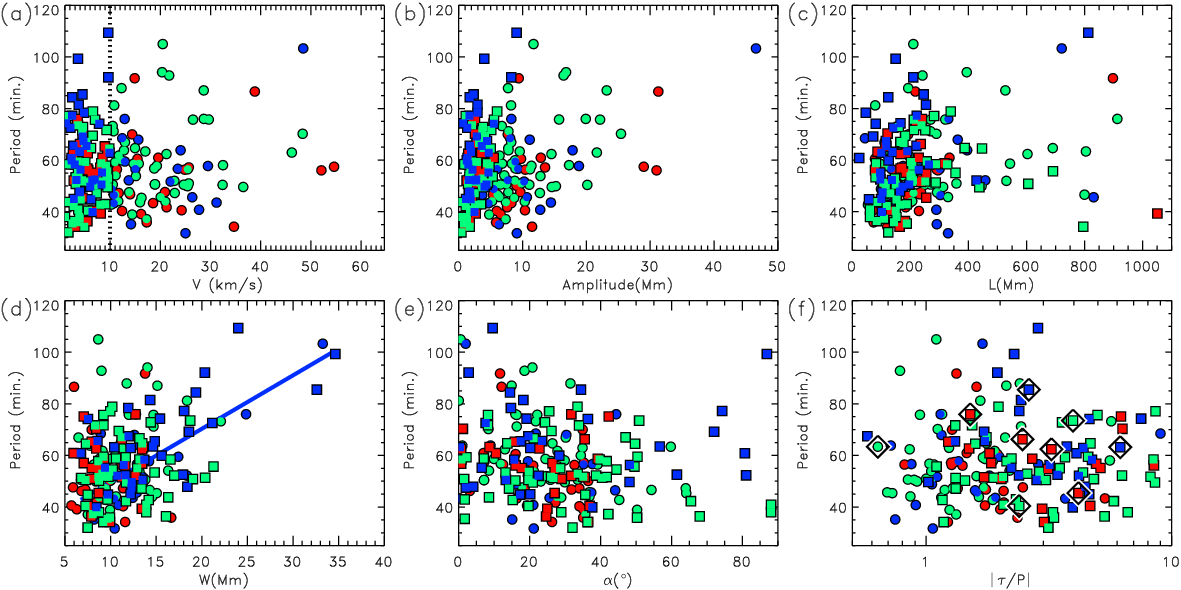}
\caption{Scatter plots of period, $P$, vs: (a) velocity amplitude, $V$. (b) displacement amplitude, $A$. (c) Length of the spine, $L$. (d) Width of the spine, $W$.  (e) Angle between the direction of motion and the spine, $\alpha$. (f) Damping time per period, $|\tau/P|$. The square symbols are for SAO events ($V<10 \kms$) and circles are for LAOs ($V\ge10 \kms$). For greater clarity the error bars are not plotted, but can be found in Tables \ref{table:table_filaments0}-\ref{table:table3}. The colors represent the filament type: active region (AR, red), intermediate (IT, green) and quiescent (QS, dark blue). The big black diamonds indicate events with negative values of $\tau/P$.\label{fig:multipanels1}}
\end{figure*}

\begin{figure}[!ht]
\includegraphics[width=0.4\textwidth]{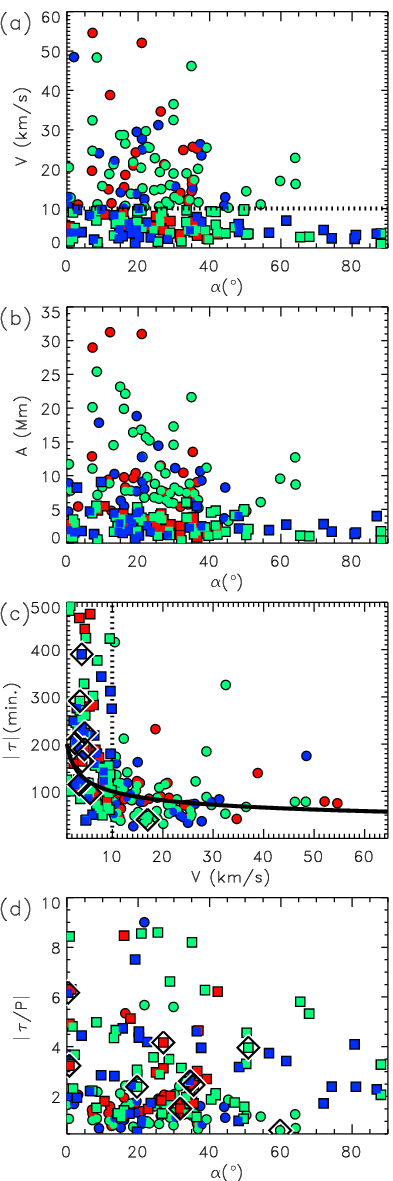}
\caption{Scatter plots of (a) damping time, $\tau$, vs $V$. (b) $|\tau/P|$ vs $\alpha$. (c) $V$ vs $\alpha$. (d) $A$ vs $\alpha$. Symbols and colors are as in Fig. \ref{fig:multipanels1}.\label{fig:multipanels2}}
\end{figure}

\begin{figure}[!ht]
\centering\includegraphics[width=0.4\textwidth]{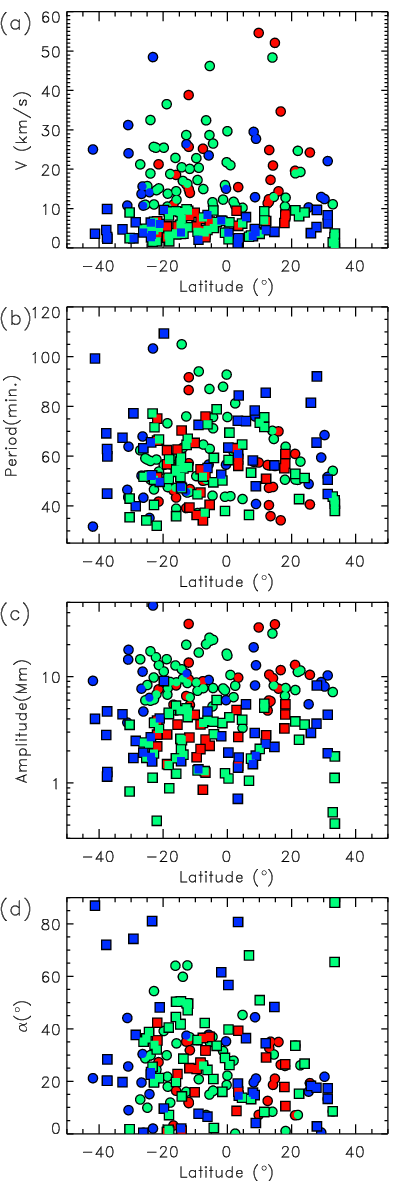}
\caption{Scatter plots of latitude vs: (a) $P$. (b)  $\alpha$. (c) $A$. (d) $V$. Symbols and colors are as in Fig. \ref{fig:multipanels1}.  \label{fig:multipanels3}}
\end{figure}

The data presented in the catalog enabled us to search for possible dependencies between pairs of filament and oscillation parameters: the velocity amplitude ($V$), oscillation period ($P$), damping time ($\tau$), damping time per period ($\tau/P$), displacement ($A$), and angle between the proper motion and the filament spine ($\alpha$). We also computed the Pearson correlation matrix \citep{neter1993} using the IDL subroutine \emph{correlate.pro}. The matrix elements are the correlations between pairs of parameters, and range from -1 to 1.  A linear correlation between two parameters yields an associated matrix element close to 1 (or -1). Although we found that the values of the matrix are small, in general, we will discuss those pairs of parameters whose correlations or lack thereof are interesting. Figures \ref{fig:multipanels1} to \ref{fig:multipanels3} show scatter plots of some pairs of these parameters.  In Figure \ref{fig:multipanels1} the scatter plots of the period, $P$, vs the other parameters are displayed in 6 panels (a-f). Figure \ref{fig:multipanels2} shows the damping parameters, $\tau$ or $\tau/P$, vs $v$ and $\alpha$ (panels a-e). Figure \ref{fig:multipanels3} plots several parameters vs solar latitude of the filament.  In these scatter plots, the LAOs and SAOs are plotted with circles and squares, respectively. 

\subsection{Velocity Amplitude, $V$}\label{subsec:velocityamplitudes}
\begin{figure*}[!ht]
\centering\includegraphics[width=0.99\textwidth]{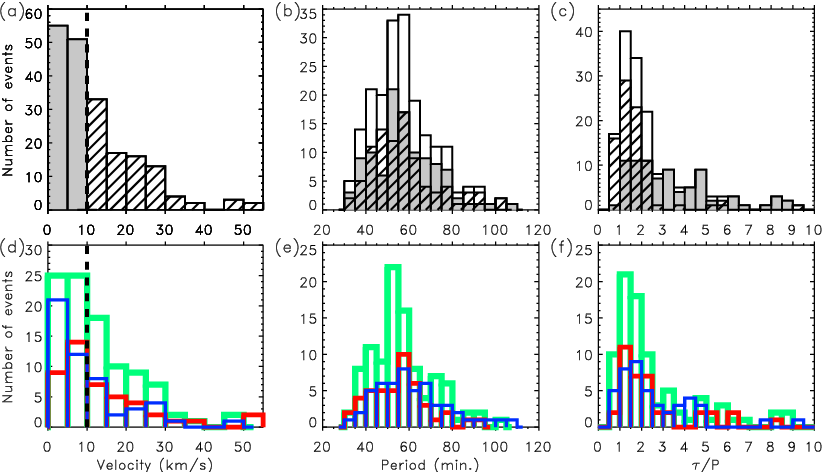}
\caption{Histograms of the number of events binned by $V$ (first column), $P$ (second column), and $|\tau/P|$ (third column). In the top row the shaded and striped areas represent SAO and LAO events, respectively, for three properties: $V$ (a), $P$ (b) and $|\tau/P|$ (c). In (a) the vertical dashed line indicates the separation between SAOs and LAOs at $V=10\kms$. In (b) and (c) the curve with a white area underneath is the histogram of the total number of events. In the bottom row, histograms of (d) $V$, (e) $P$, and (f) $|\tau/P|$, divided according to the three types of filaments: active region (AR, red), intermediate (IT, green) and quiescent (QS, blue) are shown. 
\label{fig:multihistogram1}}
\end{figure*}

In the survey we found velocity amplitudes from a few $\kms$ to $55\kms$ (see Tables \ref{table:table0} to \ref{table:table3}). Histograms of the velocity distribution for all events and the distribution according to filament type are plotted in Figure \ref{fig:multihistogram1}(a) and (d), respectively.  The vertical dotted line separates LAOs ($V>10\kms$) from SAOs ($V<10\kms$). The total number of LAO events decreases with the velocity amplitude, as expected: more energetic events are less frequent than less energetic ones. 
The velocity ranges for all filament types (AR - red, IT - green and QS - blue) are similar (Figure \ref{fig:multihistogram1}(d)), indicating that all types of filaments can support both SAOs and LAOs. The velocity distribution for each filament type follows the same trend as the total distribution except for AR filaments. The apparent rollover in the AR filament distribution below 5 $\kms$ probably reflects the difficulty in detecting small filaments and small-amplitude events by eye, implying that we have underestimated the number of SAOs. 

The histograms also do not distinguish two separate populations associated with large- and small-amplitude oscillations, regardless of the choice of LAO threshold (i.e., $10\kms$ or $20\kms$). Additionally, $\nsaosexcited$ of the $\nsao$ SAOs were clearly triggered by an identified energetic disturbance. These contradict the idea that the LAOs and SAOs have a different nature and they are triggered by different mechanisms  \citep{oliver2002,arregui2012}.

The $P-V$ scatter plot (Figure \ref{fig:multipanels1}(a)) and the small correlation $P-V$ value reveal no dependence of the velocity on the period, neither for all events nor for different filament types. In contrast, the $V-\alpha$ scatter plot (Fig. \ref{fig:multipanels2}(a)) shows a clear pattern: the $V$ range decreases with $\alpha$, and the $V$ values drop sharply for events with $\alpha$ beyond $40^\circ$. This tendency leads to no LAOs for $\alpha > 65^{\circ}$. 
The two populations can be also distinguished in the $A-\alpha$ scatter plot of Figure \ref{fig:multipanels2}(b) as we will discuss in \S \ref{subsec:amplitude}. The evident correlation between velocity amplitude and damping time will be discussed in \S \ref{subsec:dampingtime-per-period}.

\subsection{Period, $P$}\label{subsec:perids}
The period reflects the restoring force and the underlying physics of the oscillation. The period values range from 30 to 110 min for the total population, with a mean value of 58 min, a standard deviation of 15 min, and  a clear peak centered at $\sim$58 min (Figure \ref{fig:multihistogram1}(b)). The period distributions for LAOs (striped) and SAOs (shaded) have mean values and standard deviations comparable to those of the $P$ distribution for all events. This indicates that SAOs and LAOs are not two distinct populations of events with respect to their periods.



The period distributions for the three filament types do not differ significantly from each other or from the total distribution (Figure \ref{fig:multihistogram1}(e)). For IT filaments the mean period is 56 min $\pm$ 14 min; the distribution for AR filaments peaks at 57 $\pm$ 16 min; the mean period for QS filaments is 62 $\pm$ 17 min with long-period tail extending to 110 min. 
If LAOs were nonlinear, as discussed in \S \ref{sec:intro}, the period could depend on $V$ or $A$. However, Figures \ref{fig:multipanels1}(a) and \ref{fig:multipanels1}(b), together with the negligible $P-V$ and $P-A$ correlation elements, demonstrate that $P$ does not depend on either $V$ or $A$ for the catalogued events.  

Many theoretical models of MHD modes in filaments predict a relationship between the oscillation period and the filament length or width \citep[see review by][]{arregui2012}. To test this hypothesis, we plotted the oscillation period as a function of length $L$ and width $W$ in Figures \ref{fig:multipanels1}(c) and \ref{fig:multipanels1}(d), respectively. We found no correlation between $P$ and $L$ for all types. Although the period is not correlated with $W$ for AR and IT filaments, QS filament periods tend to increase with $W$. The correlation element is relatively large, 0.74, and the linear $P-W$ relationship is
\begin{equation}\label{eq:linear_trend_period_width_qs}
P_\mathrm{QS}=   23.4 \pm 0.4 + (2.31 \pm 0.02) W_\mathrm{QS} \, ,
\end{equation}where $P_\mathrm{QS}$ is in minutes and the errors in the period have been considered. The general tendency is for wider QS filaments to oscillate with longer periods than narrower prominences.
 
Figure \ref{fig:multipanels1}(e) shows that, for angles $\alpha<70^\circ$, the range of possible periods generally decreases with $\alpha$. For $\alpha<20^\circ$ the periods occupy the range from 30 to 110 min, whereas for $20^\circ<\alpha<40^\circ$ the periods range from 30 to 95 min and for $40^\circ<\alpha<70^\circ$ the range is from 30 to 80 min. Only a few cases have $\alpha>70^\circ$, and some of them do not follow this trend. $P$ decreases gradually with $\alpha$, so there is a no clear drop in $P$ for $\alpha>40^{\circ}$ as we found for $V$ (see \S\ref{subsec:velocityamplitudes}).

The decrease of $P$ with $\alpha$, in conjunction with the sharp decrease in $V$ at $\alpha>40^\circ$, suggests a connection with the polarization of the oscillations. Theoretical modeling predicts that oscillations along the field have longer periods than transverse oscillations. \citet{wang2016} and \citet{zhang2017a} observed simultaneous longitudinal and transverse oscillations in a prominence, and confirmed that the transverse oscillation period was shorter than the longitudinal period. At this point we are tempted to define longitudinal and transverse oscillations according to the $V-\alpha$ results: longitudinal for $\alpha<40^\circ$ and transverse for $\alpha>40^\circ$. High-resolution observations reveal that on-disk filaments are composed of many narrow, field-aligned threads oriented at a shallow angle to the spine \citep[e.g.][]{lin2005}, and often are composed of segments spaced along a common PIL. That is, the spine is not necessarily a coherent, magnetically continuous structure. Transverse oscillations involve coherent movement of the whole magnetic structure or magnetically linked portions thereof, whereas longitudinal oscillations involve individual thread motions at an angle with the spine. In almost all catalog events, only an small fraction of the filament oscillates, suggesting that the local, rather than global, magnetic field is engaged. Further study of individual, well-observed events is needed to resolve whether $\alpha$ is a reliable marker of the boundary between transverse and longitudinal events.

\begin{figure}[!ht]
\includegraphics[width=0.45\textwidth]{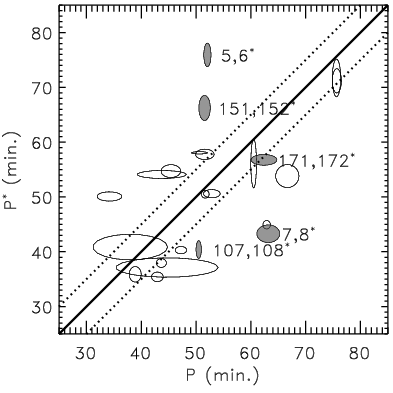}
\caption{Scatter plot of the two periods in double events in the same filament. $P$ is the first oscillation period and $P^{*}$ is the subsequent one. The data are shown as ellipses where the vertical semi-axis is the error bar for $P^{*}$ and the horizontal semi-axis is the error bar for $P$. Shaded ellipses are for double events including one amplified oscillation, with the relevant event numbers written on the side of each ellipse. \label{fig:period-duplicated}}
\end{figure}

In \S \ref{subsec:case135} we reported two consecutive oscillations in the same filament during the same data sequence. The periods of both events, $P$ and $P^{*}$, agreed, suggesting that the common period is the characteristic period of oscillation of the structure \citep{ramsey1966}. However, in the cases described in \S \ref{subsec:case141} and \S \ref{subsec:case99}, $P$ and $P^{*}$ are clearly different. In Figure \ref{fig:period-duplicated} the scatter plot of $P^{*}$ vs $P$ is shown for all the double events in the catalog. For several cases the ellipse is inside or close to the region of  $P^{*} \sim P \pm 5 \min$ (region between the two dotted-lines). For these cases, we can reasonably consider the oscillation as a characteristic of the system. The shaded ellipses correspond to the double cases with amplified oscillations, which in some events were probably associated with flaring activity near the filament (see, e.g., \S \ref{subsec:case141} and \S \ref{subsec:case99}). Thus, these oscillations were probably forced and are not characteristic motions. However, more cases exhibit significant differences between $P^{*}$ and $P$ not associated with oscillation amplification. In almost all of these cases we found that the substantial period differences were associated with reconfiguration of the filament structure. For example, in cases 174-175* and 186-187* the filament structures change with time, judging from the observed flows along the slit and movements of the equilibrium position of the filament. 

\subsection{Damping, $\tau$ and $\tau/P$}\label{subsec:dampingtime-per-period}
$|\tau/P|$ measures the number of oscillations within the characteristic damping time. The absolute value of $\tau/P$ is considered because  $\tau$ is negative when an oscillation is amplified with time, as discussed in \S\S \ref{subsec:case141} and \ref{subsec:case99}.  A large value of  $|\tau/P|$ indicates  weak damping, while a small ratio indicates strong damping.  
The $|\tau/P|$  histogram (Fig. \ref{fig:multihistogram1}(c)) for all events extends from 0.6 to 2711 (not shown in the histogram), and peaks at  $|\tau/P|=1.25$.  Most events are strongly damped ($|\tau/P|<3$), and a significant number are very strongly damped ($|\tau/P|<1$). A  value of $|\tau/P| \ge 10$ essentially signifies an undamped oscillation. 
In contrast with the $V$ and $P$ distributions considered above, the  $|\tau/P|$ distributions for SAOs and LAOs clearly differ: the SAO distribution is wide, with a peak close to 1.75, while the LAO distribution is  narrower with a peak near 1.25 and  scattered points at larger values of $|\tau/P|$. The LAO events ($V > 10 \kms$) are mainly below $|\tau/P|=3$ while SAOs cover a larger range.  The distributions for the 3 filament types appear similar (Fig. \ref{fig:multihistogram1}(f)) to the total $|\tau/P|$ distribution.


Figure \ref{fig:multipanels2}(c) shows that larger velocity amplitudes are positively correlated with stronger damping,  which indicates that the higher-speed oscillations are likely to be nonlinear. The sharp transition in the $|\tau|$ range at $V=10\kms$ divides LAOs from SAOs in Figure \ref{fig:multipanels2}(a), reflecting a distinct boundary between linear and nonlinear oscillations. The scatter plot $|\tau/P|$-$V$ is not shown but resembles that of Figure \ref{fig:multipanels2}(a) with the same trend: the damping time $|\tau/P|$ decreases as $V$ increases.  

\citet{zhang2013} found a nonlinear relationship between $\tau$ and $V$ in their simulations of prominence mass formation: $\tau \sim V^{-0.3}$.  This scaling law (solid black line in Figure \ref{fig:multipanels2}(c)) is roughly consistent with observed and derived values from our events, suggesting that LAOs may be damped through radiative cooling. In their model, each flux tube supporting a cool thread has two coronal segments that connect the thread with the chromosphere at both footpoints. The oscillations alternately compress and rarefy both segments, heating or cooling the coronal plasma. The combined density and temperature increases raise the radiative losses, thus damping the oscillations. The \citet{zhang2013} model predicted that this effect could yield a temperature variation of several hundred thousand Kelvins, which should be observable in some EUV lines. An alternative mechanism that can explain strong damping is the mass accretion associated with thermal nonequilibrium \citep{luna2012b,ruderman2016}, when evaporated chromospheric plasma continually condenses onto the prominence threads. In this model the damping is not related directly to the oscillation velocity. However, events with larger $V$ are associated with violent events, which could produce increased evaporation and consequently stronger damping. A combination of mass accretion and radiative damping is also possible.

$|\tau/P|$ and dimensions $L$ and $W$ are uncorrelated (the corresponding correlation elements are close to zero), implying that the damping is not related to the prominence size. The building blocks of prominences are cool, elongated threads aligned with the magnetic field, so the damping process is probably associated with the local magnetic or plasma characteristics and not with the global dimensions of the filament. Similarly $|\tau/P|$ is uncorrelated with $P$ or $A$.

The $\tau/P$-$\alpha$ scatter plot (Fig. \ref{fig:multipanels2}(d)) shows a decreased range of $\tau/P$ for $\alpha>40^\circ$. This behavior is similar to the $V$-$\alpha$ (\S \ref{subsec:velocityamplitudes}) and the $A$-$\alpha$ scatter plots, as we will discuss in \S \ref{subsec:amplitude}. 

Events $\listampli$ were characterized by amplified oscillations ($\tau<0$). In Figures \ref{fig:multipanels1} to \ref{fig:multipanels3} these cases are marked by symbols surrounded by a big diamond. Cases 6*, 7, 65, 171 are similar to 108* and 152*: an amplified oscillation prior to a filament reconfiguration or eruption. Case 38 is less clear but probably is associated with reconfiguration. The amplification in cases 134 and 156 is not evident in the time-distance diagrams, and might be associated with filament proper motions. These amplified oscillations are very interesting and deserve to be studied in greater depth.

\subsection{Displacement, $A$}\label{subsec:amplitude}

\begin{figure*}[!ht]
\centering\includegraphics[width=0.99\textwidth]{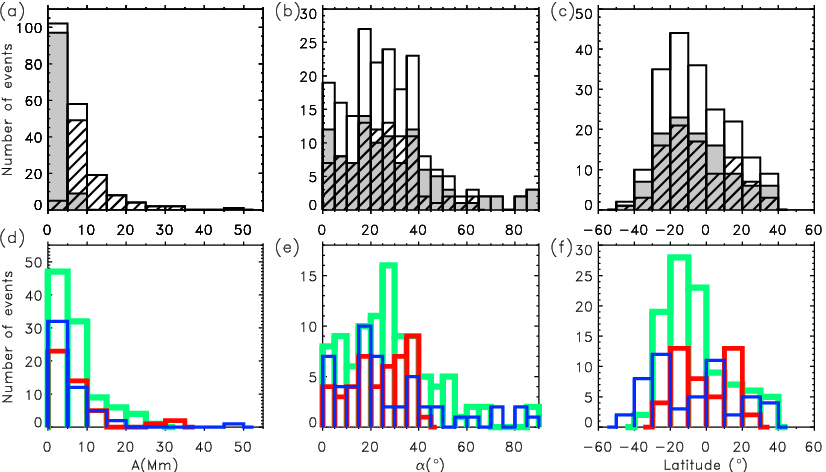}
\caption{Histograms of the number of events binned by $A$ (first column), $\alpha$ (second column), and latitude (third column). Panels and annotations are as in Fig. \ref{fig:multihistogram1}. \label{fig:multihistogram2}}
\end{figure*}

The maximum displacement of the filament mass with respect to the equilibrium position during the fitted oscillation, $A$, was derived from Equation (\ref{eq:oscillation-fit}):
\begin{equation}
A=MAX(|A_0 e^{-A_1 (t -t_0)} \cos \left[ A_2 (t - t_0) +A_3 \right]|) \, .
\end{equation}
The distributions of $A$ for SAOs and LAOs differ substantially (Figure \ref{fig:multihistogram2}(a)). For SAOs, the distribution is concentrated at the origin with a large peak in the range 0-5 Mm, many fewer events between 5-10 Mm, and no events with $A > 10$ Mm. In contrast, LAO displacements cover a larger range ($A$ = 0-50 Mm), with a peak at 7.5 Mm. The $A$ distributions for the three filament types are similar, with a maximum in the range 0-5 Mm and a decreasing number of events for increasing $A$ (Fig. \ref{fig:multihistogram2}(d)).

In the $P-A$ scatter plot (Fig. \ref{fig:multipanels1}(b)),  SAOs are concentrated at  $A <  10$ Mm while LAOs extend up to $A=46$ Mm. Note that no events have large $A$ and low $P$. Because the velocity amplitude is approximately $V\sim A/P$, the region of large $A$ and low $P$ corresponds to very large $V$ values where no events were found in our survey. Recently, we discovered an oscillation event with the largest velocity amplitude reported thus far ($100 \kms$ ) and a displacement of more than 50 Mm \citet{luna2017}, which would fit in the empty region of Figure \ref{fig:multipanels1}(b).

In the $A-\alpha$ scatter plot (Fig. \ref{fig:multipanels2}(b)) we see that the range of displacements is reduced when $\alpha$ increases. Similar to the $V-\alpha$ or $\tau/P$-$\alpha$ plots, $A$ drops significantly for $\alpha>40^{\circ}$ and there are no LAOs for $\alpha > 65^\circ$. This suggests that the oscillation or excitation mechanisms differ on either side of $\alpha=40^{\circ}$, as discussed in \S\S \ref{subsec:velocityamplitudes} and \ref{subsec:dampingtime-per-period}. Figure \ref{fig:multipanels3}(c) shows that $A$ is independent of the filament latitude.

\subsection{Direction of motion $\alpha$}\label{subsec:direction-motion}
The parameter $\alpha$ is the angle between the direction of the oscillation and the filament spine (\S \ref{sec:description-examples}).  
Within the catalog we found oscillations in any direction from 0$^\circ$ to 90$^\circ$ (Fig. \ref{fig:multihistogram2}(b)). The total distribution has a peak close to 18$^\circ$ and a mean value of  $27^\circ \pm 18^\circ$. For LAOs, the maximum is $\sim$28$^\circ$ with a mean of $25^\circ \pm  14^\circ$, while for SAOs the peak also is close to 18$^\circ$ and the mean value is 29$^\circ \pm 21^\circ$. The number of events decreases for $\alpha > 40^\circ$ and only SAOs have $\alpha > 65^\circ$, as we found for the $V-\alpha$,$|\tau/P|$-$\alpha$ and $A-\alpha$ scatter plots (\S\S \ref{subsec:velocityamplitudes}, \ref{subsec:dampingtime-per-period}, and \ref{subsec:amplitude}). Therefore we define two populations of oscillations with respect to $\alpha$: $\nlong$ events with $\alpha < 40^\circ$ and $\ntrans$ with $\alpha > 40^\circ$.

Figure \ref{fig:multihistogram2}(b) shows that LAOs and SAOs have similar $\alpha$ distributions. The mean values  are consistent with direct measurements of the angle between the filament magnetic field and its spine \citep[$\alpha \sim 25^\circ$ on average;][]{Leroy1983a,Leroy1984a,tandberg1995,trujillo2002,casini2003,Lopez-Ariste2006a}. This suggests that most of the oscillations in the catalogued events are aligned with the magnetic field (longitudinal).


The $\alpha$ distribution is clearest for IT events (Fig. \ref{fig:multihistogram2}(e)): the oscillations are aimed in all directions, but the peak coincides with the mean at 25$^\circ \pm$ 13$^\circ$. The $\alpha$ distribution for AR events has a peak at 37.5$^\circ$ with a mean value of 25$^\circ \pm$ 14$^\circ$. Interestingly, there are no oscillations in AR filaments with $\alpha>45^{\circ}$, indicating that the motions are mainly longitudinal. For QS filaments the distribution has a maximum around 18$^\circ$ and a mean value at 22$^\circ \pm$ 20$^\circ$. The QS $\alpha$-distribution  covers the entire domain, but the oscillation offsets are mainly below $40^{\circ}$. In summary, the mean $\alpha$ values for all filament types agree with the observed magnetic-field orientation relative to the spine, implying longitudinal polarization, particularly for IT filaments. 

\subsection{Latitude}\label{subsec:latitude}
Prominence oscillations may reflect the global structure of the supporting filament channels, which is intrinsically tied to the large-scale solar magnetic field. Figure \ref{fig:multipanels3} displays several oscillation properties ---  $V$, $P$, $A$, and $\alpha$ --- as functions of solar latitude in Stonyhurst Heliographic Coordinates \citep[][]{thompson2006}. $V$, $P$, and $\alpha$ generally display larger ranges of values at specific latitudes. For IT filaments, these oscillation properties largely occupy the region between $-25^{\circ}$ and $0^{\circ}$ latitudes (see Figures \ref{fig:multipanels3}(a), (b), and (d)). In contrast, AR events exhibit larger ranges of these properties around two latitudes, $-15^{\circ}$ and $15^{\circ}$. For $A$ this trend is less evident (Figure \ref{fig:multipanels3}(c)), but a 2D histogram (not shown) reveals the same trend.

The latitude distribution (Figure \ref{fig:multihistogram2}(c)) shows that all survey events were located between $50^\circ$ and $-50^\circ$, typical for solar maximum. However, a substantial fraction of events accumulated around $-15^\circ$, in the southern hemisphere, regardless of oscillation type (SAO or LAO). In Figure \ref{fig:multihistogram2}(f) the latitude distributions for the three filament types are shown. The distribution peaks at $-15^{\circ}$ and $15^{\circ}$ for AR filaments, at $-15^{\circ}$ for IT filaments, and at $-25^{\circ}$ and $5^{\circ}$ for QS filaments. It is evident From Figures \ref{fig:multihistogram2}(c) and \ref{fig:multihistogram2}(f) that the regions of a large number of events coincide with the regions of large dispersion of oscillation parameters of Figure \ref{fig:multipanels3}. This suggests that in those latitudes there are more filaments and more activity triggering oscillations. In this sense, the existence of these latitudes is not necessarily showing a latitudinal dependence of oscillation parameters or intrinsic characteristics of the filaments.

\cite{bashkirtsev1993} found a smooth, sinusoidal latitudinal dependence for 30 SAO events observed over more than 8 years, with periods of 80 min at $-20^{\circ}$ and $20^{\circ}$ latitudes and 40 min at $0^{\circ}$. We have not found a clear relationship between the periods or other properties and the filament latitude. Their study covered almost a solar cycle, so their latitudinal dependence could be related to the well-known migration of filaments from the poles toward the equator during the cycle. To determine whether this potentially profound relationship is solid, our catalog would have to be expanded significantly to include oscillation events throughout at least 1 solar cycle.

\section{Seismology}\label{sec:seismology}

Prominence seismology combines observations and theoretical modeling to infer hard-to-measure parameters such as the magnetic field (see \S \ref{sec:intro}). There are essentially three driving mechanisms for prominence oscillations: gravitational force, pressure imbalance, and magnetic Lorentz force. 

Longitudinal oscillations are driven by a combination of gravity projected along the field \citep[pendulum model,][]{luna2012b} and gas pressure gradients \citep[slow modes,][]{joarder1992b}. In the pendulum model, the period depends exclusively on the radius of curvature of the dips supporting the cool prominence plasma, $R$. \citet{luna2012c} and \citet{zhang2013} determined that gas pressure gradients contribute negligibly to the restoring force when the radius of curvature is much smaller than a limit defined by the prominence characteristics ($R \ll R_\mathrm{lim}$), where $R_\mathrm{lim}$ is
\begin{equation}\label{eq:pendlim}
R_\mathrm{lim}= 1/4 L_t (L_f-L_t) \, \kappa \, g /c_{sc}^2 . 
\end{equation}
Here $\kappa$ is the temperature contrast between the cool and adjacent hot plasmas, $L_f$ is the field line length, $L_t$ is the thread length, $g$ is the solar gravitational constant, and $c_{sc}$ is the coronal sound speed. In that case the period is
\begin{equation}\label{eq:pendulumperiod}
P=2\, \pi \sqrt{\frac{R}{g}} \, .
\end{equation}

Assuming that the magnetic tension in the dipped part of the tubes must be larger than the weight of the threads, the minimum magnetic-field strength, $B$, depends on the particle number density of the prominence thread, $n$, and the period $P$. In the absence of direct density measurements, \citet{luna2014} adopted the range of typical values $n=10^{10}-10^{11} \mathrm{cm^{-3}}$ as the main source of uncertainty and determined that   
\begin{equation}\label{eq:pendulumB}
B(G) \ge (0.28 \pm 0.15) \, P(\mathrm{min}) \, .
\end{equation}

For transverse horizontal oscillations, \citet{kleczek1969} assumed that the filament was supported by a single line-tied magnetic flux tube, and that the restoring force was supplied by magnetic tension. We assume again that $n$ takes typical prominence values, and using their Eq. (9), we find
\begin{equation}\label{eq:kkB-seismology}
B(G)= (5.5 \pm 3) \frac{L(\mathrm{Mm})}{P(\mathrm{min})} \, ,
\end{equation}
where $L$ is the length of the filament. The uncertainty in the numerical coefficient is associated with the uncertainty in $n$.

\begin{figure}[!ht]
\centering\includegraphics[width=0.45\textwidth]{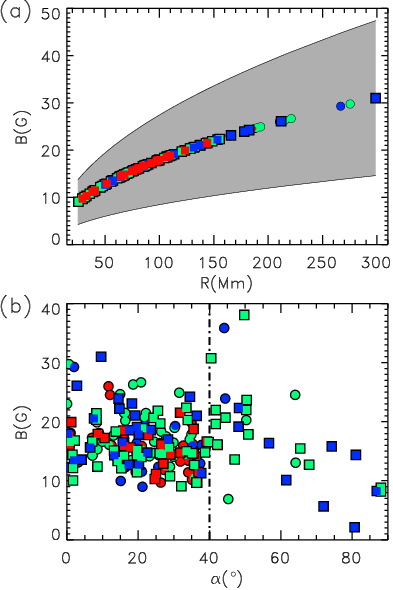}
\caption{Seismology diagnostics for longitudinal and transverse oscillations. (a) The lower limit on $B$ as a function of $R$ for longitudinal oscillations. The shaded area corresponds to the uncertainty range. (b) The estimated magnetic field strength for events with longitudinal and transverse oscillations, from Equations (\ref{eq:pendulumB}) and (\ref{eq:kkB-seismology}) respectively. The vertical dot-dashed line indicates the assumed separation between longitudinal and transverse oscillations. Symbols and colors are as in the scatter plots. \label{fig:seismology}}
\end{figure}

Without additional data analysis and field extrapolation \citep[e.g.,][]{luna2017}, it is difficult to establish which catalog events are oscillations parallel or perpendicular to the magnetic field. However, our statistical analysis revealed a clear distinction between oscillations with $\alpha < 40^{\circ}$ and those with $\alpha > 40^{\circ}$ (\S \ref{sec:statistics}). Although the two populations are not necessarily uniquely associated with different oscillation polarizations, for seismology purposes we applied the longitudinal model to the oscillations with $\alpha < 40^{\circ}$ and the transverse model to the $\alpha > 40^{\circ}$ cases. This is also justified because the two models predict approximately the same $B$ for a given event. We determined $B$ and $R$ from Equations (\ref{eq:pendulumperiod}) and (\ref{eq:pendulumB}) for the events with $\alpha < 40^{\circ}$ (Figure \ref{fig:seismology}(a)). The shaded area covers the uncertainties in $B$. The magnetic field ranges from 9 to 48 $G$, and $R$ from 25 to 300 Mm. The mean values are $B=16$ G and $R=89$ Mm. The obtained values are consistent with the rare direct measurements of prominence magnetic fields \citep[see review by][]{mackay2010}. 

The magnetic field plotted in Figure \ref{fig:seismology}(a) is a lower limit, so we expect larger values to occur. In particular the field could be significantly underestimated for small radii of curvature, $R$. The reason is that the magnetic tension is proportional to $B^2/R$ and the weight of the prominence is proportional to $n \, g$. Thus, assuming similar $n$, the $B$ necessary to balance the gravity is smaller for smaller $R$ than for larger $R$.

In order to check the validity of the pendulum model, we computed Equation \ref{eq:pendlim} and compared it with $R$ for all catalog cases. Because we do not have direct measurements of $L_f$ and $L_t$, we used $L$ and $W$, the length of the spine and width of the filament. $W$ is probably comparable to the thread lengths, but $L$ is a lower limit on the length of the sheared field lines in the filament channel for $\alpha > 0$. $c_{sc}$ is typically $\sim200 \, \kms$ and the typical temperature contrast is $\kappa=100$. The resulting $R_\mathrm{lim}$ is largely greater than $R_\mathrm{lim}$, demonstrating the applicability of the pendulum model to the catalog events.

Figure \ref{fig:seismology}(b) shows the inferred magnetic field as a function of $\alpha$. The pendulum model (Eq. (\ref{eq:pendulumB})) is used for events with $\alpha < 40^\circ$, and transverse model (Eq. (\ref{eq:kkB-seismology})) for $\alpha > 40^\circ$. For longitudinal oscillations ($\alpha < 40^\circ$) the $B$ range generally decreases with $\alpha$, reminiscent of the behavior of $P$. The same trend applies to the transverse oscillations ($\alpha > 40^\circ$), although some events reach large $B$ values ($38$ G). For transverse oscillations the $B$ values are consistent with direct measurements \citep[see, e.g.,][]{harvey1969}. Our AR events are all longitudinal, while IT and QS events occupy both categories. It is interesting to note that the minimum field strengths do not differ significantly among the filament types, although AR filaments are embedded in higher field-strength regions. This lower limit is consistent with direct measurements in AR filaments \citep{kuckein2009,sasso2010,kuckein2012,sasso2014} showing strong fields of up to several hundred Gauss.

\section{Summary and Conclusions}\label{sec:conclusions}


In this work we have surveyed prominence oscillations detected through visual inspection of the GONG network H$\alpha$ data during January - June 2014, providing an extensive sample of events close to solar maximum of cycle 24. We have catalogued a large variety of oscillations including strongly damped motions, undamped oscillations, and amplified oscillations, enabling the first statistically significant study of filament oscillations and their pertinent properties. The filament and oscillation parameters are described in the text and Tables; additional information and animations can be found in the online catalog: \url{http://www.iac.es/galeria/mluna/pages/gong-catalogue-of-laos.php}. 

We have found $\totalevents$ oscillation events, including $\nsao$ SAOs and $\nlao$ LAOs. In 85 cases we have identified the triggering agents of the oscillations as flares, prominence eruptions, a jet, and a Moreton wave. For the remaining $\nunk$ events the triggering agent is not identified. The occurrence rate of one LAO event every two days implies that LAOs are common phenomena on the Sun, as are SAOs. 

We have parametrized the oscillations by fitting an exponentially decaying sinusoid, and statistically the distributions and correlations of key physical parameters. 
The fitted velocity amplitudes, $V$, are in the range $1.6 - 55 \kms$, and show a clear tendency to occur less frequently with increased $V$. This indicates that the LAOs are less common than SAOs, particularly since we probably underestimated the number of SAOs approaching the small-amplitude limit. The $V$ range decreases with $\alpha$, dropping sharply for events beyond $40^\circ$, and there are no LAOs for $\alpha > 65^{\circ}$.

The oscillation periods, $P$, range from 32 to 110 min. Surprisingly, the periods of both LAOs and SAOs have well-defined distributions centered at $P=58 \pm 15$ min. This indicates that LAOs and SAOs are not two distinct populations of events with respect to their periods. For all three filament types the mean oscillation period is around 1 hour. The $P$ range decreases with the angle between the oscillation displacement and the filament spine, $\alpha$. In general, we have not found strong correlations between $P$ and other oscillation parameters.


The damping time per period, $\tau/P$ covers a large range, including some cases with negative values (amplification). The $\tau/P$ distribution for LAOs peaks at 1.25, and most of the events exhibit very strong damping. For SAOs, the range of observed $\tau/P$ values is wider, peaking at 1.75. The three filament types behave similarly. For LAOs $\tau$ and $\tau/P$ decrease with $V$, regardless of filament type, confirming that LAOs involve nonlinear motions with velocity-dependent damping. This is a very interesting result because the kinetic energy involved in large-amplitude oscillations is enormous, due to the combination of large thread masses and large velocities. Therefore the physical mechanism must be efficient enough to damp the substantial motion in a few oscillations. Our earlier theoretical studies showed that reasonable rates of mass accretion could explain the observed damping rates. On the other hand, the observed relation between $\tau$ and $V$ is consistent with the \citet{zhang2013} scaling law, $\tau \sim V^{-0.3}$, which suggests that the damping is associated with radiative cooling. More observational and theoretical work needs to be done to understand the damping process more thoroughly.

For the catalog events, the direction of the motion with respect to the filament spine, $\alpha$, covers all possible angles between $0^\circ$ and $90^\circ$, and the $\alpha$ distributions for LAOs and SAOs exhibit no clear peak. However, the mean $\alpha$ value is the same for all three filament types: $27^\circ$, which agrees with previous direct measurements of $\alpha \sim 25^\circ$ on average \citep{Leroy1983a,Leroy1984a,tandberg1995,trujillo2002,casini2003,Lopez-Ariste2006a}. Thus, most of the oscillation displacements are probably aligned with the filament magnetic fields. 

We have not found evidence of any relationships between the oscillation parameters and the solar latitude, in contrast to the findings of \citet{bashkirtsev1993}. However, their study covered almost a solar cycle, and their latitudinal dependence could be associated with the well-known migration of filaments from poles to equator. To determine whether this profound relationship is solid, our catalog must be expanded to include events throughout at least 1 solar cycle .


We have applied seismological techniques to the entire catalog. For the longitudinally oscillating cases, we determined the radius of curvature of the magnetic dips hosting the prominence, $R$, and the minimum field strength, $B$, required to support the mass against gravity. $R=25-300$ Mm and $B=2 - 38$ G with mean values of $R=89$ Mm and $B=17$ G. For transverse oscillations, the magnetic field strength derived from the magnetic restoring force yields a wider range of $B=2 - 38$ G but a similar mean value.


Most of the oscillations are longitudinal, with the motion directed along the local magnetic field. Surprisingly, the period distributions for both SAOs and LAOs have a strong peak centered at 58 min, which implies that most solar filaments share a common structure. Namely, their structure is composed of dipped flux tubes with a radius of curvature of $\sim$90 Mm and an angle between the threads and the spine of $\sim$30$^\circ$. The magnetic-field strength is probably larger than the minimum estimate of 16 G. We also found that many SAOs are initiated by energetic disturbances, which contradicts the idea that SAOs are exclusively driven by photospheric or chromospheric waves. On the other hand, \citet{ning2009} and \citet{hillier2013} studied numerous oscillations in small prominence features, and found velocities in general below 10 $\kms$ and periods of the order of minutes. These localized versions of SAOs are more consistent with wave driving than our SAOs, which affect large portions or the entire filament.

In future research we will extend the catalog to events near the solar minimum of the same cycle 24, to augment our statistics and explore the possibility that oscillation parameters and filament properties evolve during the solar cycle. We invite the community to utilize this catalog for other research projects and to aid in expanding its contents, in order to advance our understanding of the fundamental structure and evolution of solar prominences. 

\acknowledgements

The Global Oscillation Network Group (GONG) Program is managed by the NSO and operated by AURA, Inc. under a cooperative agreement with the NSF. The data are acquired by instruments operated by the Big Bear Solar Observatory, High Altitude Observatory, Learmonth Solar Observatory, Udaipur Solar Observatory, Instituto de Astrof\'isica de Canarias, and Cerro Tololo Interamerican Observatory. The operation of Big Bear Solar Observatory is supported by NJIT, US NSF AGS-1250818, and NASA NNX13AG14G grants. This paper made use of the IAC Supercomputing facility HTCondor (\url{http://research.cs.wisc.edu/htcondor/}), partly financed by the Ministry of Economy and Competitiveness with FEDER funds, code IACA13-3E-2493. This research also made use of NASA Astrophysics Data System. 

This work was initiated during International Space Science Institute (ISSI) team 314 meetings in Bern led by M. Luna on ``Large-Amplitude Oscillations in Solar Prominences''. M. Luna acknowledges the support by the Spanish Ministry of Economy and Competitiveness through project AYA2014-55078-P. H. Gilbert, J. Karpen, T. Kucera and K. Muglach acknowledge support by the NASA Heliophysics Guest Investigator program. J. Terradas and J. L. Ballester want to thank the financial support from MINECO AYA2014-54485-P and FEDER Funds, and the Conselleria d'Innovaci\'o, Recerca i Turisme del Govern Balear to IAC3.

\appendix 

\section{Event Catalog}\label{annex:list-events}

\begin{table}[H]
\centering
    \begin{tabular}{| c | c |c | c | c | c | c | c | c |}
    \hline
    Event \# & Time & Tel. & Pos. (x,y) & Type & L (Mm) & W (Mm) & Trigger & Erupts\\ \hline 
1 &  1-Jan-2014 16:50 & C & -18, -76 & IP & 269 & 15 & FLARE &  \\ \hline
2* &  1-Jan-2014 16:50 & C & -18, -76 & IP & 269 & 15 & FLARE &  \\ \hline
3 &  4-Jan-2014 22:04 & B & -251, 302 & IP & 172 & 7 & FLARE &  \\ \hline
4 &  5-Jan-2014 17:03 & C & -399, 48 & IP & 87 & 14 & FLARE &  \\ \hline
5 &  5-Jan-2014 17:03 & C & -308, -116 & AR & 240 & 13 &   &  \\ \hline
6* &  5-Jan-2014 17:03 & C & -308, -116 & AR & 240 & 13 &   &  \\ \hline
7 &  5-Jan-2014 17:03 & C & 730, -430 & QS & 139 & 10 & PE &  \\ \hline
8* &  5-Jan-2014 17:03 & C & 730, -430 & QS & 139 & 10 & PE &  \\ \hline
9 &  5-Jan-2014 22:11 & B & 806, -79 & IP & 206 & 11 &   &  \\ \hline
10 &  6-Jan-2014 08:59 & U & 84, 492 & QS & 44 & 9 & FLARE &  \\ \hline
11 &  6-Jan-2014 07:11 & U & -132, -261 & IP & 185 & 8 &   &  \\ \hline
12 &  6-Jan-2014 16:56 & C & -268, -466 & QS & 102 & 8 &   &  \\ \hline
13 &  6-Jan-2014 16:56 & C & -184, 57 & AR & 78 & 12 &   &  \\ \hline
14* &  6-Jan-2014 16:56 & C & -184, 57 & AR & 78 & 12 &   &  \\ \hline
15 &  6-Jan-2014 08:59 & U & 57, 304 & IP & 119 & 8 & FLARE &  \\ \hline
16 &  7-Jan-2014 09:13 & U & -335, 78 & IP & 387 & 13 & FLARE &  \\ \hline
17 &  7-Jan-2014 09:13 & U & -325, -282 & IP & 190 & 7 & FLARE &  \\ \hline
18 &  7-Jan-2014 09:13 & U & 143, -370 & IP & 445 & 8 &   &  \\ \hline
19 &  7-Jan-2014 16:58 & C & -100, -501 & QS & 143 & 12 & PE &  \\ \hline
20 &  8-Jan-2014 08:59 & U & -139, -316 & IP & 313 & 14 & PE &  \\ \hline
21 &  8-Jan-2014 08:59 & U & 8, -494 & IP & 187 & 10 &   &  \\ \hline
22* &  8-Jan-2014 08:59 & U & 8, -494 & IP & 187 & 10 &   &  \\ \hline
23 &  8-Jan-2014 08:59 & U & 260, -367 & IP & 543 & 9 & FLARE &  \\ \hline
24 &  9-Jan-2014 17:10 & C & 155, -306 & IP & 287 & 20 & FLARE &  \\ \hline
25 &  9-Jan-2014 17:10 & C & 242, -505 & IP & 303 & 12 &   &  \\ \hline
26 & 10-Jan-2014 14:03 & T & 433, 80 & IP & 54 & 9 &   &  \\ \hline
27 & 11-Jan-2014 16:43 & C & 815, -393 & IP & 221 & 9 & PE &  \\ \hline
28 & 15-Jan-2014 20:00 & B & 502, -217 & AR & 125 & 6 &   &  \\ \hline
29 & 16-Jan-2014 07:37 & U & 268, 125 & AR & 80 & 15 &   &  \\ \hline
30 & 16-Jan-2014 19:59 & B & -624, -298 & IP & 228 & 9 &   &  \\ \hline
31 & 24-Jan-2014 17:00 & C & -311, -385 & IP & 281 & 10 & FLARE &  \\ \hline
32 & 24-Jan-2014 19:11 & M & 284, -151 & IP & 394 & 14 &   &  \\ \hline
33 & 25-Jan-2014 07:40 & U & -184, -376 & IP & 251 & 8 &   &  \\ \hline
34 & 25-Jan-2014 16:41 & C & -110, -129 & AR & 163 & 10 & FLARE &  \\ \hline
35* & 25-Jan-2014 16:41 & C & -110, -129 & AR & 163 & 10 & FLARE &  \\ \hline
36 & 25-Jan-2014 16:41 & C & 508, -95 & IP & 527 & 15 & FLARE &  \\ \hline
37 & 26-Jan-2014 08:27 & U & 850, 99 & IP & 81 & 8 &   &  \\ \hline
38 & 27-Jan-2014 16:41 & C & 335, -142 & AR & 255 & 9 &   &  \\ \hline
39* & 27-Jan-2014 16:41 & C & 335, -142 & AR & 255 & 9 &   &  \\ \hline
40 & 27-Jan-2014 16:41 & C & 449, -402 & IP & 209 & 17 &   &  \\ \hline
41 & 28-Jan-2014 20:23 & B & 635, -420 & IP & 247 & 9 &   &  \\ \hline
42 & 29-Jan-2014 07:23 & U & 445, -304 & IP & 82 & 11 &   &  \\ \hline
43 & 29-Jan-2014 16:49 & C & 556, -428 & IP & 251 & 9 &   &  \\ \hline
44 & 29-Jan-2014 19:59 & B & -822, 11 & IP & 182 & 11 & FLARE &  \\ \hline
45 & 30-Jan-2014 07:23 & U & 702, 33 & IP & 195 & 9 & PE &  \\ \hline
46 & 31-Jan-2014 15:20 & C & -557, 16 & IP & 201 & 15 &   &  \\ \hline
47 &  1-Feb-2014 07:17 & U & -404, 58 & AR & 317 & 9 & FLARE &  \\ \hline
48 &  5-Feb-2014 17:07 & C & -812, -129 & IP & 206 & 10 & FLARE & Y\\ \hline
49 &  5-Feb-2014 17:07 & C & -212, -269 & IP & 205 & 10 & FLARE &  \\ \hline
50 &  6-Feb-2014 17:05 & C & -626, -433 & IP & 363 & 13 & FLARE &  \\ \hline
51 &  6-Feb-2014 17:05 & C & -388, -266 & IP & 231 & 10 & FLARE &  \\ \hline
52 &  7-Feb-2014 13:24 & T & -506, 243 & IP & 111 & 6 &   &  \\ \hline
53 &  8-Feb-2014 20:10 & B & 428, -271 & AR & 211 & 9 & FLARE &  \\ \hline
54 &  8-Feb-2014 13:24 & T & -193, 114 & IP & 84 & 16 &   &  \\ \hline
55 &  8-Feb-2014 13:24 & T & -122, -162 & AR & 181 & 8 &   &  \\ \hline
56 &  9-Feb-2014 16:52 & C & -82, -93 & IP & 162 & 11 & FLARE &  \\ \hline
57 &  9-Feb-2014 16:52 & C & -192, -119 & IP & 109 & 8 & FLARE &  \\ \hline
58 &  9-Feb-2014 16:52 & C & -390, -197 & AR & 232 & 7 & FLARE &  \\ \hline
59 & 11-Feb-2014 17:39 & C & 517, -203 & AR & 101 & 9 & FLARE &  \\ \hline
60 & 12-Feb-2014 13:28 & T & 368, 297 & AR & 184 & 8 & FLARE &  \\ \hline
    \end{tabular}
\caption{Table of the observation details and parameters of the filamet for events 1 to 60. In first column the asterisk indicates that the oscillation is in the same time-distance diagram than in previous case.}\label{table:table_filaments0}
\end{table}
\begin{table}[H]
\centering
    \begin{tabular}{| c | c |c | c | c | c | c | c | c |}
    \hline
    Event \# & Time & Tel. & Pos. (x,y) & Type & L (Mm) & W (Mm) & Trigger & Erupts\\ \hline 
61 & 12-Feb-2014 19:42 & B & 571, -105 & IP & 160 & 14 & FLARE &  \\ \hline
62 & 13-Feb-2014 17:05 & C & 601, 238 & AR & 236 & 6 & FLARE &  \\ \hline
63 & 13-Feb-2014 20:04 & B & 708, -382 & QS & 721 & 33 & FLARE &  \\ \hline
64 & 14-Feb-2014 13:27 & T & 723, 248 & AR & 295 & 9 & FLARE &  \\ \hline
65 & 14-Feb-2014 19:34 & B & -592, -233 & IP & 805 & 8 & FLARE &  \\ \hline
66 & 16-Feb-2014 19:51 & B & -17, -93 & AR & 132 & 7 & FLARE &  \\ \hline
67 & 17-Feb-2014 07:28 & U & -110, -202 & IP & 898 & 14 & FLARE &  \\ \hline
68 & 17-Feb-2014 18:52 & B & 3, -199 & IP & 1051 & 13 &   &  \\ \hline
69 & 19-Feb-2014 19:47 & B & 380, -215 & IP & 913 & 13 & FLARE &  \\ \hline
70 & 22-Feb-2014 07:28 & U & -16, -378 & IP & 359 & 9 &   &  \\ \hline
71 & 23-Feb-2014 16:55 & C & -448, 277 & AR & 120 & 12 & FLARE &  \\ \hline
72 & 24-Feb-2014 16:59 & C & 44, -311 & IP & 219 & 22 &   &  \\ \hline
73 & 24-Feb-2014 16:59 & C & 330, -267 & IP & 200 & 10 &   &  \\ \hline
74 & 25-Feb-2014 16:38 & C & 548, -213 & IP & 447 & 7 &   &  \\ \hline
75 & 25-Feb-2014 16:38 & C & 366, -172 & AR & 114 & 8 &   &  \\ \hline
76 & 25-Feb-2014 16:38 & C & 526, 144 & QS & 46 & 12 &   & Y\\ \hline
77 & 27-Feb-2014 13:22 & T & -1, 164 & IP & 209 & 9 & FLARE &  \\ \hline
78 & 27-Feb-2014 16:57 & C & -203, 350 & IP & 147 & 9 &   &  \\ \hline
79* & 27-Feb-2014 16:57 & C & -203, 350 & IP & 147 & 9 &   &  \\ \hline
80 & 28-Feb-2014 13:22 & T & 125, -328 & QS & 813 & 24 &   &  \\ \hline
81 &  7-Mar-2014 17:16 & C & 590, -74 & IP & 437 & 12 & FLARE &  \\ \hline
82 & 12-Mar-2014 07:29 & U & -567, -237 & IP & 210 & 9 & FLARE &  \\ \hline
83 & 14-Mar-2014 16:58 & C & -812, 265 & AR & 192 & 8 & FLARE &  \\ \hline
84 & 16-Mar-2014 07:22 & U & -539, -213 & IP & 831 & 13 & PE &  \\ \hline
85 & 20-Mar-2014 16:59 & C & -315, -4 & IP & 242 & 9 &   &  \\ \hline
86 & 21-Mar-2014 19:21 & B & 521, 347 & AR & 335 & 11 & PE &  \\ \hline
87 & 23-Mar-2014 16:54 & C & 398, 0 & IP & 320 & 12 & JET &  \\ \hline
88 & 24-Mar-2014 18:57 & B & 586, -13 & IP & 152 & 9 &   &  \\ \hline
89* & 24-Mar-2014 18:57 & B & 586, -13 & IP & 152 & 9 &   &  \\ \hline
90 & 28-Mar-2014 18:09 & B & 683, 41 & IP & 79 & 18 &   &  \\ \hline
91 & 29-Mar-2014 16:57 & C & 329, 372 & IP & 120 & 9 & MW &  \\ \hline
92 & 29-Mar-2014 16:57 & C & 99, -201 & AR & 217 & 6 & FLARE &  \\ \hline
93 & 30-Mar-2014 07:10 & U & 83, -405 & IP & 95 & 10 &   &  \\ \hline
94 & 31-Mar-2014 18:49 & B & 379, -390 & IP & 289 & 14 & FLARE &  \\ \hline
95* & 31-Mar-2014 18:49 & B & 379, -390 & IP & 289 & 14 & FLARE &  \\ \hline
96 &  7-Apr-2014 18:33 & B & 522, 416 & AR & 72 & 6 & FLARE &  \\ \hline
97 &  9-Apr-2014 12:42 & T & 469, 216 & IP & 139 & 9 & FLARE &  \\ \hline
98 &  9-Apr-2014 18:30 & M & 520, 200 & AR & 111 & 8 &   & Y\\ \hline
99 & 10-Apr-2014 05:07 & L & 653, -431 & QS & 458 & 18 &   &  \\ \hline
100 & 14-Apr-2014 16:44 & C & 146, -110 & IP & 248 & 10 &   & Y\\ \hline
101 & 17-Apr-2014 05:01 & L & -648, -161 & IP & 531 & 9 & FLARE &  \\ \hline
102 & 17-Apr-2014 05:01 & L & 297, 447 & QS & 210 & 20 &   &  \\ \hline
103 & 17-Apr-2014 16:52 & C & 5, -324 & IP & 177 & 8 & PE &  \\ \hline
104 & 17-Apr-2014 16:52 & C & 731, -150 & QS & 141 & 21 &   &  \\ \hline
105 & 19-Apr-2014 04:48 & L & 602, 466 & QS & 240 & 17 &   &  \\ \hline
106 & 19-Apr-2014 16:47 & C & -185, -125 & IP & 691 & 11 & FLARE &  \\ \hline
107 & 21-Apr-2014 18:20 & B & 701, -371 & IP & 149 & 7 & FLARE & Y\\ \hline
108* & 21-Apr-2014 18:20 & B & 701, -371 & IP & 149 & 7 & FLARE & Y\\ \hline
109 & 22-Apr-2014 16:41 & C & 61, 296 & IP & 229 & 6 &   &  \\ \hline
110 & 23-Apr-2014 16:49 & C & 264, 294 & IP & 270 & 7 & FLARE &  \\ \hline
111 & 24-Apr-2014 07:05 & U & -262, -263 & AR & 197 & 9 &   &  \\ \hline
112 & 24-Apr-2014 07:05 & U & 205, 148 & IP & 289 & 11 &   &  \\ \hline
113 & 25-Apr-2014 07:10 & U & 424, 132 & IP & 264 & 14 &   &  \\ \hline
114 & 26-Apr-2014 13:27 & T & 465, -370 & IP & 800 & 14 & PE &  \\ \hline
115 &  1-May-2014 18:11 & B & -380, 231 & IP & 290 & 8 & FLARE &  \\ \hline
116 &  1-May-2014 18:11 & B & 753, -385 & IP & 428 & 11 & FLARE &  \\ \hline
117 &  1-May-2014 07:10 & U & -625, -379 & IP & 137 & 12 &   &  \\ \hline
118 &  2-May-2014 07:04 & U & -296, 190 & AR & 212 & 9 & FLARE &  \\ \hline
119* &  2-May-2014 07:04 & U & -296, 190 & AR & 212 & 9 & FLARE &  \\ \hline
120 &  2-May-2014 13:42 & T & 36, -342 & IP & 173 & 12 &   &  \\ \hline
    \end{tabular}
\caption{Same as Table \ref{table:table_filaments0} for events 61 to 120.}\label{table:table_filaments1}
\end{table}
\begin{table}[H]
\centering
    \begin{tabular}{| c | c |c | c | c | c | c | c | c |}
    \hline
    Event \# & Time & Tel. & Pos. (x,y) & Type & L (Mm) & W (Mm) & Trigger & Erupts\\ \hline 
121 &  5-May-2014 17:22 & B & 92, -391 & IP & 291 & 8 &   &  \\ \hline
122 & 10-May-2014 06:31 & U & -499, 417 & IP & 254 & 14 &   &  \\ \hline
123 & 10-May-2014 13:07 & T & -181, -581 & QS & 71 & 19 & PE &  \\ \hline
124 & 11-May-2014 18:06 & B & -610, -486 & QS & 395 & 13 & FLARE &  \\ \hline
125 & 12-May-2014 04:11 & L & -142, 443 & QS & 117 & 15 &   &  \\ \hline
126 & 12-May-2014 04:11 & L & 77, -22 & IP & 192 & 12 & FLARE &  \\ \hline
127 & 12-May-2014 04:11 & L & -494, -634 & QS & 330 & 10 &   &  \\ \hline
128 & 12-May-2014 07:21 & U & 281, -102 & IP & 96 & 9 &   &  \\ \hline
129 & 12-May-2014 18:01 & M & -346, 361 & IP & 164 & 14 &   &  \\ \hline
130 & 13-May-2014 18:07 & B & 323, -26 & IP & 115 & 9 &   &  \\ \hline
131 & 13-May-2014 18:07 & B & -86, -464 & IP & 222 & 18 &   &  \\ \hline
132 & 14-May-2014 07:22 & U & 413, -626 & QS & 149 & 35 &   &  \\ \hline
133 & 15-May-2014 16:43 & C & 285, 142 & IP & 330 & 25 &   &  \\ \hline
134 & 16-May-2014 06:47 & U & 354, 197 & QS & 246 & 33 &   &  \\ \hline
135 & 16-May-2014 13:07 & T & -514, 355 & IP & 142 & 9 & FLARE &  \\ \hline
136 & 17-May-2014 04:54 & L & 521, -70 & IP & 202 & 12 &   &  \\ \hline
137 & 17-May-2014 13:07 & T & 649, 135 & QS & 175 & 14 &   &  \\ \hline
138 & 18-May-2014 19:54 & B & 455, -96 & IP & 133 & 13 &   &  \\ \hline
139 & 23-May-2014 12:58 & T & -101, 240 & IP & 228 & 11 &   &  \\ \hline
140 & 23-May-2014 13:07 & T & 269, -302 & IP & 292 & 9 &   &  \\ \hline
141 & 23-May-2014 19:27 & B & 658, 146 & QS & 164 & 18 &   &  \\ \hline
142 & 23-May-2014 18:10 & B & -577, -192 & IP & 329 & 9 & FLARE &  \\ \hline
143 & 26-May-2014 03:39 & L & 243, -432 & IP & 603 & 11 & FLARE &  \\ \hline
144 & 26-May-2014 17:08 & C & 743, -32 & QS & 97 & 18 &   &  \\ \hline
145 & 26-May-2014 16:45 & C & -348, -260 & IP & 143 & 10 & FLARE &  \\ \hline
146* & 26-May-2014 16:45 & C & -348, -260 & IP & 143 & 10 & FLARE &  \\ \hline
147 & 27-May-2014 04:26 & L & 280, -233 & AR & 121 & 8 &   &  \\ \hline
148 & 27-May-2014 11:24 & T & 552, -404 & IP & 796 & 9 &   &  \\ \hline
149 & 28-May-2014 04:21 & L & -26, -352 & AR & 219 & 7 & FLARE &  \\ \hline
150 & 29-May-2014 04:37 & L & 184, -346 & AR & 183 & 8 &   &  \\ \hline
151 & 30-May-2014 05:31 & L & 371, -353 & AR & 235 & 11 & FLARE &  \\ \hline
152* & 30-May-2014 05:31 & L & 371, -353 & AR & 235 & 11 & FLARE &  \\ \hline
153 & 30-May-2014 11:42 & T & 765, -153 & IP & 152 & 8 & FLARE &  \\ \hline
154* & 30-May-2014 11:42 & T & 765, -153 & IP & 152 & 8 & FLARE &  \\ \hline
155 &  1-Jun-2014 04:36 & L & 692, -354 & AR & 123 & 7 & FLARE & Y\\ \hline
156 &  2-Jun-2014 13:25 & T & 291, 166 & IP & 239 & 19 &   &  \\ \hline
157 &  2-Jun-2014 17:45 & B & -203, 59 & QS & 124 & 19 &   &  \\ \hline
158 &  3-Jun-2014 04:32 & L & -85, 54 & QS & 166 & 15 &   &  \\ \hline
159 &  4-Jun-2014 17:42 & B & -628, -55 & IP & 339 & 10 &   &  \\ \hline
160 &  5-Jun-2014 13:09 & T & 97, -236 & IP & 256 & 9 &   &  \\ \hline
161 &  5-Jun-2014 19:42 & M & -174, 404 & QS & 98 & 12 &   &  \\ \hline
162 &  6-Jun-2014 13:14 & T & 323, -235 & IP & 104 & 12 &   & Y\\ \hline
163 &  7-Jun-2014 04:45 & L & 23, -204 & IP & 108 & 15 &   &  \\ \hline
164 &  7-Jun-2014 13:09 & T & -592, -239 & IP & 206 & 9 &   &  \\ \hline
165* &  7-Jun-2014 13:09 & T & -592, -239 & IP & 206 & 9 &   &  \\ \hline
166 &  8-Jun-2014 13:09 & T & 3, 205 & AR & 290 & 11 &   &  \\ \hline
167 &  8-Jun-2014 17:56 & B & -345, -224 & IP & 355 & 13 &   &  \\ \hline
168 &  9-Jun-2014 13:11 & T & 242, 212 & AR & 229 & 11 &   &  \\ \hline
169* &  9-Jun-2014 13:11 & T & 242, 212 & AR & 229 & 11 &   &  \\ \hline
170 &  9-Jun-2014 13:11 & T & 379, 6 & IP & 326 & 17 &   &  \\ \hline
171 &  9-Jun-2014 13:11 & T & 45, -304 & AR & 166 & 13 &   &  \\ \hline
172* &  9-Jun-2014 13:11 & T & 45, -304 & AR & 166 & 13 &   &  \\ \hline
173 & 10-Jun-2014 18:52 & M & 499, 220 & AR & 73 & 17 & FLARE & Y\\ \hline
174 & 12-Jun-2014 04:33 & L & -328, -575 & QS & 99 & 12 &   &  \\ \hline
175* & 12-Jun-2014 04:33 & L & -328, -575 & QS & 99 & 12 &   &  \\ \hline
176 & 12-Jun-2014 04:46 & U & 289, -344 & IP & 176 & 13 &   &  \\ \hline
177* & 12-Jun-2014 04:46 & U & 289, -344 & IP & 176 & 13 &   &  \\ \hline
178 & 12-Jun-2014 04:33 & L & 438, -202 & IP & 166 & 11 &   &  \\ \hline
179 & 12-Jun-2014 12:51 & T & 469, 490 & QS & 196 & 13 &   &  \\ \hline
180* & 12-Jun-2014 12:51 & T & 469, 490 & QS & 196 & 13 &   &  \\ \hline
    \end{tabular}
\caption{Same as Table \ref{table:table_filaments0} for events 121 to 180.}\label{table:table_filaments2}
\end{table}
\begin{table}[H]
\centering
    \begin{tabular}{| c | c |c | c | c | c | c | c | c |}
    \hline
    Event \# & Time & Tel. & Pos. (x,y) & Type & L (Mm) & W (Mm) & Trigger & Erupts\\ \hline 
181 & 13-Jun-2014 04:30 & L & -163, -574 & QS & 125 & 16 &   &  \\ \hline
182 & 13-Jun-2014 13:11 & T & 599, -297 & IP & 606 & 19 &   &  \\ \hline
183 & 14-Jun-2014 04:40 & L & 672, -294 & IP & 288 & 13 &   &  \\ \hline
184 & 14-Jun-2014 13:10 & T & 745, 55 & IP & 23 & 7 &   &  \\ \hline
185 & 14-Jun-2014 13:10 & T & -527, 490 & QS & 58 & 10 & FLARE &  \\ \hline
186 & 15-Jun-2014 17:51 & M & -345, 512 & IP & 86 & 16 & FLARE &  \\ \hline
187* & 15-Jun-2014 17:51 & M & -345, 512 & IP & 86 & 16 & FLARE &  \\ \hline
188 & 16-Jun-2014 13:11 & T & -836, 5 & IP & 189 & 15 &   &  \\ \hline
189 & 16-Jun-2014 13:11 & T & -214, 521 & IP & 61 & 14 &   &  \\ \hline
190* & 16-Jun-2014 13:11 & T & -214, 521 & IP & 61 & 14 &   &  \\ \hline
191 & 17-Jun-2014 17:56 & B & -334, -510 & QS & 116 & 12 &   &  \\ \hline
192 & 17-Jun-2014 13:12 & T & -757, -7 & IP & 301 & 13 & FLARE &  \\ \hline
193 & 17-Jun-2014 13:12 & T & -39, 520 & IP & 119 & 12 &   &  \\ \hline
194 & 17-Jun-2014 13:12 & T & 695, 385 & IP & 107 & 7 & FLARE &  \\ \hline
195 & 19-Jun-2014 18:44 & M & 241, -231 & IP & 131 & 13 &   &  \\ \hline
196 & 29-Jun-2014 16:50 & C & 130, -200 & IP & 691 & 21 &   &  \\ \hline
    \end{tabular}
\caption{Same as Table \ref{table:table_filaments0} for events 181 to 195.}\label{table:table_filaments3}
\end{table}

\begin{table}[H]
\centering
    \begin{tabular}{| c | c | c | c | c | c | c | c |}
    \hline
    \# & Time Ini. & $\alpha$(${}^\circ$) & $P$ (min.) & $\tau$ (min.) & $\tau/P$ & A (Mm) & V (km s$^{-1}$)\\ \hline 
1 &  1-Jan-2014 13:50 & 16 &     76$\pm$2 & 121$\pm$15 &    1.6$\pm$0.2 &     22$\pm$2&  26.5$\pm$3.6 \\ \hline
2* &  1-Jan-2014 18:56 & 16 &     71$\pm$3 & 173$\pm$107 &    2.4$\pm$2. &      8$\pm$3&  10.7$\pm$7.0 \\ \hline
3 &  4-Jan-2014 20:30 & 26 &     57$\pm$2 & 105$\pm$33 &    1.8$\pm$0.6 &      4$\pm$1&   8.0$\pm$3.5 \\ \hline
4 &  5-Jan-2014 11:43 & 9 &     63$\pm$4 & --- & --- &      5$\pm$3&   9.4$\pm$9.2 \\ \hline
5 &  5-Jan-2014 12:01 & 32 &     52$\pm$1 & --- & --- &      1$\pm$1&   2.7$\pm$6.0 \\ \hline
6* &  5-Jan-2014 17:05 & 32 &     76$\pm$3 & -115$\pm$38 &   -1.5$\pm$0.6 &      3$\pm$1&   3.5$\pm$6.0 \\ \hline
7 &  5-Jan-2014 10:44 & 0 &     63$\pm$3 & -390$\pm$545 &   -6.2$\pm$9. &      2$\pm$1&   3.9$\pm$6.7 \\ \hline
8* &  5-Jan-2014 21:31 & 0 &     43$\pm$3 & 160$\pm$329 &    3.7$\pm$8. &      5$\pm$3&  10.7$\pm$11.1 \\ \hline
9 &  5-Jan-2014 20:13 & 37 &     44$\pm$2 & 38$\pm$9 &    0.9$\pm$0.2 &      5$\pm$2&  11.6$\pm$6.2 \\ \hline
10 &  6-Jan-2014 06:58 & 22 &     68$\pm$1 & 617$\pm$367 &    9.0$\pm$5. &      8$\pm$1&  12.4$\pm$1.9 \\ \hline
11 &  6-Jan-2014 08:02 & 23 &     63$\pm$2 & 94$\pm$23 &    1.5$\pm$0.4 &      4$\pm$1&   5.5$\pm$3.8 \\ \hline
12 &  6-Jan-2014 11:07 & 38 &     40$\pm$3 & 158$\pm$141 &    4.0$\pm$4. &      2$\pm$1&   6.5$\pm$7.9 \\ \hline
13 &  6-Jan-2014 12:27 & 16 &     61$\pm$1 & 118$\pm$18 &    1.9$\pm$0.3 &     10$\pm$1&  15.4$\pm$2.1 \\ \hline
14* &  6-Jan-2014 16:26 & 16 &     56$\pm$6 & 475$\pm$1313 &    8.5$\pm$20 &      3$\pm$2&   5.6$\pm$11.5 \\ \hline
15 &  6-Jan-2014 06:45 & 1 &     64$\pm$1 & --- & --- &      8$\pm$1&  12.8$\pm$1.7 \\ \hline
16 &  7-Jan-2014 05:58 & 14 &     65$\pm$2 & 302$\pm$158 &    4.7$\pm$3. &      4$\pm$1&   6.2$\pm$3.6 \\ \hline
17 &  7-Jan-2014 04:04 & 3 &     46$\pm$1 & 96$\pm$24 &    2.1$\pm$0.5 &      5$\pm$1&  11.0$\pm$4.3 \\ \hline
18 &  7-Jan-2014 04:45 & 23 &     50$\pm$1 & 46$\pm$5 &    0.9$\pm$0.1 &     15$\pm$2&  25.5$\pm$4.7 \\ \hline
19 &  7-Jan-2014 14:16 & 26 &     44$\pm$1 & 83$\pm$8 &    1.9$\pm$0.2 &     14$\pm$2&  31.2$\pm$4.5 \\ \hline
20 &  8-Jan-2014 05:47 & 30 &     50$\pm$2 & 67$\pm$15 &    1.4$\pm$0.3 &     15$\pm$3&  36.5$\pm$8.8 \\ \hline
21 &  8-Jan-2014 04:49 & 2 &     43$\pm$2 & 100$\pm$45 &    2.3$\pm$1. &      4$\pm$1&   9.0$\pm$4.7 \\ \hline
22* &  8-Jan-2014 08:18 & 2 &     35$\pm$2 & --- & --- &      1$\pm$1&   2.5$\pm$9.3 \\ \hline
23 &  8-Jan-2014 03:28 & 24 &     59$\pm$1 & 102$\pm$14 &    1.7$\pm$0.2 &      7$\pm$1&  14.7$\pm$2.9 \\ \hline
24 &  9-Jan-2014 16:56 & 40 &     51$\pm$2 & 97$\pm$29 &    1.9$\pm$0.6 &      5$\pm$2&   9.3$\pm$5.7 \\ \hline
25 &  9-Jan-2014 12:31 & 44 &     46$\pm$3 & 73$\pm$33 &    1.6$\pm$0.8 &      4$\pm$2&  10.8$\pm$7.5 \\ \hline
26 & 10-Jan-2014 11:23 & 45 &     43$\pm$2 & 80$\pm$26 &    1.9$\pm$0.7 &      3$\pm$1&  10.2$\pm$5.5 \\ \hline
27 & 11-Jan-2014 12:37 & 24 &     66$\pm$3 & 114$\pm$51 &    1.7$\pm$0.8 &      2$\pm$1&   2.7$\pm$7.5 \\ \hline
28 & 15-Jan-2014 17:38 & 35 &     39$\pm$1 & 78$\pm$13 &    2.0$\pm$0.3 &      7$\pm$2&  16.2$\pm$5.8 \\ \hline
29 & 16-Jan-2014 08:29 & 13 &     53$\pm$2 & 232$\pm$168 &    4.4$\pm$3. &      2$\pm$1&   3.1$\pm$5.4 \\ \hline
30 & 16-Jan-2014 20:46 & 12 &     57$\pm$2 & 58$\pm$10 &    1.0$\pm$0.2 &      9$\pm$2&  18.8$\pm$6.1 \\ \hline
31 & 24-Jan-2014 17:17 & 26 &     77$\pm$4 & 663$\pm$1546 &    8.6$\pm$20 &      1$\pm$1&   1.6$\pm$9.5 \\ \hline
32 & 24-Jan-2014 21:00 & 21 &     94$\pm$2 & --- & --- &     17$\pm$2&  20.4$\pm$3.5 \\ \hline
33 & 25-Jan-2014 07:16 & 34 &     72$\pm$2 & 94$\pm$20 &    1.3$\pm$0.3 &      4$\pm$1&   4.4$\pm$3.0 \\ \hline
34 & 25-Jan-2014 12:19 & 37 &     34$\pm$3 & 103$\pm$185 &    3.0$\pm$6. &      1$\pm$1&   3.1$\pm$12.9 \\ \hline
35* & 25-Jan-2014 14:47 & 37 &     50$\pm$2 & 46$\pm$6 &    0.9$\pm$0.1 &      9$\pm$2&  25.2$\pm$6.5 \\ \hline
36 & 25-Jan-2014 16:06 & 15 &     87$\pm$1 & 185$\pm$26 &    2.1$\pm$0.3 &     23$\pm$2&  28.6$\pm$3.1 \\ \hline
37 & 26-Jan-2014 08:09 & 36 &     74$\pm$1 & 343$\pm$117 &    4.6$\pm$2. &      5$\pm$1&   7.8$\pm$2.7 \\ \hline
38 & 27-Jan-2014 14:28 & 27 &     45$\pm$3 & -189$\pm$216 &   -4.2$\pm$5. &      2$\pm$1&   4.5$\pm$10.7 \\ \hline
39* & 27-Jan-2014 20:02 & 27 &     55$\pm$2 & 172$\pm$112 &    3.1$\pm$2. &      4$\pm$1&   7.4$\pm$5.2 \\ \hline
40 & 27-Jan-2014 16:04 & 39 &     52$\pm$1 & 327$\pm$154 &    6.3$\pm$3. &      2$\pm$1&   4.4$\pm$3.8 \\ \hline
41 & 28-Jan-2014 20:14 & 29 &     59$\pm$3 & 101$\pm$35 &    1.7$\pm$0.6 &      6$\pm$2&  11.0$\pm$5.1 \\ \hline
42 & 29-Jan-2014 08:23 & 7 &     59$\pm$2 & 131$\pm$34 &    2.2$\pm$0.6 &      4$\pm$1&   8.2$\pm$3.2 \\ \hline
43 & 29-Jan-2014 19:36 & 35 &     52$\pm$2 & 424$\pm$617 &    8.2$\pm$10 &      2$\pm$1&   4.7$\pm$8.1 \\ \hline
44 & 29-Jan-2014 18:35 & 26 &     69$\pm$2 & --- & --- &      4$\pm$1&   6.5$\pm$4.2 \\ \hline
45 & 30-Jan-2014 05:20 & 28 &     62$\pm$1 & 117$\pm$14 &    1.9$\pm$0.2 &      6$\pm$1&  11.7$\pm$2.6 \\ \hline
46 & 31-Jan-2014 13:40 & 28 &     44$\pm$1 & 72$\pm$10 &    1.6$\pm$0.2 &     10$\pm$2&  21.0$\pm$3.7 \\ \hline
47 &  1-Feb-2014 07:55 & 39 &     59$\pm$10 & 159$\pm$264 &    2.7$\pm$5. &      1$\pm$1&   2.4$\pm$15.4 \\ \hline
48 &  5-Feb-2014 13:53 & 26 &     52$\pm$2 & 150$\pm$58 &    2.9$\pm$1. &      4$\pm$2&   7.7$\pm$6.4 \\ \hline
49 &  5-Feb-2014 20:03 & 44 &     54$\pm$3 & --- & --- &      2$\pm$1&   3.7$\pm$8.7 \\ \hline
50 &  6-Feb-2014 16:45 & 30 &     68$\pm$2 & 78$\pm$13 &    1.2$\pm$0.2 &     11$\pm$2&  15.6$\pm$4.5 \\ \hline
51 &  6-Feb-2014 17:26 & 34 &     60$\pm$2 & 105$\pm$20 &    1.8$\pm$0.4 &      9$\pm$2&  12.2$\pm$4.2 \\ \hline
52 &  7-Feb-2014 09:44 & 30 &     67$\pm$2 & 140$\pm$29 &    2.1$\pm$0.5 &      8$\pm$2&  13.0$\pm$3.5 \\ \hline
53 &  8-Feb-2014 21:28 & 64 &     47$\pm$1 & 61$\pm$5 &    1.3$\pm$0.01 &     13$\pm$1&  22.9$\pm$2.5 \\ \hline
54 &  8-Feb-2014 10:57 & 68 &     36$\pm$1 & 194$\pm$140 &    5.3$\pm$4. &      1$\pm$1&   2.8$\pm$5.9 \\ \hline
55 &  8-Feb-2014 10:34 & 25 &     36$\pm$1 & --- & --- &      2$\pm$1&   5.5$\pm$7.0 \\ \hline
56 &  9-Feb-2014 14:10 & 35 &     63$\pm$1 & 78$\pm$8 &    1.2$\pm$0.1 &     22$\pm$3&  46.2$\pm$5.0 \\ \hline
57 &  9-Feb-2014 15:04 & 42 &     37$\pm$1 & --- & --- &      1$\pm$1&   3.9$\pm$7.0 \\ \hline
58 &  9-Feb-2014 11:48 & 32 &     47$\pm$1 & 82$\pm$20 &    1.8$\pm$0.4 &      5$\pm$2&  14.0$\pm$5.0 \\ \hline
59 & 11-Feb-2014 17:13 & 35 &     57$\pm$1 & 72$\pm$6 &    1.3$\pm$0.01 &     14$\pm$2&  25.7$\pm$3.3 \\ \hline
60 & 12-Feb-2014 14:19 & 18 &     55$\pm$1 & 283$\pm$170 &    5.1$\pm$3. &      3$\pm$1&   6.3$\pm$4.7 \\ \hline
    \end{tabular}
\caption{Table of oscillation best fit parameters for 1 to 60. In first column the asterisk indicates that the oscillation is in the same time-distance diagram than in previous case.}\label{table:table0}
\end{table}
\begin{table}[H]
\centering
    \begin{tabular}{| c | c | c | c | c | c | c | c |}
    \hline
    \# & Time Ini. & $\alpha$(${}^\circ$) & $P$ (min.) & $\tau$ (min.) & $\tau/P$ & A (Mm) & V (km s$^{-1}$)\\ \hline 
61 & 12-Feb-2014 17:45 & 24 &     73$\pm$3 & 260$\pm$144 &    3.6$\pm$2. &      4$\pm$1&   4.6$\pm$3.9 \\ \hline
62 & 13-Feb-2014 18:05 & 19 &     48$\pm$1 & 118$\pm$24 &    2.5$\pm$0.5 &     10$\pm$2&  20.9$\pm$4.2 \\ \hline
63 & 13-Feb-2014 20:39 & 2 &    103$\pm$1 & 175$\pm$12 &    1.7$\pm$0.1 &     47$\pm$2&  48.5$\pm$2.4 \\ \hline
64 & 14-Feb-2014 08:38 & 21 &     56$\pm$1 & 79$\pm$6 &    1.4$\pm$0.1 &     31$\pm$2&  52.1$\pm$3.7 \\ \hline
65 & 14-Feb-2014 20:06 & 60 &     63$\pm$3 & -40$\pm$7 &   -0.6$\pm$0.1 &     10$\pm$1&  17.0$\pm$11.5 \\ \hline
66 & 16-Feb-2014 17:26 & 37 &     40$\pm$2 & 188$\pm$255 &    4.7$\pm$7. &      2$\pm$2&   6.7$\pm$11.9 \\ \hline
67 & 17-Feb-2014 02:42 & 12 &     92$\pm$3 & 122$\pm$18 &    1.3$\pm$0.2 &      9$\pm$1&  14.9$\pm$3.1 \\ \hline
68 & 17-Feb-2014 19:59 & 25 &     39$\pm$3 & 117$\pm$149 &    3.0$\pm$4. &      3$\pm$2&   8.1$\pm$11.6 \\ \hline
69 & 19-Feb-2014 15:38 & 16 &     76$\pm$2 & 88$\pm$13 &    1.2$\pm$0.2 &     20$\pm$3&  28.7$\pm$6.1 \\ \hline
70 & 22-Feb-2014 09:04 & 23 &     51$\pm$2 & 51$\pm$13 &    1.0$\pm$0.3 &     10$\pm$3&  25.7$\pm$8.2 \\ \hline
71 & 23-Feb-2014 16:04 & 26 &     34$\pm$1 & 42$\pm$8 &    1.2$\pm$0.2 &     11$\pm$2&  34.7$\pm$6.1 \\ \hline
72 & 24-Feb-2014 15:03 & 22 &     73$\pm$1 & 416$\pm$127 &    5.7$\pm$2. &      7$\pm$1&  10.5$\pm$2.9 \\ \hline
73 & 24-Feb-2014 19:13 & 20 &     54$\pm$2 & 155$\pm$81 &    2.9$\pm$2. &      5$\pm$2&   9.3$\pm$7.9 \\ \hline
74 & 25-Feb-2014 15:23 & 50 &     65$\pm$2 & 102$\pm$21 &    1.6$\pm$0.3 &      5$\pm$1&   9.8$\pm$3.5 \\ \hline
75 & 25-Feb-2014 15:28 & 16 &     65$\pm$3 & 80$\pm$27 &    1.2$\pm$0.5 &      4$\pm$2&   8.9$\pm$6.5 \\ \hline
76 & 25-Feb-2014 12:37 & 15 &     78$\pm$2 & --- & --- &      3$\pm$1&   4.8$\pm$3.3 \\ \hline
77 & 27-Feb-2014 09:48 & 7 &     57$\pm$1 & 75$\pm$8 &    1.3$\pm$0.2 &     29$\pm$3&  54.6$\pm$6.3 \\ \hline
78 & 27-Feb-2014 15:59 & 34 &     53$\pm$2 & 136$\pm$90 &    2.6$\pm$2. &      4$\pm$2&   9.6$\pm$6.2 \\ \hline
79* & 27-Feb-2014 19:41 & 34 &     51$\pm$1 & 87$\pm$20 &    1.7$\pm$0.4 &      7$\pm$2&  18.9$\pm$5.9 \\ \hline
80 & 28-Feb-2014 12:08 & 10 &    109$\pm$2 & 312$\pm$106 &    2.9$\pm$1. &      9$\pm$2&   9.6$\pm$2.8 \\ \hline
81 &  7-Mar-2014 16:44 & 21 &     49$\pm$1 & 423$\pm$345 &    8.6$\pm$7. &      4$\pm$1&   9.5$\pm$3.7 \\ \hline
82 & 12-Mar-2014 09:02 & 1 &    105$\pm$10 & 115$\pm$52 &    1.1$\pm$0.6 &     12$\pm$3&  20.5$\pm$11.1 \\ \hline
83 & 14-Mar-2014 13:31 & 35 &     70$\pm$2 & 89$\pm$14 &    1.3$\pm$0.2 &      8$\pm$2&  14.4$\pm$3.0 \\ \hline
84 & 16-Mar-2014 03:31 & 37 &     46$\pm$1 & 74$\pm$15 &    1.6$\pm$0.4 &     11$\pm$2&  26.4$\pm$5.9 \\ \hline
85 & 20-Mar-2014 16:02 & 19 &     93$\pm$2 & 73$\pm$8 &    0.8$\pm$0.09 &     16$\pm$2&  21.8$\pm$3.8 \\ \hline
86 & 21-Mar-2014 17:23 & 7 &     61$\pm$2 & 89$\pm$25 &    1.5$\pm$0.4 &     13$\pm$2&  19.6$\pm$5.1 \\ \hline
87 & 23-Mar-2014 16:05 & 22 &     76$\pm$1 & 71$\pm$5 &    0.9$\pm$0.07 &     16$\pm$2&  29.6$\pm$3.5 \\ \hline
88 & 24-Mar-2014 19:17 & 29 &     44$\pm$2 & 50$\pm$8 &    1.2$\pm$0.2 &      7$\pm$1&  16.0$\pm$4.0 \\ \hline
89* & 24-Mar-2014 20:57 & 29 &     38$\pm$2 & 251$\pm$269 &    6.6$\pm$7. &      1$\pm$1&   3.8$\pm$6.5 \\ \hline
90 & 28-Mar-2014 17:52 & 0 &     81$\pm$3 & 135$\pm$32 &    1.7$\pm$0.4 &      8$\pm$2&  10.8$\pm$3.9 \\ \hline
91 & 29-Mar-2014 18:23 & 26 &     58$\pm$1 & 108$\pm$12 &    1.9$\pm$0.2 &     11$\pm$1&  19.3$\pm$2.3 \\ \hline
92 & 29-Mar-2014 15:50 & 12 &     87$\pm$2 & 139$\pm$20 &    1.6$\pm$0.3 &     31$\pm$3&  38.8$\pm$4.4 \\ \hline
93 & 30-Mar-2014 03:53 & 10 &     59$\pm$5 & --- & --- &      8$\pm$2&  15.7$\pm$9.2 \\ \hline
94 & 31-Mar-2014 16:11 & 30 &     51$\pm$2 & 43$\pm$11 &    0.8$\pm$0.2 &      6$\pm$2&  18.9$\pm$7.0 \\ \hline
95* & 31-Mar-2014 19:38 & 30 &     58$\pm$1 & 325$\pm$62 &    5.6$\pm$1. &     17$\pm$2&  32.5$\pm$2.5 \\ \hline
96 &  7-Apr-2014 21:46 & 19 &     41$\pm$1 & 70$\pm$9 &    1.7$\pm$0.2 &     10$\pm$1&  24.2$\pm$2.8 \\ \hline
97 &  9-Apr-2014 11:21 & 33 &     56$\pm$1 & 71$\pm$11 &    1.3$\pm$0.2 &     10$\pm$2&  24.9$\pm$4.5 \\ \hline
98 &  9-Apr-2014 18:19 & 29 &     51$\pm$4 & 103$\pm$110 &    2.0$\pm$2. &      3$\pm$2&   6.8$\pm$10.6 \\ \hline
99 & 10-Apr-2014 06:14 & 20 &     52$\pm$1 & 94$\pm$13 &    1.8$\pm$0.3 &      8$\pm$1&  13.7$\pm$2.9 \\ \hline
100 & 14-Apr-2014 12:16 & 7 &     50$\pm$1 & 53$\pm$4 &    1.0$\pm$0.09 &     20$\pm$2&  32.4$\pm$2.7 \\ \hline
101 & 17-Apr-2014 05:20 & 25 &     52$\pm$1 & 83$\pm$8 &    1.6$\pm$0.2 &      9$\pm$1&  22.8$\pm$2.8 \\ \hline
102 & 17-Apr-2014 02:21 & 3 &     92$\pm$2 & 179$\pm$35 &    1.9$\pm$0.4 &      8$\pm$1&   9.7$\pm$2.4 \\ \hline
103 & 17-Apr-2014 17:41 & 39 &     56$\pm$1 & 61$\pm$7 &    1.1$\pm$0.1 &     11$\pm$2&  25.5$\pm$3.7 \\ \hline
104 & 17-Apr-2014 18:38 & 8 &     73$\pm$4 & --- & --- &      1$\pm$1&   2.1$\pm$7.4 \\ \hline
105 & 19-Apr-2014 02:57 & 0 &     59$\pm$2 & 118$\pm$34 &    2.0$\pm$0.6 &      9$\pm$2&  12.9$\pm$4.4 \\ \hline
106 & 19-Apr-2014 12:44 & 26 &     65$\pm$1 & 114$\pm$20 &    1.8$\pm$0.3 &      8$\pm$2&  15.0$\pm$4.3 \\ \hline
107 & 21-Apr-2014 18:20 & 20 &     50$\pm$1 & 155$\pm$34 &    3.1$\pm$0.7 &      4$\pm$1&   6.6$\pm$2.2 \\ \hline
108* & 21-Apr-2014 21:29 & 20 &     40$\pm$3 & -97$\pm$70 &   -2.4$\pm$2. &      2$\pm$1&   5.6$\pm$9.3 \\ \hline
109 & 22-Apr-2014 12:17 & 11 &     61$\pm$1 & 110$\pm$19 &    1.8$\pm$0.3 &      5$\pm$1&   8.5$\pm$3.4 \\ \hline
110 & 23-Apr-2014 14:32 & 26 &     51$\pm$4 & 45$\pm$15 &    0.9$\pm$0.3 &      7$\pm$2&  10.5$\pm$7.4 \\ \hline
111 & 24-Apr-2014 04:04 & 16 &     43$\pm$1 & 232$\pm$42 &    5.3$\pm$0.1 &      8$\pm$1&  18.5$\pm$3.3 \\ \hline
112 & 24-Apr-2014 10:53 & 21 &     41$\pm$2 & 37$\pm$6 &    0.9$\pm$0.1 &     13$\pm$2&  27.7$\pm$6.2 \\ \hline
113 & 25-Apr-2014 03:17 & 16 &     58$\pm$2 & 276$\pm$189 &    4.7$\pm$3. &      2$\pm$1&   3.0$\pm$6.4 \\ \hline
114 & 26-Apr-2014 12:40 & 54 &     47$\pm$3 & 55$\pm$21 &    1.2$\pm$0.5 &      6$\pm$2&  11.0$\pm$8.7 \\ \hline
115 &  1-May-2014 17:40 & 8 &     70$\pm$1 & 78$\pm$4 &    1.1$\pm$0.06 &     25$\pm$2&  48.4$\pm$2.8 \\ \hline
116 &  1-May-2014 16:17 & 23 &     52$\pm$1 & 218$\pm$103 &    4.2$\pm$2. &      3$\pm$1&   6.4$\pm$4.9 \\ \hline
117 &  1-May-2014 03:08 & 81 &     52$\pm$2 & 125$\pm$44 &    2.4$\pm$0.9 &      2$\pm$1&   3.3$\pm$4.8 \\ \hline
118 &  2-May-2014 02:17 & 8 &     76$\pm$1 & 377$\pm$144 &    5.0$\pm$2. &      6$\pm$1&   7.0$\pm$2.4 \\ \hline
119* &  2-May-2014 09:08 & 8 &     72$\pm$5 & 110$\pm$89 &    1.5$\pm$1. &      5$\pm$3&   9.3$\pm$9.8 \\ \hline
120 &  2-May-2014 11:04 & 48 &     50$\pm$4 & 51$\pm$19 &    1.0$\pm$0.4 &      4$\pm$2&   6.1$\pm$8.6 \\ \hline
    \end{tabular}
\caption{Same as Table \ref{table:table0} for events 61 to 120. }\label{table:table1}
\end{table}
\begin{table}[H]
\centering
    \begin{tabular}{| c | c | c | c | c | c | c | c |}
    \hline
    \# & Time Ini. & $\alpha$(${}^\circ$) & $P$ (min.) & $\tau$ (min.) & $\tau/P$ & A (Mm) & V (km s$^{-1}$)\\ \hline 
121 &  5-May-2014 16:13 & 15 &     35$\pm$3 & 26$\pm$8 &    0.7$\pm$0.3 &      6$\pm$2&  14.2$\pm$9.1 \\ \hline
122 & 10-May-2014 08:01 & 18 &     81$\pm$3 & 198$\pm$104 &    2.4$\pm$1. &      3$\pm$1&   3.7$\pm$5.8 \\ \hline
123 & 10-May-2014 09:14 & 72 &     69$\pm$3 & 119$\pm$58 &    1.7$\pm$0.9 &      3$\pm$2&   4.6$\pm$6.5 \\ \hline
124 & 11-May-2014 18:47 & 9 &     64$\pm$3 & 46$\pm$8 &    0.7$\pm$0.2 &     18$\pm$3&  24.0$\pm$5.8 \\ \hline
125 & 12-May-2014 02:40 & 18 &     65$\pm$6 & 86$\pm$52 &    1.3$\pm$0.9 &      4$\pm$2&   5.3$\pm$9.1 \\ \hline
126 & 12-May-2014 01:19 & 31 &     88$\pm$2 & 212$\pm$55 &    2.4$\pm$0.7 &      8$\pm$2&  12.3$\pm$3.3 \\ \hline
127 & 12-May-2014 06:20 & 21 &     32$\pm$2 & 34$\pm$9 &    1.1$\pm$0.3 &      9$\pm$3&  25.0$\pm$9.1 \\ \hline
128 & 12-May-2014 08:22 & 7 &     55$\pm$2 & 123$\pm$35 &    2.2$\pm$0.7 &      5$\pm$1&   8.4$\pm$4.0 \\ \hline
129 & 12-May-2014 01:26 & 15 &     43$\pm$2 & 50$\pm$16 &    1.2$\pm$0.4 &      3$\pm$1&   8.0$\pm$5.8 \\ \hline
130 & 13-May-2014 18:26 & 16 &     50$\pm$3 & 72$\pm$25 &    1.4$\pm$0.5 &      4$\pm$1&   6.8$\pm$5.6 \\ \hline
131 & 13-May-2014 14:46 & 74 &     77$\pm$2 & 185$\pm$83 &    2.4$\pm$1. &      2$\pm$1&   2.4$\pm$3.5 \\ \hline
132 & 14-May-2014 07:13 & 87 &     99$\pm$5 & 227$\pm$138 &    2.3$\pm$1. &      4$\pm$2&   3.6$\pm$5.6 \\ \hline
133 & 15-May-2014 18:25 & 44 &     76$\pm$2 & 87$\pm$12 &    1.1$\pm$0.2 &      8$\pm$1&  12.9$\pm$2.7 \\ \hline
134 & 16-May-2014 02:25 & 35 &     85$\pm$5 & -224$\pm$195 &   -2.6$\pm$2. &      2$\pm$1&   4.5$\pm$9.0 \\ \hline
135 & 16-May-2014 11:49 & 7 &     50$\pm$1 & --- & --- &     11$\pm$1&  24.7$\pm$3.2 \\ \hline
136 & 17-May-2014 02:37 & 2 &     59$\pm$3 & 218$\pm$219 &    3.7$\pm$4. &      3$\pm$2&   5.1$\pm$7.4 \\ \hline
137 & 17-May-2014 10:35 & 20 &     58$\pm$2 & 78$\pm$12 &    1.4$\pm$0.2 &     19$\pm$2&  29.5$\pm$5.1 \\ \hline
138 & 18-May-2014 18:22 & 38 &     57$\pm$1 & 49$\pm$6 &    0.9$\pm$0.1 &      9$\pm$2&  23.5$\pm$4.3 \\ \hline
139 & 23-May-2014 15:06 & 48 &     56$\pm$2 & 179$\pm$111 &    3.2$\pm$2. &      2$\pm$1&   4.1$\pm$5.4 \\ \hline
140 & 23-May-2014 14:58 & 18 &     48$\pm$2 & 82$\pm$20 &    1.7$\pm$0.4 &      8$\pm$2&  13.8$\pm$5.2 \\ \hline
141 & 23-May-2014 14:14 & 4 &     48$\pm$3 & 213$\pm$216 &    4.4$\pm$5. &      2$\pm$1&   4.0$\pm$8.5 \\ \hline
142 & 23-May-2014 18:01 & 37 &     46$\pm$2 & 32$\pm$7 &    0.7$\pm$0.2 &      6$\pm$2&  20.1$\pm$7.2 \\ \hline
143 & 26-May-2014 02:38 & 13 &     62$\pm$2 & 53$\pm$6 &    0.9$\pm$0.01 &     15$\pm$2&  21.2$\pm$2.8 \\ \hline
144 & 26-May-2014 17:52 & 62 &     53$\pm$2 & 180$\pm$145 &    3.4$\pm$3. &      3$\pm$1&   6.9$\pm$6.2 \\ \hline
145 & 26-May-2014 13:31 & 50 &     39$\pm$2 & 48$\pm$13 &    1.2$\pm$0.4 &      5$\pm$2&  14.4$\pm$6.1 \\ \hline
146* & 26-May-2014 17:38 & 50 &     36$\pm$2 & 119$\pm$146 &    3.3$\pm$4. &      1$\pm$1&   3.4$\pm$9.4 \\ \hline
147 & 27-May-2014 03:16 & 1 &     70$\pm$2 & 444$\pm$385 &    6.3$\pm$6. &      3$\pm$1&   4.5$\pm$5.4 \\ \hline
148 & 27-May-2014 09:16 & 36 &     34$\pm$3 & 40$\pm$22 &    1.2$\pm$0.7 &      2$\pm$1&   4.8$\pm$10.0 \\ \hline
149 & 28-May-2014 00:34 & 42 &     75$\pm$6 & 467$\pm$1263 &    6.2$\pm$20 &      2$\pm$2&   3.5$\pm$9.2 \\ \hline
150 & 29-May-2014 01:16 & 27 &     42$\pm$1 & 86$\pm$13 &    2.1$\pm$0.3 &      8$\pm$1&  21.2$\pm$4.3 \\ \hline
151 & 30-May-2014 00:01 & 36 &     52$\pm$2 & 89$\pm$23 &    1.7$\pm$0.5 &      4$\pm$1&   6.8$\pm$5.1 \\ \hline
152* & 30-May-2014 05:37 & 36 &     66$\pm$3 & -163$\pm$105 &   -2.5$\pm$2. &      3$\pm$1&   4.0$\pm$7.8 \\ \hline
153 & 30-May-2014 11:43 & 36 &     45$\pm$12 & --- & --- &      1$\pm$3&   4.1$\pm$30.4 \\ \hline
154* & 30-May-2014 13:42 & 36 &     37$\pm$3 & 49$\pm$25 &    1.3$\pm$0.7 &      7$\pm$4&  17.2$\pm$15.4 \\ \hline
155 &  1-Jun-2014 02:06 & 32 &     32$\pm$13 & 101$\pm$859 &    3.1$\pm$30 &      0$\pm$2&   1.7$\pm$52.6 \\ \hline
156 &  2-Jun-2014 07:05 & 51 &     73$\pm$3 & -291$\pm$236 &   -4.0$\pm$3. &      2$\pm$1&   3.6$\pm$5.9 \\ \hline
157 &  2-Jun-2014 18:26 & 14 &     84$\pm$2 & --- & --- &      2$\pm$1&   2.4$\pm$3.9 \\ \hline
158 &  3-Jun-2014 02:33 & 19 &     74$\pm$7 & 557$\pm$2260 &    7.5$\pm$30 &      1$\pm$1&   1.2$\pm$13.5 \\ \hline
159 &  4-Jun-2014 17:30 & 33 &     79$\pm$2 & 155$\pm$29 &    2.0$\pm$0.4 &      6$\pm$1&   6.6$\pm$2.6 \\ \hline
160 &  5-Jun-2014 10:20 & 50 &     60$\pm$1 & 272$\pm$73 &    4.6$\pm$1. &      2$\pm$1&   3.6$\pm$3.2 \\ \hline
161 &  5-Jun-2014 19:57 & 16 &     49$\pm$2 & 51$\pm$13 &    1.1$\pm$0.3 &      5$\pm$2&  10.3$\pm$6.1 \\ \hline
162 &  6-Jun-2014 10:46 & 2 &     48$\pm$3 & --- & --- &      2$\pm$1&   3.9$\pm$7.0 \\ \hline
163 &  7-Jun-2014 00:36 & 64 &     45$\pm$3 & 33$\pm$9 &    0.7$\pm$0.2 &      9$\pm$3&  16.2$\pm$13.1 \\ \hline
164 &  7-Jun-2014 07:26 & 42 &     52$\pm$3 & 106$\pm$62 &    2.1$\pm$1. &      3$\pm$1&   5.9$\pm$6.4 \\ \hline
165* &  7-Jun-2014 11:26 & 42 &     58$\pm$2 & 123$\pm$39 &    2.1$\pm$0.7 &      5$\pm$2&   9.1$\pm$4.9 \\ \hline
166 &  8-Jun-2014 10:17 & 13 &     56$\pm$5 & 46$\pm$20 &    0.8$\pm$0.4 &      5$\pm$4&  11.3$\pm$11.9 \\ \hline
167 &  8-Jun-2014 16:55 & 29 &     55$\pm$3 & 193$\pm$167 &    3.5$\pm$3. &      2$\pm$1&   3.6$\pm$8.0 \\ \hline
168 &  9-Jun-2014 07:32 & 34 &     47$\pm$2 & 64$\pm$16 &    1.3$\pm$0.4 &      6$\pm$2&  11.7$\pm$5.2 \\ \hline
169* &  9-Jun-2014 11:55 & 34 &     40$\pm$1 & 84$\pm$21 &    2.1$\pm$0.5 &      6$\pm$1&  12.4$\pm$3.5 \\ \hline
170 &  9-Jun-2014 14:46 & 40 &     58$\pm$4 & --- & --- &      1$\pm$1&   2.2$\pm$17.5 \\ \hline
171 &  9-Jun-2014 09:34 & 1 &     62$\pm$3 & -202$\pm$226 &   -3.2$\pm$4. &      1$\pm$1&   2.6$\pm$9.3 \\ \hline
172* &  9-Jun-2014 13:59 & 1 &     57$\pm$2 & 280$\pm$189 &    4.9$\pm$3. &      3$\pm$1&   5.5$\pm$5.3 \\ \hline
173 & 10-Jun-2014 18:19 & 35 &     36$\pm$1 & 84$\pm$20 &    2.4$\pm$0.6 &      6$\pm$1&  17.3$\pm$4.8 \\ \hline
174 & 12-Jun-2014 00:01 & 20 &     63$\pm$1 & 275$\pm$118 &    4.4$\pm$2. &      5$\pm$1&   9.9$\pm$3.3 \\ \hline
175* & 12-Jun-2014 04:31 & 20 &     45$\pm$1 & 207$\pm$134 &    4.6$\pm$3. &      1$\pm$1&   2.8$\pm$5.1 \\ \hline
176 & 12-Jun-2014 01:50 & 21 &     67$\pm$3 & --- & --- &      3$\pm$1&   5.1$\pm$5.7 \\ \hline
177* & 12-Jun-2014 05:37 & 21 &     54$\pm$3 & 56$\pm$35 &    1.0$\pm$0.7 &     13$\pm$5&  15.1$\pm$9.5 \\ \hline
178 & 12-Jun-2014 00:30 & 47 &     67$\pm$2 & 118$\pm$31 &    1.8$\pm$0.5 &      4$\pm$1&   5.3$\pm$5.1 \\ \hline
179 & 12-Jun-2014 13:11 & 13 &     52$\pm$1 & 57$\pm$8 &    1.1$\pm$0.2 &     10$\pm$2&  22.1$\pm$5.0 \\ \hline
180* & 12-Jun-2014 14:10 & 13 &     51$\pm$1 & 142$\pm$48 &    2.8$\pm$0.1 &      4$\pm$1&   8.3$\pm$4.1 \\ \hline
    \end{tabular}
\caption{Same as Table \ref{table:table0} for events 121 to 180. }\label{table:table2}
\end{table}
\begin{table}[H]
\centering
    \begin{tabular}{| c | c | c | c | c | c | c | c |}
    \hline
    \# & Time Ini. & $\alpha$(${}^\circ$) & $P$ (min.) & $\tau$ (min.) & $\tau/P$ & A (Mm) & V (km s$^{-1}$)\\ \hline 
181 & 13-Jun-2014 05:38 & 28 &     60$\pm$5 & --- & --- &      1$\pm$1&   2.3$\pm$12.9 \\ \hline
182 & 13-Jun-2014 16:37 & 2 &     51$\pm$3 & 245$\pm$353 &    4.8$\pm$7. &      1$\pm$1&   3.2$\pm$8.5 \\ \hline
183 & 14-Jun-2014 01:44 & 1 &     58$\pm$2 & 493$\pm$673 &    8.4$\pm$10 &      1$\pm$1&   1.7$\pm$7.4 \\ \hline
184 & 14-Jun-2014 11:38 & 81 &     61$\pm$4 & 249$\pm$313 &    4.1$\pm$5. &      1$\pm$1&   2.3$\pm$8.5 \\ \hline
185 & 14-Jun-2014 09:46 & 17 &     45$\pm$3 & --- & --- &      2$\pm$2&   6.3$\pm$10.0 \\ \hline
186 & 15-Jun-2014 02:37 & 9 &     44$\pm$6 & --- & --- &      1$\pm$1&   1.4$\pm$19.3 \\ \hline
187* & 15-Jun-2014 21:04 & 9 &     54$\pm$1 & 118$\pm$22 &    2.2$\pm$0.4 &      7$\pm$1&  11.1$\pm$2.6 \\ \hline
188 & 16-Jun-2014 11:43 & 57 &     63$\pm$8 & 237$\pm$356 &    3.7$\pm$6. &      2$\pm$1&   3.9$\pm$12.6 \\ \hline
189 & 16-Jun-2014 08:50 & 88 &     38$\pm$9 & 125$\pm$501 &    3.3$\pm$10 &      0$\pm$1&   1.0$\pm$28.2 \\ \hline
190* & 16-Jun-2014 15:59 & 88 &     41$\pm$3 & 84$\pm$67 &    2.1$\pm$2. &      2$\pm$1&   3.7$\pm$10.1 \\ \hline
191 & 17-Jun-2014 21:14 & 20 &     67$\pm$6 & 39$\pm$9 &    0.6$\pm$0.2 &      4$\pm$2&   4.7$\pm$7.4 \\ \hline
192 & 17-Jun-2014 15:50 & 35 &     57$\pm$1 & 162$\pm$51 &    2.9$\pm$0.9 &      6$\pm$2&  15.0$\pm$4.6 \\ \hline
193 & 17-Jun-2014 09:50 & 65 &     42$\pm$1 & 246$\pm$166 &    5.8$\pm$4. &      1$\pm$1&   2.8$\pm$4.5 \\ \hline
194 & 17-Jun-2014 12:36 & 27 &     51$\pm$2 & 68$\pm$16 &    1.3$\pm$0.3 &      4$\pm$1&   9.0$\pm$4.0 \\ \hline
195 & 19-Jun-2014 18:10 & 14 &     47$\pm$3 & 62$\pm$21 &    1.3$\pm$0.5 &      3$\pm$1&   7.3$\pm$5.1 \\ \hline
196 & 29-Jun-2014 16:21 & 50 &     56$\pm$4 & 61$\pm$28 &    1.1$\pm$0.6 &      5$\pm$3&   9.4$\pm$10.6 \\ \hline
    \end{tabular}
\caption{Same as Table \ref{table:table0} for events 181 to 195. }\label{table:table3}
\end{table}

\section{Time-distance diagrams in Curved slits}\label{annex:curved-slits}

The GONG network telescopes offer fairly good spatial resolution of around 1 arcsec per pixel. However, the seeing conditions at the network telescope locations often limit  the quality of the images, yielding poor effective spatial resolution greater than 1 arcsec. As we discussed in \S \ref{sec:gong-description}, it was necessary to follow the motion of the large-amplitude displacements with curved paths in order to accurately track the entire motion of the filament. 

A time-distance diagram is constructed to follow the motion along the path defined by the artificial slit. In many cases \citep[e.g.,][]{luna2014}, straight slits consisting of rectangles of length $l$ and width $w$ in pixels are placed lengthwise along the path of the motion studied. In order to increase the signal-to-noise ratio, the intensity is averaged along the width $w$, which essentially projects the intensity onto the axis of the slit. The resulting intensity along the slit as a function of time is the time-distance diagram. Using a curved slit is theoretically similar to using a straight slit. In a curved slit the projection is defined along the normal lines to the curved slit axis. Thus, for each pixel, there is a normal line intersecting the slit axis at ($x_{q}$, $y_{q}$) and the distance between the pixel and the slit is $d$ (i.e. between the pixel and ($x_{q}$, $y_{q}$)). A pixel belongs to the slit if $d\le w/2$. 

\begin{figure*}[!h]
\centering\includegraphics[width=0.999\textwidth]{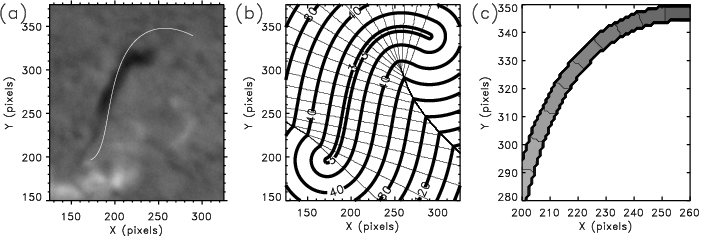}
\caption{(a) H$\alpha$ image of case 1. The filament is located in the center of the image. The white curve is the axis of the slit, $S$. (b) Isocontours of $Dist(i,j)$ and $Int(i,j)$ (Equation \ref{eq:min-dist}) for the image in (a).  (c) Close-up view  of a region of the slit showing the bins used to construct the time-distance diagram. All positions inside the bins have $B(i,j)=q=$constant (Equation \ref{eq:bin-array}). The grey gradient highlights the different bins, with white corresponding to $q=1$ and black to $q=N_{pix}$. \label{fig:anex_f1}}
\end{figure*}

In general an image is described by the 2D function $I(x,y)$, where $I$ is the intensity in the filter considered (H$\alpha$ in our situation) and $x$ and $y$ are the coordinates of each position in the image. We assume, without loss of generality,  that the origin of the coordinates $(x,y)=(0,0)$ is at the left-bottom boundary of the image. These coordinates take entire values of the resolution $\delta$ of the image, then $x=i \, \delta$ and $y= j \, \delta$ where $i$ and $j$ define the position within the image in pixels. Alternatively, the image can be described in pixels $I(i,j)$. 

We first define a sufficiently smooth curve,  $S$, that represents the axis of the slit, by clicking repetitively on the image along the path of the oscillatory motion and fitting these points with a polynomial function of 4th degree. The white line in Figure \ref{fig:anex_f1}(a) shows the curve $S$ obtained for event 1 of the catalog. We divide this curve into segments of length $\delta$ in order to pixelate the curve as the image. The coordinate along the slit axis is then $s=\delta \, q$ where $q$ is a one dimensional array with $N_{pix}=l/\delta$ elements. 

The time-distance diagram consists of $I(t,i_q,j_q) = \langle I(t,A) \rangle$ where $(i_q,j_q)$ is the position of the $q$-segment of the axis of the slit, $t$ is time, $A$ is an area surrounding $(i_q,j_q)$, and the $\langle \dots \rangle$ means the average of the intensity over $A$. The main difficulty is how A is defined. Some authors just defines a square area centered at $(i_q,j_q)$. However, this mixes the intensities from points that are not projected perpendicularly to the slit, and some pixels are projected twice in consecutive segments of the slit. We will define A as the area enclosed by the normal lines between both ends of slit segment, $q$ and within the slit, $d \le w/2$.

For this end, we define two matrices, $Dist(i,j)$ is the distance from any point $(i,j)$ to the closest point along the slit, and $Ind(i,j)$ is the index of that point along the slit. These are
\begin{eqnarray}\label{eq:min-dist}
Dist(i,j)&=&MIN \left(\frac{1}{\delta}\sqrt{(x_i-x_q)^2+(y_j-y_q)^2}\right) \\ \, \label{eq:index-array}
Ind(i,j)&=&q_\mathrm{min} \, .
\end{eqnarray}
where the $MIN$ is the minimum over the $q$ index. We construct these arrays by computing the distance $\sqrt{(x_i-x_q)^2+(y_j-y_q)^2}$ between each pixel of the image, $(i,j)$, and all the positions over the slit, $q$. This is equivalent to computing the distance, $d$, between the pixel and the slit axis. However, this way is much more computationally effective. To calculate $Ind(i,j)$ we then find the value of $q_\mathrm{min}$ that minimizes the distance $\sqrt{(x_i-x_q)^2+(y_j-y_q)^2}$.  This is the position over the slit where the intensity of the image pixel will be projected. We repeat this process for all image pixels and obtain the arrays defined by Equations (\ref{eq:min-dist}) and (\ref{eq:index-array}). Thus, $Dist(i,j)$ is the distance measured in pixels from $(i,j)$ to the curve $S$. The closed thick lines in Figure \ref{fig:anex_f1}(b) are the isolines of the $Dist$ function over the image. We see that each isoline represents the positions of the pixels that are equidistant to the curve segment $S$, i.e. the slit axis. In the example of Figure \ref{fig:anex_f1}, the slits has $w=6$, which corresponds to the area inside the most internal isoline with a distance to $S$ of 3 pixels. In general we define the slit as the set of pixels $(i,j)$ that fulfill the condition $Dist(i,j) \le w/2$. Thus, to select the pixels inside the slit, we define a masking function
\begin{equation}\label{eq:bin-array}
Mask(i,j)=
\begin{cases}
1 \, , & \text{if } Dist(i,j) \le w/2 \\
0 \, , &\text{if } Dist(i,j) > w/2 .\\
\end{cases}
\end{equation}
The thin straight lines in Figure \ref{fig:anex_f1}(b) plot are the isolines of $Ind$ matrix. This isolines coincide with the normal lines to the slit axis.
We define a new function
\begin{equation}
B(i,j)=Mask(i,j) \times Ind(i,j) \, .
\end{equation}
The values in this array are zero outside the slit and range from 1 to $N_{pix}$ in the slit. In this way we have binned the regions of the image that are going to be averaged over the $q$-position of the slit. We clearly see these bins of constant $B(i,j)$  in Figure \ref{fig:anex_f1}(c), as well as the bins defined by the area inside the region formed by the isolines of $Dist$ and $Int$. Then the intensity over the slit, $I(q)$, is the average of the intensity over the bin where $B(i,j)=q$, that is
\begin{equation}
I(q)=\langle I\left(B(i,j)=q\right) \rangle \, .
\end{equation}
This technique can also be used for straight slits to reduce the computational time,  because  the images do not need to be rotated in order to align the $x$- or $y$-axis with the direction of the slit. Repeating this procedure for each image within the temporal sequence, we obtain the time-distance diagram, $I(t,q)$.

One problem with this technique is that the function $Ind$ is multivalued in some regions; i.e., one pixel is equidistant to several points within the slit. These problematic regions are located at the intersections of the isolines of $Ind$. Figure \ref{fig:anex_f1}(b) reveals two lines of multivalued points which coincide with the center of curvature of one segment of the slit. However, we avoid these regions by selecting sufficiently smooth curves with the center of curvature outside the slit area.

\end{document}